\documentstyle[12pt,preprint,aps,floats,epsf]{revtex}
\tighten

\begin{document}

\newcommand{\newc}{\newcommand}

\newc{\gsim}{\lower.7ex\hbox{$\;\stackrel{\textstyle>}{\sim}\;$}}
\newc{\lsim}{\lower.7ex\hbox{$\;\stackrel{\textstyle<}{\sim}\;$}}

\preprint{
\noindent
\hfill
\begin{minipage}[t]{3in}
\begin{flushright}
CERN--TH/98--383 \\
NYU/98-12-02     \\
RU--98--46       \\
SU--ITP 98--69   \\
hep-ph/9902443   \\
\vspace*{1.5cm}
\end{flushright}
\end{minipage}
}

\title {Soft Yukawa couplings in supersymmetric theories}
\author{Francesca Borzumati$^{\,a}$, Glennys R.~Farrar$^{\,b}$,
 Nir Polonsky$^{\,c}$, and Scott Thomas$^{\,d}$}
\address{$^{a}$
  Theory Division, CERN, CH--1211 Geneva 23, Switzerland}
\address{$^{b}$
 Department of Physics, New York University,
 New York, NY 10003, USA}
\address{$^{c}$
 Department of Physics and Astronomy, Rutgers University,
 Piscataway, NJ 08854--8019, USA}
\address{$^{d}$
 Department of Physics, Stanford University,
 Stanford, CA 94305, USA}

\maketitle

\begin{abstract}
The possibility of radiatively generated fermion masses arising from
chiral flavor violation in soft supersymmetry-breaking terms is
explored.  Vacuum stability constraints are considered in various
classes of models, and allow in principle all of the first- and
second-generation quarks and leptons and the $b$-quark to obtain
masses radiatively.  Radiatively induced Higgs--fermion couplings have
non-trivial momentum-dependent form factors, which at low momentum are
enhanced with respect to the case of tree-level Yukawa couplings.
These form factors may be probed by various sum rules and relations
among Higgs boson decay widths and branching ratios to fermion final
states.  An apparent, large, hard violation of supersymmetry also
results for Higgsino couplings. Mixing between left- and right-handed
scalar superpartners is enhanced. A radiative muon mass is shown to
lead to a relatively large and potentially measurable contribution to
the muon anomalous magnetic moment.  If the light-quark masses arise
radiatively, the neutron electric dipole moment is suppressed by a
natural phase alignment between the masses and dipole moment, and is
below the current experimental bound.  The possibility of neutrino
masses arising from softly broken lepton number, and concomitant
enhanced sneutrino--antisneutrino oscillations, is briefly discussed.
\end{abstract}


\newpage
\setlength{\parskip}{1.2ex}

\section{Introduction}
\label{sec:intro}

Fermion masses are signals of broken chiral flavor symmetries.  In the
standard model of electroweak and strong interactions, this flavor
violation arises from tree-level Yukawa interactions between the Higgs
boson and fermions of opposite chirality.  With these interactions,
the Higgs boson vacuum expectation value, which breaks electroweak
symmetry, then gives rise to the fermion masses.  However, the Yukawa
couplings are arbitrary parameters, and it is widely accepted that
they represent an effective low-energy description of a full theory of
flavor.  In this paper we explore the possibility, which presents
itself in supersymmetric theories, that chiral flavor symmetries are
broken predominantly by soft, dimension-three supersymmetry-breaking
terms rather than hard dimension-four superpotential Yukawa couplings.
This scenario has a number of interesting phenomenological
consequences, including enhanced Higgs couplings; apparent, large,
hard violations of supersymmetry in Higgsino couplings; relatively
large contributions to anomalous magnetic moments; suppression of
electric dipole moments; and enhanced mixing between left- and
right-handed scalar partners.

In many extended frameworks for flavor, the standard model Yukawa
couplings are understood as spurious degrees of freedom in the low
energy theory, which parametrize flavor symmetry breaking.  In the
Froggatt--Nielsen mechanism these spurions are related to powers of
scalar expectation values which spontaneously break the flavor
symmetries in the underlying fundamental theory~\cite{FN}.  In this
case the texture of hierarchies that appear in the Yukawa coupling
matrices can then be related to ``horizontal'' flavor symmetries,
which restrict the powers of the flavor-breaking expectation values.
In supersymmetric theories, such flavor-breaking scalar expectation
values preserve supersymmetry and therefore induce superpotential
couplings, which give tree-level fermion Yukawa couplings directly.
Enforcing particular flavor textures in the Yukawa matrix via
continuous or discrete horizontal
symmetries~\cite{HORIZONTAL,ANOMALOUS} is natural because of the 
holomorphy of the superpotential.

Supersymmetric theories also allow the interesting possibility that
chiral flavor symmetries are broken by auxiliary rather than scalar
expectation values.  Such expectation values break supersymmetry and
cannot induce superpotential couplings directly.  In the low-energy
theory these breakings appear as either holomorphic or non-holomorphic
scalar tri-linear terms involving a Higgs field and the scalar
partners of the left- and right-handed fermions.  These soft
tri-linear terms differ in $U(1)_R$ charge from fermion Yukawa
couplings.  Contributions to Yukawa couplings from these terms
therefore also require breaking of $U(1)_R$ symmetry.  With a
non-vanishing gaugino mass providing the $U(1)_R$ breaking, soft
chiral flavor breaking leads to radiative fermion masses at one loop.
In this mechanism the violation of fermionic chirality required for a
fermion mass is provided by the massive gaugino, while the violation
of chiral flavor symmetry originates in the scalar tri-linear terms.
These quantum contributions to the fermion masses are finite and in
principle calculable in terms of the parameters of the low-energy
theory.  The necessity of both $U(1)_R$ and chiral flavor breaking,
along with non-renormalization of the superpotential, give a natural
context in which a fermion mass can be purely radiative.  The
tree-level Yukawa coupling can vanish, and yet receive a {\it finite}
radiative contribution.  This may be enforced by continuous or
discrete horizontal flavor $R$-symmetries.  In this scenario some of
the flavor symmetries are broken in the supersymmetry-breaking sector,
either spontaneously or explicitly by interactions with a messenger
sector.  Since supersymmetry breaking requires non-vanishing auxiliary
expectation values, this mechanism for fermion masses amounts to an
auxiliary field version of the Froggatt--Nielsen mechanism.

It is well known that in supersymmetric theories there can be
significant quantum contributions to fermion masses in certain regions
of parameter space~\cite{MSSMCAL,MSSMUSE}.  However, these
contributions are usually assumed to be proportional to the tree-level
Yukawa coupling.  Various versions of radiative fermion masses arising
predominantly from soft terms have been considered in the following
scenarios: softly broken $N=2$ theories as a solution of the fermion
mass problem in these models~\cite{NTWO}, for the first two
generations of quarks and leptons in the softly broken supersymmetric
standard model~\cite{BANKS}, including radiative contributions from an
exotic sector near the electroweak scale~\cite{MA}, and for the first
generation of quarks and leptons within the context of a grand unified
theory~\cite{KRASNIKOV}.  The relatively large soft tri-linear scalar
terms required, in particular to obtain radiative second-generation
fermion masses, can lead to charge or color breaking vacua along
certain directions in field space~\cite{AHCH}.  However, as discussed
below, metastability of the charge and color preserving vacuum on
cosmological time scales is in general possible.  In addition, we
identify several classes of theories in which the global minimum
preserves color and charge.

The magnitude of masses and mixings that emerges in such models depend
on the specific textures for both soft and hard chiral flavor
breaking, and on details of the supersymmetric particle spectrum.
While these are model-dependent, some general features are
characteristic.  Since radiative masses are intrinsically suppressed
by a loop factor, purely radiative generation of masses for the first
and second generations leads to quarks and leptons in general much
lighter than the third-generation fermions, and with suppressed
mixings.  In addition, since flavor and supersymmetry breaking are
intimately linked in this scenario for radiative masses, interesting
levels of supersymmetric contributions to low-energy flavor-changing
processes can occur. These however depend on specific model-dependent
textures and the over-all scale of the superpartner masses, and may be
avoided in some models.  In this paper we concentrate on
model-independent consequences and signatures of radiative fermion
masses from soft chiral flavor violation, and leave the study of
specific textures and flavor violating processes to future work.

Radiative fermion masses have a number of striking phenomenological
consequences, many related either directly or indirectly to the
softness of the fermion mass.  Chief among these is that the chirality
violating fermion mass and Yukawa couplings are momentum-dependent,
with non-trivial form factors.  The Higgs--fermion coupling turns out
to be enhanced at low momentum wth respect to a tree-level Yukawa
coupling.  In addition, if the soft chirality violation arises in
non-holomorphic scalar tri-linear terms, the corresponding fermions
receive mass from the ``wrong Higgs'' field.  The dependence of the
physical Higgs boson couplings on the Higgs vacuum expectation values
then differs drastically from the minimal case.  For fermions with
radiative masses, the corresponding Higgsino couplings also arise
radiatively with non-trivial form factors, but with different
parametric dependences on gauge couplings.  This leads to an apparent,
large, hard violation of supersymmetry in the low-energy theory.  The
leading low-energy momentum dependence of these radiatively generated
couplings can be represented as effective radii for the appropriate
couplings.  All these deviations from the minimal expectations for
these couplings are potentially measurable with high-precision
collider measurements of Higgs couplings, as well as from Higgsino
branching ratios.  Another important feature of this scenario for soft
Yukawa couplings is that both the mass and anomalous magnetic moments
arise at one loop.  This has the consequence that supersymmetric
contributions to magnetic moments are effectively a loop factor larger
than with tree-level Yukawa couplings.  The current experimental bound
on the muon anomalous magnetic moment represents the best probe for a
radiative muon mass, and already constrains part of the parameter
space.  CP-violating electric dipole moments (EDMs) are suppressed by
a natural phase alignement between the masses and dipole moment in
interesting regions of parameter space.  With radiative light-quark
masses, the neutron EDM is easily below the current experimental
bound, but could be measurable in future experiments.  For a radiative
electron mass, the electron EDM is more model-dependent.  The current
experimental bound may already be used to infer preferred regions of
neutralino/selectron parameter space if the electron mass is
radiative.  Finally, radiative fermion masses imply enhanced mixing
between the associated left- and right-handed scalar partners, resepct
to a tree-level Yukawa.  These mixings can significantly modify the
production cross sections and branching ratios of scalar
superpartners.

In the next section, the radiative contributions to fermion masses in
softly-broken supersymmetric models are discussed.  Non-trivial
momentum dependence of the Higgs coupling, the Higgs Yukawa 
radius, and enhancement of the Higgs coupling at zero momentum
transfer are introduced and calculated.  The apparent violation of
supersymmetry in the relatively large difference between Higgs and
Higgsino couplings is introduced.  In section~\ref{sec:possible} the
magnitude of the soft tri-linear terms required to obtain particular
fermion masses are presented.  Enhanced left--right mixing of scalar
superpartners, the origin of the Cabibbo--Kobayashi--Maskawa (CKM)
quark mixing terms, and various scenarios for supersymmetric
contributions to flavor-changing processes are also discussed.  A
stability analysis of the scalar potential is presented in
section~\ref{sec:stability}.  A number of classes of models for
introducing stabilizing quartic scalar couplings are discussed.  These
can lead to either a globally stable charge- and color-preserving
vacuum, or metastability on cosmological time scales.  The
(meta)stability bounds allow in principle all of the
first-two-generation quarks and leptons, as well as the $b$-quark, to
obtain masses radiatively.  A classification of the effective
operators that give rise to the auxiliary spurion version of the
Froggatt--Nielson mechanism for either holomorphic or non-holomorphic
scalar tri-linear terms is presented in section~\ref{sec:class}.  It
is pointed out that non-holomorphic soft tri-linear terms require a
low scale of supersymmetry breaking.  In section~\ref{musection} the
relatively large contribution to the muon anomalous magnetic moment in
this scenario is compared with current experimental bounds.  This
already constrains some of the parameter space in this scenario, and
is shown to place an upper limit on the enhancement of the Higgs--muon
coupling compared to the tree-level case.  If both muon and $b$-quark
masses are radiative, an analogous upper limit on the Higgs--$b$-quark
coupling can be extracted from the muon anomalous magnetic moment
bound, assuming gaugino universality.  CP-violating Higgs--fermion
couplings and fermion EDMs are investigated in
section~\ref{sec:cpviol}.  With radiative masses for the light quarks,
the neutron EDM is shown to be comfortably below the current
experimental bound due to a natural high degree of phase alignment
between the quark masses and dipole moment.  The electron EDM is shown
to be more model-dependent, but sufficiently suppressed in certain
regions of the neutralino/selectron parameter space.  In
section~\ref{sec:probing} a number of sum rules, relations among
Higgs-boson dec ay widths and branching ratios to fermion final states
are presented.  These hold in large classes of supersymmetric theories
including the minimal supersymmetric standard model.  For radiative
masses, deviations from minimal expectations for these sum rules and
relations due to non-trivial form factors and possible non-holomorphic
Higgs couplings may be measurable at future colliders.  The
possibility of softly broken lepton number as a source of neutrino
mass and concomitant enhanced sneutrino--antisneutrino oscillation is
discussed in section~\ref{leptonsec}.  Conventions and one-loop
integrals for fermion masses, Higgs and Higgsino couplings, and
anomalous moments are given in the appendices.  Preliminary results
were presented previously in ref.~\cite{TALKS}.


\section{Soft Yukawa Couplings}
\label{sec:soft}

In a supersymmetric theory the scalar superpartners of the quarks and
leptons carry the same flavor symmetries as the associated quarks and
leptons.  Flavor symmetries may therefore be broken by terms that
involve only scalar fields.  Quark and lepton masses require terms
that break flavor symmetries for both left- and right-handed fields.
The most general softly broken supersymmetric Lagrangian contains
holomorphic scalar tri-linear operators of the form
\begin{equation}
{\cal L} \supset AH_{\alpha}\phi_{L}\phi_{R} ~+~{\rm h.c.}
\label{ophol}
\end{equation}
as well as non-holomorphic operators~\cite{HR} of the form
\begin{equation}
{\cal L} \supset A^{\prime}H_{\alpha}^{*}\phi_{L}\phi_{R}~+~{\rm h.c.}\,,
\label{opnonhol}
\end{equation}
where flavor indices on the scalar fields and the complex $A$- and
$A^{\prime}$-parameters are suppressed.  For leptons and
down-type quarks, the holomorphic (non-holomorphic) operator contains
the Higgs doublet $H_{\alpha}=H_1$ $(H_2)$ with $U(1)_Y$ hypercharge
$Y=-1$ $(1)$.  For up-type quarks the holomorphic(non-holomorphic)
operators contain $H_{\alpha}=H_2$ $(H_1)$.  The $A$- and
$A^{\prime}$-parameters break the chiral flavor symmetries carried by
the scalars.  Even though the scalar superpartners carry the same
flavor symmetries as the associated fermion, they differ by one unit
of $R$-charge under $U(1)_R$ symmetry.  Inducing a fermion mass at the
quantum level from the flavor-violating $A$- or
$A^{\prime}$-parameters therefore requires additional terms that
violate $U(1)_R$ symmetry.  At lowest order this $U(1)_R$ breaking can
be provided by a gaugino mass, which along with the flavor-violating
$A$-parameters give rise to fermion masses at one loop, as given
explicitly below.  The insertion of a gaugino mass can also be
understood in terms of the necessity of fermionic chirality violation
for a quark or lepton mass.  Since the dimensionless fermion Yukawa
coupling arises quantum mechanically from dimensionful parameters, the
radiative contributions are finite and calculable.  The fermion mass
and the Higgs and Higgsino Yukawa couplings discussed below are
therefore soft in the technical sense that no counter-terms are
required in the high energy theory.  The couplings are also soft in
the more colloquial sense that the quantum contributions arise at the
superpartner scale from soft supersymmetry-breaking terms (which do
require counter-terms but do not introduce quadratic divergences into
the low-energy theory).

For a quark or lepton that receives a radiative mass, Higgs and
Higgsino couplings also arise radiatively.  However, since the chiral
flavor breaking arises in supersymmetry-breaking terms, there is no
symmetry that enforces the equality of the effective Higgs and
Higgsino couplings.  This is unlike the case of a tree-level
superpotential Yukawa coupling, where equality of these couplings is
enforced by supersymmetry.  In fact, as discussed below, even the
parametric dependence on gauge coupling constants of the effective
Higgsino coupling differs from that of the fermion mass and effective
Higgs coupling.  This leading order difference between Higgs and
Higgsino Yukawa couplings amounts to a large violation of
supersymmetry in a dimensionless coupling at low energy, i.e. an
apparent hard violation of supersymmetry.

The over-all magnitude and parametric dependence of radiative fermion
masses and Higgs and Higgsino couplings are discussed in the following
subsections.  In this section flavor-changing effects are suppressed
and scalar masses generally taken to be flavor-independent for
simplicity.

%

\subsection{The fermion mass}
\label{sec:mass}

The one-loop scalar partner--gaugino diagram that dresses the fermion
propagator
%
%
\begin{figure}[h]
\begin{center}
\epsfxsize= 6.5 cm
\leavevmode
\epsfbox{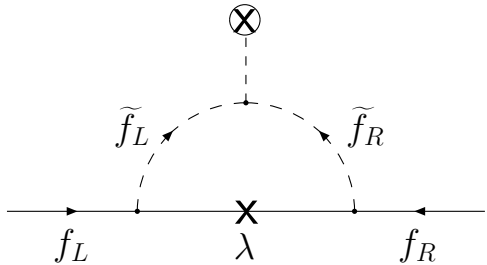}
\end{center}
\caption[f1]{One-loop contribution to fermion mass from soft
chiral flavor breaking and gaugino mass.}
\label{massdiag}
\end{figure}
generates a finite contribution to the fermion mass, as shown in
fig.~\ref{massdiag}.  The $A$- or $A^{\prime}$-parameter breaks both
left and right flavor symmetries, while the gaugino mass insertion
connects left and right chiral fermions.  For 
$m_f^2 \ll m_{\tilde{f}}^2, m_{\lambda}^2$, with $m_{\tilde{f}}$ the
physical scalar mass eigenvalues, the one-loop radiative fermion mass
may be evaluated at $p^2=0$, and is given by
\begin{equation}
 m_f =  - m^{2}_{LR}
\left\{ \frac{\alpha_s}{2 \pi} \,C_f  \,m_{\tilde{g}}
\,I(m^{2}_{\tilde{f}_1}, m^{2}_{\tilde{f}_2}, m^2_{\tilde{g}})
       +\frac{\alpha^\prime}{2\pi} \sum_{j} K_f^{j}\,
                                       m_{\tilde{\chi}_j^0}
\,I(m^{2}_{\tilde{f}_1}, m^{2}_{\tilde{f}_2},
                                       m^2_{\tilde{\chi}_{j}^0})
\right\}\,,
\label{mass}
\end{equation}
where $C_{f} = 4/3, 0$ for quarks and leptons, respectively.  The sum
is over neutralino eigenstates with coupling coefficients $K_f^j$ to
the fermions, and 
$I(m^{2}_{\tilde{f}_{1}}, m^{2}_{\tilde{f}_{2}}, m^2)$ 
is a loop function discussed below.  The dependence on the left--right
squark or slepton mixing,
$m_{LR}^2 = A \langle H_\alpha \rangle$ or
$A^{\prime} \langle H_\alpha \rangle$,
and on the chiral violation arising from the gaugino masses,
$m_{\tilde{g}}$, and/or $m_{\tilde{\chi}^0_j}$, is displayed
explicitly in~(\ref{mass}).  Note that without hard tree-level Yukawa
couplings in the superpotential, there are no left--right mixing terms
arising from interference with the superpotential Higgsino mass,
$W \supset \mu H_1 H_2$.  The first and second terms in~(\ref{mass})
correspond to the strong (gluino) and hypercharge/weak (neutralino)
contributions, respectively.  The neutralino contributions include
both a pure $\widetilde{B}$--$\widetilde{B}$ and  mixed
$\widetilde{B}$--$\widetilde{W}_{3}$ propagators.  The coupling
coefficients $K_{f}^{j} $ are
\begin{equation}
 K_{f}^{j} =
 \frac{Y_{f_L}}{2}\,N_{jB}
 \left[\frac{Y_{f_R}}{2} \,N_{jB} + \cot\theta_{W} N_{jW}
 \right]\,,
\label{Kfactor}
\end{equation}
where $N_{ij}$ is the neutralino eigenvector mixing matrix, and the 
hypercharge is normalized as $Q = T_3 + {1 \over 2} Y$.  In the mostly
gaugino or mostly Higgsino region of parameter space, the neutralino
propagator is well approximated by pure
$\widetilde{B}$--$\widetilde{B}$ exchange, since the
$\widetilde{B}$--$\widetilde{W}_{3}$ contributions are suppressed by
gaugino--Higgsino mixing;
$N_{jB} N_{jW} \sim {\cal O} (m_Z^2 \mu /
(\mu^2\!-\!m_{\widetilde{W}}^2)
(m_{\widetilde{W}}\!-\!m_{\widetilde{B}}))$
for a given mass eigenstate.  Chargino contributions to the mass are
forbidden by gauge invariance in the absence of hard tree-level Yukawa
couplings.  Hereafter only the strong contribution to the quark masses
will be retained, except in the discussion of CP-violating effects
given in section~\ref{sec:cpviol}.  In general the parameters
$m_{LR}^2$, $m_{\tilde{g}}$, $m_{\tilde{\chi}_j^0}$, and the coupling
coefficients $K_f^j$ appearing in~(\ref{mass}) may be complex, but the
masses appearing in the loop functions are understood to be the real
positive mass eigenvalues.  In the absence of a tree-level Yukawa
coupling, it is always possible to work in a basis in which the
radiative fermion mass is real.  Note that throughout, all
calculations are performed using mass eigenstates.  Insertion
approximation is not employed; the over-all factors of 
$m_{LR}^2 m_{\tilde{g}}$ and 
$m_{LR}^2 K_f^j m_{\tilde{B}}$ in~(\ref{mass}) appear as a consequence
of algebraic relationships between the factors in mixing coefficients.

%
%
\begin{figure}[ht]
\begin{center}
\epsfxsize= 12.5 cm
\leavevmode
\epsfbox{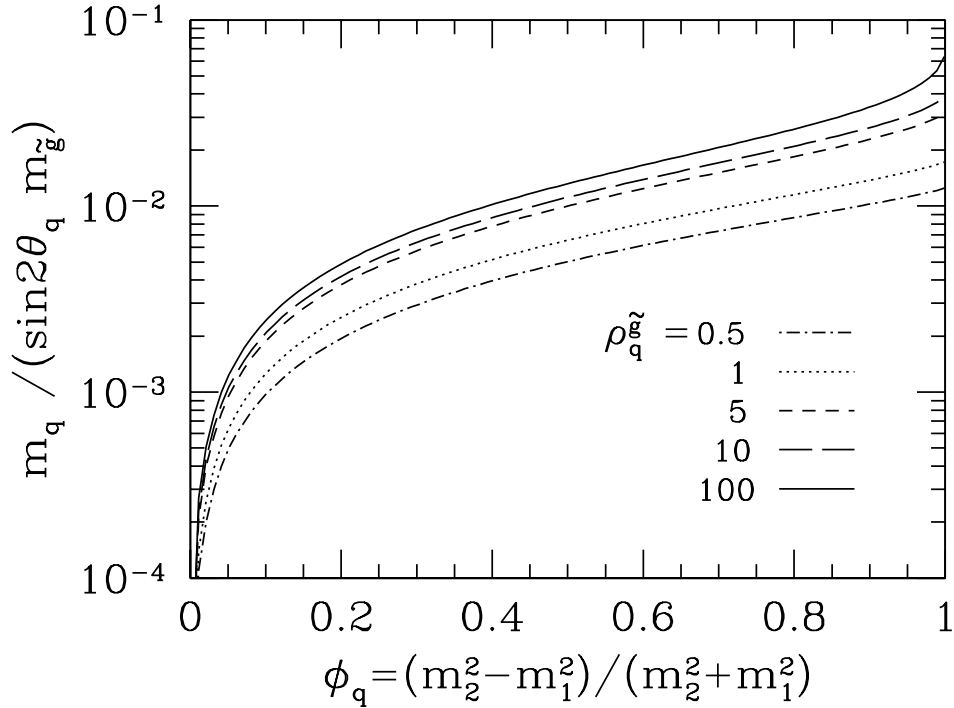}
\end{center}
\caption[f1]{Radiative quark mass, $m_{q}$,
 in units of the squark mixing angle times the gluino mass,
 $\sin2\theta_q \, m_{\widetilde{g}}$,
 as a function of the fractional squark mass splitting
 $\phi_q $, where $m_i^2$ here indicates $m_{\tilde{q}_i}^2$.
 The mass is shown for various values of the ratio
 $\rho_{q}^{\tilde{g}}$.
 For a radiative lepton mass, the squark parameters are replaced by
 slepton parameters, the gluino mass by the Bino mass, and
 the over-all result is scaled by a factor of
 $(3  / 8) (\alpha^{\prime} / \alpha_s) \simeq 0.03$ for
 the hypercharge coupling and charges.}
\label{massfig}
\end{figure}
The loop function $I(m^{2}_{\tilde{f}_{1}},
m^{2}_{\tilde{f}_{2}},m^2_{\lambda})$, with $\lambda$ denoting
generically a gaugino, is given explicitly in appendix~\ref{sec:defI},
along with definitions of the scalar mass eigenstates and mixing
matrices.  The dominant momentum in the loop is controlled by the
largest mass scale.  The loop function is then typically bounded by
$\widetilde{m}^{\,2} I( m^{2}_{\tilde{f}_{1}},
m^{2}_{\tilde{f}_{2}},m^2_{\lambda}) \lesssim {\cal{O}}(1)$, where
$\widetilde{m} =\max(m_{\tilde{f}_{1}},m_{\tilde{f}_{2}},m_{\lambda})$ 
(see, e.g., ref.~\cite{CARENA}).  This bound is typically saturated
for $ m_{\tilde{f}} \simeq m_{\lambda}$ (but is evaded in certain
limits -- see appendix~\ref{sec:defI}).  The radiative fermion masses
then scale as
$m_f \sim (\alpha /2\pi) \,m_{LR}^2 m_{\lambda} /\widetilde{m}^{\,2}$,
where $\alpha = (4/3) \alpha_s$ or $\alpha^{\prime}$, for quarks or
leptons respectively.  With this scaling and the form of the
left--right mixing given above, it is apparent that, if the
superpartners are decoupled, by taking all soft supersymmetry-breaking
parameters simultaneously large, including the flavor-breaking
$A$-parameters, the resulting radiative fermion mass becomes
independent of the supersymmetry-breaking scale.  This can be seen
directly by considering the effective mass Yukawa coupling, $m_f / |
\langle H_{\alpha} \rangle |$.  This dimensionless one-loop finite
coupling must approach a constant as the superpartners are decoupled,
since the fermion mass must be proportional to the electroweak
symmetry-breaking scale in this limit.

In order to present the numerical results for the one-loop radiative
mass and associated couplings, it is convenient to trade the three
parameters that describe the scalar partner sector, $m_{LL}^2$,
$m_{RR}^2$, and $m_{LR}^2$ used in appendix~\ref{sec:sfermion}, for
\begin{equation}
\phi_f \equiv
\frac{m_{\tilde{f}_2}^2 \!-\! m_{\tilde{f}_1}^2}
     {m_{\tilde{f}_2}^2 \!+\! m_{\tilde{f}_1}^2} \,;
\hspace*{0.5truecm}
\rho_{f}^\lambda \equiv
\frac{m_{\tilde{f}_1}^2 \!+\! m_{\tilde{f}_2}^2}
     {2m_{\lambda}^2} \,;
\hspace*{0.5truecm}
 \sin 2 \theta_f \!=\!
 -\frac{ 2 m^2_{LR}}{m^2_{\tilde{f}_2}\!-\!m^{2}_{\tilde{f}_1}} \,,
\end{equation}
which are respectively the fractional scalar partner mass-squared
splitting, the average scalar partner mass squared, normalized to the
relevant gaugino mass squared, $m_{\tilde g}^2$, 
$m_{\widetilde B}^2$, or $m_{\tilde{\chi}_j^0}^2$ generically
indicated as $m_\lambda^2$, and the scalar partner mixing angle.
In terms of these parameters, and in the pure gaugino
limit for which $\widetilde{B}$--$\widetilde{W}_{3}$
vanishes, the radiative mass~(\ref{mass}) is
\begin{equation}
 m_f =  - \sin 2 \theta_f \,\phi_f
\left\{ \frac{\alpha_s}{2 \pi} \,C_f
        \,\rho_f^{\tilde{g}} \,m_{\tilde{g}}^3
\,I(m^{2}_{\tilde{f}_1}, m^{2}_{\tilde{f}_2}, m^2_{\tilde{g}})
       +\frac{\alpha^\prime}{2\pi}
        \,\rho_f^{\tilde{B}} \,m_{\widetilde{B}}^3
\,I(m^{2}_{\tilde{f}_1}, m^{2}_{\tilde{f}_2}, m^2_{\widetilde{B}})
\right\}\,.
\label{massnew}
\end{equation}
%
The on-shell quark mass radiatively generated at one loop through
gluino exchange, is plotted in fig.~\ref{massfig} as a function of
$\phi_q$. The mass is displayed in units of the squark mixing angle
times the gluino mass, $\sin2\theta_q \,m_{\widetilde{g}}$, for
various values of $\rho_{q}^{\tilde{g}}$.  The lepton mass,
radiatively generated in the pure gaugino limit, can be obtained from
the same figure, by replacing the squark parameters by slepton
parameters, the gluino mass by the Bino mass, and by rescaling the
over-all result by a factor of 
${3 \over 8} (\alpha^{\prime} / \alpha_s) \simeq 0.03$ for the 
hypercharge coupling.


\subsection{The Higgs coupling}
\label{sec:higgs}

\begin{figure}[ht]
\begin{center}
\epsfxsize= 6.5 cm
\leavevmode
\epsfbox{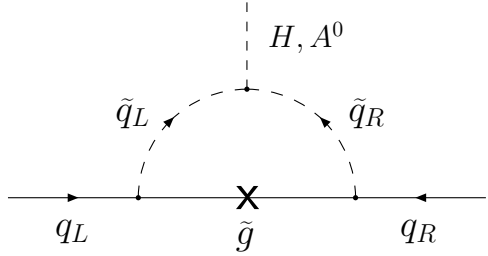}
\end{center}
\caption[f1]{One-loop radiative couplings of quarks to
 neutral scalar Higgs bosons $H$ ($=h^0,H^0$) or the neutral pseudoscalar
 $A^0$ from chiral flavor violation and gaugino mass.}
\label{higgsdiag}
\end{figure}

The momentum-dependent coupling of the physical Higgs bosons to
fermions arises radiatively in a manner similar to the fermion masses,
as shown in fig.~\ref{higgsdiag}.  For simplicity, the discussion here
is restricted to the neutral Higgs bosons, $h^0$, $H^0$, and $A^0$,
which accompany a single pair of Higgs doublets.

The generic neutral Higgs--fermion--fermion operator obtained
at the one-loop level, has the form
$ H_{\alpha}(q) \overline{f_L}(q_1) f_R(q_2)$,
where $q\!=\!q_1\!-\!q_2$ and $\alpha=1,2$.
%
%
Consider first the couplings of the physical scalar Higgs bosons,
$h^0$ and $H^0$, which differ somewhat from that of the pseudoscalar
Higgs $A^0$.  The physical mass eigenstates couple to the fermions
with amplitude
$\lambda_{h^0,H^0} = \Theta \{ \cos \beta, \sin \beta \} \bar{h}_{f,H} /
\sqrt{2}$, where $\Theta$ is the Higgs mixing matrix between
physical and interaction eigenstates.  The mixing matrix $\Theta$ is
non-trivial and differs in the holomorphic and non-holomorphic cases
as discussed in section~\ref{sec:probing}.  The first (second) term in
the curly brackets corresponds to $\alpha=1(2)$, and $\sqrt{2}$ is the
standard normalization factor for a real scalar field.
With all external particles on-shell, relevant to, for instance, the
decay  $ H(q) \to f_L(q_1) {\overline{f_R}}(-q_2)$ or the resonant
production
$f_L(q_1) {\overline{f_R}}(-q_2) \to H(q)$ with $H=h^0,H^0$, then
in the approximation $q_1^2 = q_2^2 = 0$, as is appropriate for
$m_f^2 \ll m_H^2$, the radiative Higgs Yukawa coupling to $h^0$ or
$H^0$ through an $A$- or $A^{\prime}$-term may be related to the
corresponding fermion mass Yukawa coupling,
$m_f / | \langle H_{\alpha} \rangle |$.
Summing over the scalar partner mass eigenstates and retaining only
the gluino contribution for quarks and the Bino contribution for
leptons:
\begin{equation}
 \bar{h}_{f,H}(m_H^2) = \frac{m_f}{| \langle H_{\alpha} \rangle |}
 \left(\sin^2 2\theta_f \,
   J_1(m_H^2; m_{\tilde{f}_1}^2, m_\lambda^2,m_{\tilde{f}_2}^2 ) +
       \cos^2 2\theta_f \,
   J_2(m_H^2; m_{\tilde{f}_1}^2, m_\lambda^2,m_{\tilde{f}_2}^2 )
 \right) \,,
\label{vertex1}
\end{equation}
where $\alpha=1$ or $2$.  With this definition, $\bar{h}_{f,H}$ is the
effective Yukawa coupling for $h^0$ or $H^0$ with Higgs mixing effects
factored out.  The loop functions with $q^2=m_H^2$ are defined as
\begin{eqnarray}
 J_1 (m_H^2; m_{\tilde{f}_1}^2, m_\lambda^2,m_{\tilde{f}_2}^2 )
  & = &
 \frac{1}{2}
 \frac{ \sum_{i=1,2}
        C_0( 0, 0, m_H^2;
          m_{\tilde{f}_i}^2, m_\lambda^2, m_{\tilde{f}_i}^2 )}
      {I(m^2_{\tilde{f}_1}, m^2_{\tilde{f}_2},m^2_\lambda)},
  \nonumber \\[1.1ex]
 J_2 (m_H^2; m_{\tilde{f}_1}^2, m_\lambda^2,m_{\tilde{f}_2}^2 )
   & = &
 \frac{
        C_0( 0,0,m_H^2;
          m_{\tilde{f}_1}^2, m_\lambda^2,m_{\tilde{f}_2}^2 )}
      {I( m_{\tilde{f}_1}^2, m^2_{\tilde{f}_2},m^2_\lambda )}\,,
\label{Jfunct}
\end{eqnarray}
where the conventions for the three-point functions
$C_0(0,0,m_H^2;m_a^2, m_b^2, m_c^2)$ are defined in
appendix~\ref{defC}.

The expression~(\ref{vertex1}) is significantly simplified in the
heavy superpartner limit $m_{\tilde{f}}^2, m_{\lambda}^2 \gg m_H^2$,
which is particularly relevant to the case of the light Higgs boson,
$h^{0}$, for which $m_{h^0} \leq m_Z$ at tree level.  In this limit
the vertex loop functions may be evaluated at $q^2=0$, and reduce to
the mass loop function 
$C_0 (0,0,0;m_a^2,m_b^2,m_c^2) = I(m_a^2,m_b^2,m_c^2)$ 
(see appendix~\ref{defC}).  Making this approximation also isolates
intrinsic coupling effects from momentum-dependent effects at finite
$q^2$ discussed below.  The coupling~(\ref{vertex1}) in this limit
becomes
\begin{equation}
  \bar{h}_{f,H}(0) =
 \frac{m_{f}}{|\langle H_{\alpha} \rangle|}
 \left\{
 \sin^{2}2\theta_f \left[\frac{1}{2}
 \frac{\sum_{i}
  I(m_{\tilde{f}_{i}}^{2},m_{\tilde{f}_{i}}^{2},m^{2}_{\lambda})}
 {I(m^{2}_{\tilde{f}_{1}},m^{2}_{\tilde{f}_{2}},m^2_\lambda)}   - 1
 \right]    + 1
 \right\} \,.
\label{vertex2}
\end{equation}

\begin{figure}[ht]
\begin{center}
\epsfxsize=12.5 cm
\leavevmode
\epsfbox{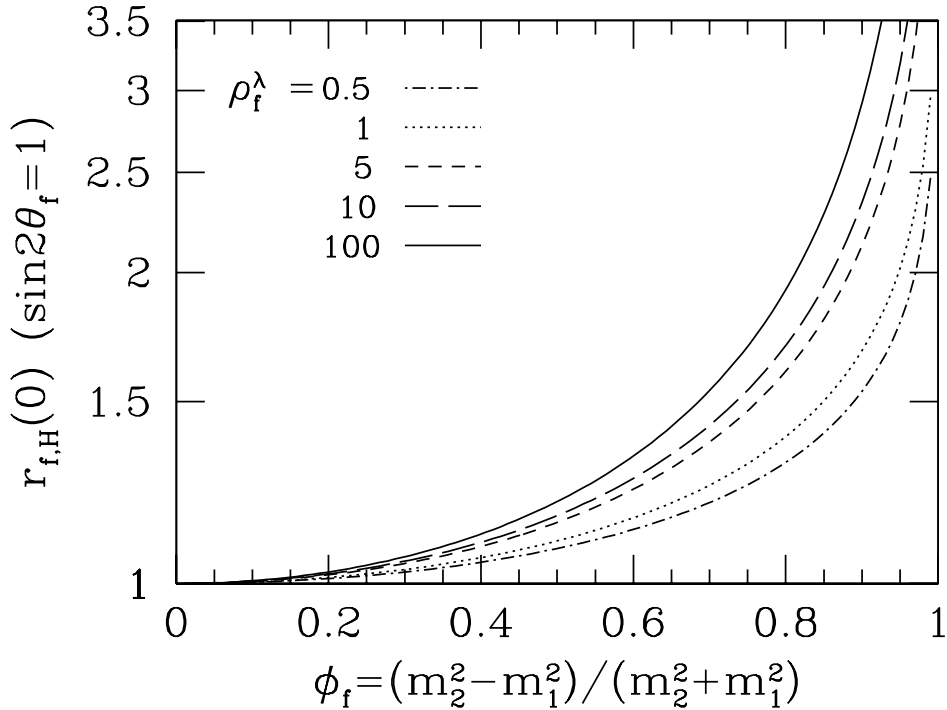}
\end{center}
\caption[f1]{The ratio of $h^0$ and $H^0$ Higgs--fermion Yukawa
 couplings to the mass Yukawa coupling,
 $r_{f,H}\! =\! \bar{h}_{f,H}/\bar{h}_{f,m}$,
 as a function of the fractional scalar mass splitting
 $\phi_f $
 for different values
 of $\rho_{f}^\lambda$.
 The ratio is shown for $q^2=0$ and
 for maximal scalar mixing, $\sin2\theta_f = 1$.}
\label{higgscoupling}
\end{figure}

In order to characterize the magnitude of the intrinsic coupling
of the Higgs bosons to fermions it is useful to
define the ratio of the effective Higgs Yukawa coupling
to the effective mass Yukawa coupling with the Higgs mixing
effects factored out:
\begin{equation}
  r_{f,H}(m_H^2)
\ \equiv \
 {\bar{h}_{f,H}(m_H^2) \over (m_f / | \langle H_{\alpha} \rangle | )}
\ \equiv \
 {\bar{h}_{f,H}(m_H^2) \over \bar{h}_{f,m}(0)} \,.
\label{higgsratio}
\end{equation}
{}From eqs.~(\ref{vertex1}) and~(\ref{Jfunct}) it can be shown that
$r_{f,H} \geq 1$.  This compares with $r_{f,H}=1$ at lowest order for
a fermion mass arising from a hard tree-level Yukawa coupling.  The
ratio $r_{f,H}$ for a soft radiative mass is plotted in
fig.~\ref{higgscoupling} as a function of the fractional scalar mass
splitting $\phi_f$ for $q^2=0$.  For 
$m_{LR}^2 \sim m_{\tilde{f}}^2, m_{\lambda}^2$ the enhancement 
$r_{f,H}(0) \geq 1$ can be sizeable, and increases with the scalar
mass splitting.
%
%
The divergence in $r_{f,H}(0)$ for $ \phi_f \to 1$ is due to a
vanishing eigenvalue in the scalar partner mass squared matrix in this
limit.  The physical Higgs coupling is infrared-singular in this limit
from diagrams in fig.~\ref{higgsdiag} with two massless scalar
propagators.  The fermion mass however remains finite in this limit,
with a single massless scalar propagator in the diagrams of
fig.~\ref{massdiag}.

The disparity between the effective Higgs and mass Yukawa coupling 
is due to the difference in the loop momentum integrations.
For $m_{\tilde{f}}^2, m_{\lambda}^2 \gg m_H^2$, the difference may be
understood as arising from the differing combinatorics for chiral
insertions, proportional to $m_{LR}^2$, between figs.~\ref{massdiag}
and~\ref{higgsdiag} (the scalar mass eigenstate propagators implicitly
contain a resummation of an arbitrary number of chiral insertions).
For $q^2=0$ and at first order in $m_{LR}^2/\widetilde{m}^2$, where
$\widetilde{m} = {\rm max}( m_{\tilde{f}}, m_{\lambda})$, the diagrams in
figs.~\ref{massdiag} and~\ref{higgsdiag} are identical, with a single
chiral insertion, yielding $r_{f,H}(0)=1$ at this order.  Since an odd
number of $m_{LR}^2$ left--right mixing insertions are required by
chirality, the mass and Higgs Yukawa diagrams only differ through
diagrams with three or more such insertions.  The $q^2=0$ ratio in the
heavy superpartner limit is thus parametrically
$r_{f,H}(0) = 1 + {\cal O}(m_{LR}^4 / \widetilde{m}^{\,4})$.
This behavior may also be obtained directly from~(\ref{vertex2}), with
the limiting values of the mixing angle and loop functions given in
appendix~\ref{defC}.  In the strict superpartner decoupling limit,
$r_{f,H}(0) \to 1$, so that the effective Higgs and mass Yukawa
coupling coincide.

The parametric dependence of the Higgs coupling
ratio~(\ref{higgsratio}) at $q^2=0$ can also be understood from the
general form of the effective Higgs Yukawa coupling.  Promoting all
parameters to background fields, and requiring invariance with respect
to all background and gauge symmetries, the functional dependence of
the radiative fermion masses at one loop may be written
\begin{equation}
m_f = \bar{h}_{f,m}(0) H_{\alpha} =
 {\cal F} \left(
  m_{\tilde{f}}^2(
    A^* A H_{\alpha}^* H_{\alpha} ,
    H_{\beta}^* H_{\beta})  ,
  m_{\lambda}^* m_{\lambda},
  \frac{\mu H_1 H_2}{ m_{\lambda}^{*}}
          \right) ~
  {m_{\lambda}^* A \over  m_{\tilde{f}}^2}
  H_{\alpha},
\label{mfcn}
\end{equation}
where $\bar{h}_{f,m}(0)$ is the radiative mass Yukawa coupling defined
by~(\ref{higgsratio}),
$A^* A H_{\alpha}^* H_{\alpha} = m_{LR}^4$,
$m_{\tilde{f}}$ are the physical mass eigenvalues, the Higgs doublets
$H_{\alpha,\beta}$ are understood as expectation values, and here
$\beta=1,2$ is summed while $\alpha = 1$ or $2$ is determined by the
scalar tri-linear coupling and is not summed.  ${\cal F}$ is a
dimensionless function that depends only on the particular
combinations of parameters indicated in~(\ref{mfcn}).  These
combinations of parameters are invariant under the gauge and
background symmetries.  In the heavy superpartner limit, ${\cal F}$ is
an analytic function with power series expansion and non-vanishing
first derivatives with respect to any of the invariant combinations of
dependent variables.  The non-trivial dependence on $H_{\alpha,\beta}$
represents, in the heavy superpartner limit, non-renormalizable
operators with multiple Higgs expectation values, which contribute to
the fermion mass.  All such one-loop operators, coupling through the
$A$- or $A^{\prime}$-terms, are implicitly included in the effective
Yukawa coupling~(\ref{vertex1}).  Additional non-trivial dependence on
$H_{\beta}^* H_{\beta}$ arises implicitly through 
${\cal O}(g^2 H_{\beta}^* H_{\beta})$ $D$-term contributions to the
scalar partner mass-squared eigenvalues $m_{\tilde{f}}^2$.  The
dependence on $H_1 H_2$ arises from
${\cal O}(g g^{\prime} H_1 H_2 \mu /
 (m_{\widetilde{B}}(\mu^2\!-m_{\widetilde{W}}^2)))$
neutralino mixing effects between gaugino and Higgsino eigenstates
through Higgs expectation values.  This functional dependence vanishes
in the pure gaugino limit.

The effective Yukawa coupling at $q^2=0$ of a Higgs boson to a
fermion, again with mixing effects factored out, is very generally
related to the mass by
\begin{equation}
\bar{h}_{f,H}(0) \equiv
{ \partial m_f \over \partial |H_{\alpha}| }\,.
\end{equation}
With this definition and the general form of the radiative fermion
mass~(\ref{mfcn}), the ratio of physical Higgs to mass Yukawa
couplings~(\ref{higgsratio}) at $q^2=0$ is given by
\begin{equation}
r_{f,H}(0) = 1
 + 2 m_{LR}^4\, {\partial \bar{h}_{f,m}(0) \over \partial m_{LR}^4}
 + 2 m_Z^2 \left\{ \cos^2 \beta , \sin^2 \beta \right\}
 \left(
 { \partial m_{\tilde{f}}^2  \over \partial m_Z^2 }
 { \partial \bar{h}_{f,m}(0) \over \partial m_{\tilde{f}}^2 }
+{ \partial \bar{h}_{f,m}(0) \over \partial m_{Z}^2 }  \right)\,.
\end{equation}
The first term on the right-hand side arises from
differentiating~(\ref{mfcn}) with respect to the Higgs doublet
$H_{\alpha}$ multiplying $\bar{h}_{f,m}(0)$.  The second term comes
from $A$- or $A^{\prime}$-term contributions to $\bar{h}_{f,m}(0)$
through the $m_{LR}^4$ dependence.  The third term comes from $D$-term
contributions through $m_{\tilde{f}}^2$ where the first (second) term
in the brackets is for $\alpha=1$ $(2)$.  The fourth term represents
contributions through neutralino mixing effects.  In the heavy
superpartner limit, the mass scale for variations of $\bar{h}_{f,m}$
are controlled by the largest superpartner mass,
${\partial \bar{h}_{f,m}(0) / \partial m_{LR}^4}$
is of 
$  {\cal O} (1/\widetilde{m}^{\,4})$, which agrees with the scaling
given above based on chiral insertions in the one-loop radiative mass
diagram.  Likewise, for the $D$-term effects
$\partial \bar{h}_{f,m}(0) / \partial m_{\tilde{f}}^2$
is of ${\cal O}(1/ \widetilde{m}^{\,2})$,
$\partial  m_{\tilde{f}}^2 / \partial m_Z^2$ of ${\cal O}(1)$,
and for the neutralino mixing effects
$\partial \bar{h}_{f,m}(0) / \partial m_Z^2$ is of 
${\cal O}(1/ \widetilde{m}^{\,2})$.
In the heavy superpartner limit the full parametric dependence of the
Higgs coupling ratio is then
$r_{f,H}(0) = 1  +
{\cal O}(m_{LR}^4/ {\widetilde{m}}^{\,4})
+ {\cal O}(m_Z^2 / {\widetilde{m}}^{\,2})
+ {\cal O}(\mu m_Z^2 /
  ( m_{\widetilde{B},\tilde{g}} \widetilde{m}^{\,2}))$.
The $D$-term and neutralino mixing contributions to the effective
Higgs Yukawa couplings are not explicitly included in~(\ref{vertex1}),
and are only important respect to the non-linear dependence on the
$A$- or $A^{\prime}$-terms if $A \lsim m_Z$ (and are therefore not
dominant for second or third generation radiative masses, which
require $A$ or $A^{\prime} \sim \widetilde{m}$ as discussed in
section~\ref{sec:possible}).  An effect of the $D$-terms
and neutralino mixings is to introduce additional 
``wrong Higgs'' couplings to
fermions through the functional dependence of~(\ref{mfcn}).
Such couplings, however, are
suppressed in the heavy superpartner limit respect to the dominant
coupling through the over-all Higgs doublet $H_{\alpha}$ multiplying
$\bar{h}_{f,m}$ in~(\ref{mfcn}) by
${\cal O}(m_Z^2 / {\widetilde{m}}^{\,2})$ and
$ {\cal O}(\mu m_Z^2 /
 ( m_{\widetilde{B},\tilde{g}} \widetilde{m}^{\,2}))$
respectively.  In the strict superpartner decoupling limit the only
operator that couples Higgs doublets to fermions is the renormalizable
effective Yukawa coupling.  All other operators are
non-renormalizable, and vanish in this limit.  So $\bar{h}_{f,m}(0)$
necessarily approaches an $H_{\alpha}$-independent function with
$r_{f,H}(0) \to 1$ in this limit.

It is apparent from fig.~\ref{higgscoupling} that the scalar Higgs
couplings generated radiatively through the $A$- or $A^{\prime}$-terms
enhances the ratio $r_{f,H}(0)$ in the heavy superpartner limit.  The
enhancement $r_{f,H} \geq 1$ persists for finite superpartner masses,
$ m_{\tilde{f}}^2, m_{\lambda}^2 \sim m_H^2$.  In this case the
Higgs--fermion Yukawa couplings have non-trivial form factors,
$\bar{h}_{f,H}=\bar{h}_{f,H}(q^2)$, in addition to the usual
renormalization group evolution of a Yukawa coupling.  For
$m_{\tilde{f}}^2, m_{\lambda}^2 \sim m_H^2$ the fractional change in
the loop functions between $q^2=0$ and $q^2=m_H^2$, relevant to the
coupling to a physical Higgs boson, can lead to significant variation
in the radiative Yukawa couplings.  Neglecting Higgs couplings through
$D$-terms and neutralino mixing effects (which are subdominant for
second and third generation radiative masses as mentioned above), the
general expression for the $q^2=m_H^2$ momentum dependence of the
radiative Higgs Yukawa couplings can be obtained from~(\ref{vertex1})
and~(\ref{Jfunct}).  These expressions are considerably simplified in
the limit of degenerate superpartners
$m_{\tilde{f}_1} = m_{\tilde{f}_2} =
m_\lambda \equiv \widetilde{m}$ for which
$J_1 (m_H^2; \widetilde{m}^{\,2},
             \widetilde{m}^{\,2},\widetilde{m}^{\,2} ) =
 J_2 (m_H^2; \widetilde{m}^{\,2},\widetilde{m}^{\,2},
             \widetilde{m}^{\,2} ) $.
The leading $q^2=m_H^2$ dependence of the ratio of physical Higgs to
mass Yukawa couplings~(\ref{higgsratio}) in the heavy degenerate
superpartner limit, $\widetilde{m}^{\,2} \gg m_H^2$, can then be
obtained from the limiting forms of the loop integrals given in
appendix~\ref{defC}.  For $\widetilde{m}^{\,2} \gg m_H^2$
\begin{equation}
r_{f,H}(m_H^2) = 1 + {1 \over 12} {m_H^2 \over \widetilde{m}^{\,2} }
 + \cdots
 \label{rqdep}
\end{equation}
where $+ \cdots$ represents higher-order dependence.  (The full
functional momentum dependence, is given below in the context of the
pseudoscalar Higgs radiative Yukawa coupling.)  For finite
superpartner masses it is possible to characterize the leading
low-energy momentum dependence in terms of a finite Higgs Yukawa 
radius,
\begin{equation}
\bar{R}_{f,H}^2 \equiv   {6 \over \bar{h}_{f,H}(0) }
{ \partial \bar{h}_{f,H}(0) \over  \partial q^2 }
\label{yukawaradius}
\end{equation}
analogous to the charge radius of an electromagnetic coupling.
{}From~(\ref{rqdep}) the Higgs Yukawa radius for a radiative fermion
mass in the heavy degenerate superpartner limit is
$\bar{R}_{f,H}^2(m_H^2) \simeq 1 / (2 \widetilde{m}^2)$.  In real
space this radius is determined by the Compton wavelength of the
virtual superpartners, and vanishes in the superpartner decoupling
limit.  Note that since both the Higgs Yukawa radius and the radiative
fermion mass arise from the same diagrams, the Higgs Yukawa radius is
effectively not suppressed by a loop factor.  This is unlike the case
of a tree-level Yukawa coupling, for which the Higgs Yukawa radius is
smaller by a loop factor than the Compton wavelength of the
contributing virtual particles.  Analogous radii can also be defined
from the form factors for the Higgsino couplings discussed in the next
subsection, and, in the case of an off-shell fermion, for the fermion
mass itself.

The effective radiative Yukawa couplings for the pseudoscalar Higgs,
$A^0$, differ somewhat from those of the scalar Higgs bosons.  The
pseudoscalar Higgs couplings in fig.~\ref{higgsdiag} only connect one
scalar partner mass eigenstate with the other mass eigenstate, unlike
the scalar Higgs couplings, which connect all possible combinations of
mass eigenstates.  Summing over scalar partner mass eigenstates, the
one-loop $A^0$ effective Yukawa coupling in the pure gaugino limit can
then be related to the fermion mass Yukawa by
\begin{equation}
 \bar{h}_{f,A}(m_{A^0}^2) = \frac{m_f}{| \langle H_{\alpha} \rangle |}
  ~J_2(m_{A^0}^2; m_{\tilde{f}_1}^2, m_\lambda^2,m_{\tilde{f}_2}^2 )\,,
\label{vertexA}
\end{equation}
where
$\lambda_{A^0} = \Theta \{\cos \beta, \sin \beta\} \bar{h}_{f,A}
    / \sqrt{2}$
for $\alpha = 1,2$ is the amplitude for coupling of the physical $A^0$
eigenstate to fermions (all considered on-shell), and $\Theta$
represents the projection of $A^0$ onto the Higgs doublets.  A
definition of the ratio of the effective $A^0$ Yukawa coupling to the
effective mass coupling may be introduced
\begin{equation}
  r_{f,A}(m_{A^0}^2)
\ \equiv \
 {\bar{h}_{f,A}(m_{A^0}^2) \over \bar{h}_{f,m}(0)} \,,
\label{pseudoscratio}
\end{equation}
analogous to that for the scalar Higgs coupling~(\ref{higgsratio}).

The one-loop result for the effective pseudoscalar Higgs Yukawa
coupling~(\ref{vertexA}) has a number of interesting properties.  It
is independent of the scalar partner mixing angles and only depends on
the superpartner spectrum and momentum transfer $q^2=m_{A^0}^2$.  In
addition, in the heavy superpartner decoupling limit,
$m_{\tilde{f}}^2, m_{\lambda}^2 \gg m_{A^0}^2$, the loop function may
be evaluated at $q^2=0$, for which the vertex loop function reduces to
the mass loop function, as discussed above, yielding 
$J_2(0; m_{\tilde{f}_1}^2, m_\lambda^2,m_{\tilde{f}_2}^2 )=1$.  The
intrinsic one-loop pseudoscalar radiative couplings evaluated at
$q^2=0$ therefore do not differ from the effective Yukawa coupling,
unlike the scalar Higgs--fermion couplings.  This result may also be
obtained from the general form of the radiative fermion
mass~(\ref{mfcn}).  The effective Yukawa coupling at $q^2=0$ of the
physical $A^0$ Higgs boson to a fermion, again with mixing effects
factored out, is very generally related to the mass by
\begin{equation}
\bar{h}_{f,A}(0) = {1 \over |H_{\alpha}|}
  {\partial m_f \over \partial \theta_{\alpha}} \,,
\end{equation}
where $H_{\alpha} = |H_{\alpha}| e^{i \theta_{\alpha}}$.  Since in the
pure gaugino limit, $\bar{h}_{f,A}$ is a function of the Higgs
doublets at one loop only through the combination 
$H_{\beta}^* H_{\beta}$, which is independent of $\theta_{\beta}$, the
pseudoscalar coupling arises only through the over-all Higgs doublet,
$H_{\alpha}$, multiplying $\bar{h}_{f,m}$ in~(\ref{mfcn}).  The ratio
of pseudoscalar Higgs to mass Yukawa couplings at $q^2=0$ neglecting
neutralino mixing is therefore
\begin{equation}
 r_{f,A}(0) = 1 \,,
\label{Aone}
\end{equation}
in agreement with the explicit calculation~(\ref{vertexA}) in the
$q^2=0$ limit.  This result amounts to a very general low-energy
theorem for the $q^2=0$ coupling of $A^0$ to fermions in any theory
with a single pair of Higgs doublets, and with only one doublet
coupling to the given fermion.  It can only be spoiled by effects that
allow an $H_1 H_2$ dependence of the effective Yukawa coupling.  In the
present context if radiative masses arise from either $A$- or
$A^{\prime}$-terms, but not both, such a dependence arises in the
one-loop diagram only the from gaugino--Higgsino mixing effects 
discussed above.  Inclusion of these effects does give 
${\cal O}(\mu m_Z^2 / (m_{\widetilde{B},\tilde{g}} \widetilde{m}^2))$ 
corrections to~(\ref{Aone}).  In contrast, inclusion of $D$-term
contributions to the scalar masses does not modify the ratio.

\begin{figure}[ht]
\begin{center}
\epsfxsize=12.0 cm
\leavevmode
\epsfbox{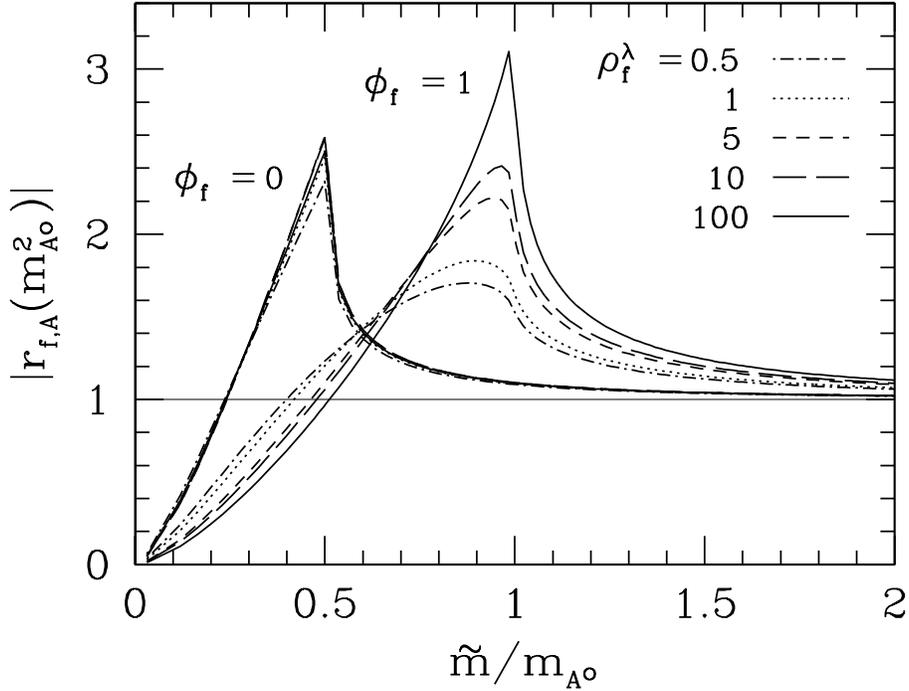}
\end{center}
\caption[f1]{Magnitude of the ratio $r_{f,A}(m_{A^0}^2)$ in the limit of
 degenerate scalar partners
 $ m_{\tilde{f}_1} \!=\! m_{\tilde{f}_2} \! =\! \widetilde{m}$
 ($\phi_f\! =\!0$),
 and of maximal splitting among superpartner masses
 $ m_{\tilde{f}_1}\! = \!0$, $ m_{\tilde{f}_2} \!=\! \widetilde{m}$
 ($\phi_f\! =\!1$),
 for different values of
 $\rho_f^\lambda$.
 In both cases, the ratio is shown as a function of $\widetilde{m}$
 divided by the pseudoscalar Higgs mass $m_{A^0}$. The horizontal line
 indicates the value $r_{f,A}(m_{A^0}^2) =1$.}
\label{rfafig}
\end{figure}
Since the $A^0$--fermion couplings are independent of scalar partner
mixing effects and $D$-term Higgs couplings, and are only modified by
finite-momentum effects in the pure gaugino limit, it is instructive
to consider the $q^2=m_{A^0}^2$ dependence explicitly.  {}From the
expressions for the loop functions given in appendix~\ref{defC}, in
the pure gaugino limit with degenerate scalar partners and gaugino of
mass
$\widetilde{m} = m_{\tilde{f}_1} \!=\! m_{\tilde{f}_2} \! =\!
\widetilde{m_{\lambda}}$,
the ratio of pseudoscalar Higgs Yukawa coupling to mass Yukawa coupling
is given by
\begin{equation}
r_{f,A}(m_{A^0}^2) =
\left\{ \begin{array}{ll}
 \displaystyle
  -  {\widetilde{m}^{\,2} \over m_{A^0}^2} ~
  \left[\log
 \left(\frac{1+\sqrt{1\!-\!4 \widetilde{m}^{\,2}/m_{A^0}^2 }}
     {1-\sqrt{1\!-\!4 \widetilde{m}^{\,2}/m_{A^0}^2 }} \right)
  - i \pi   \right]^2
         &   \quad \widetilde{m}  < {1 \over 2} m_{A^0}   \\[1.7ex]
 \displaystyle
 +  {4 \widetilde{m}^{\,2} \over m_{A^0}^2}
  \arcsin^2\left( {m_{A^0} \over 2 \widetilde{m} } \right)
         &   \quad \widetilde{m} \ge {1 \over 2} m_{A^0}
\end{array} \right. \,.
\label{rfamassdep}
\end{equation}
For $\widetilde{m} < {1 \over 2} m_{A^0}$ the imaginary piece arises
from the branch cut for physical intermediate states in the loop.  The
ratio $r_{f,A}(m_{A^0}^2) \ge 1$ for all
$\widetilde{m} > {1 \over 2} m_{A^0}$, and
$r_{f,A}(m_{A^0}^2) \to 0$ as $\widetilde{m} \to 0$.  The finite
momentum enhancement can be sizeable for $\widetilde{m} \sim m_{A^0}$,
and persists away from the degenerate superpartner limit.  The
$\widetilde{m} \gg m_{A^0} $ behavior is identical to the scalar Higgs
ratio~(\ref{rqdep}), with Higgs Yukawa radius vanishing in the strict
superpartner decoupling limit.
Figure~\ref{rfafig} shows the ratio $r_{f,A}(m_{A^0}^2)$ as a function
of $\widetilde{m}/ m_{A^0}$ for two extreme choices of the scalar
partner mass spectrum.  In the first case, $\phi_f = 0$, corresponding
to a degenerate scalar spectrum
$ m_{\tilde{f}_1} \!=\! m_{\tilde{f}_2} \! =\! \widetilde{m}$,
for different values of $\rho_f^\lambda$,
with $\rho_f^\lambda = \widetilde{m}^{\,2} /m_\lambda^2$ in this case.
The analytic expression given in eq.~(\ref{rfamassdep}) corresponds to
this case, with $\rho_f^\lambda =1$.  In the second case $\phi_f=1$,
corresponding to maximal splittings among superpartner masses,
$m_{\tilde{f}_1}\! = \!0$, $ m_{\tilde{f}_2} \!=\! \widetilde{m}$.
Again, $r_{f,A}(m_{A^0}^2)$ is shown for different values of
$\rho_f^\lambda$, which for this case
$\rho_f^\lambda = \widetilde{m}^{\,2} /(2 m_\lambda^2)$.
A simple, analytic expression can be obtained also in this case with
$\widetilde{m} = m_\lambda$, i.e.  $\rho_f^\lambda = {1 \over 2}$, by
using results listed in appendices~\ref{sec:defI} and~\ref{defC}.  For
a numerical evaluation of the loop function that appears in
$r_{f,A}(m_{A^0}^2)$, shown in fig.~\ref{rfafig}, see
ref.~\cite{DUTCH}.

As discussed in section~\ref{sec:probing} below,
measurements of $h^0$--, $H^0$--, and $A^0$--fermion couplings
offer the possibility of isolating intrinsic coupling effects from
momentum-dependent form factors.


\subsection{The Higgsino coupling}
\label{sec:higgsino}

\begin{figure}[ht]
\begin{center}
\leavevmode
\epsfxsize= 6.5 truecm
\epsfbox{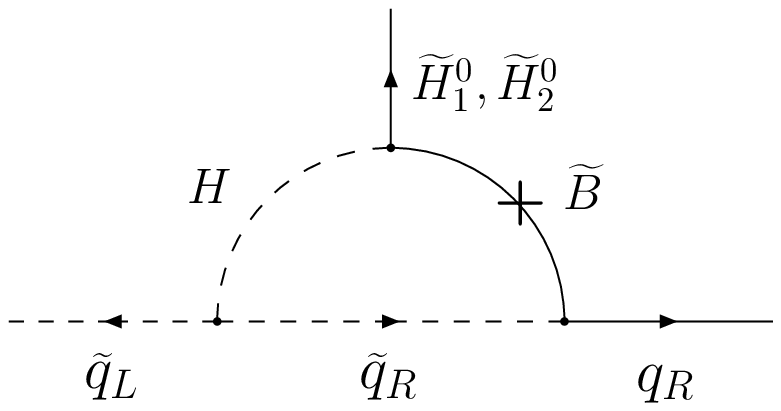}
\hspace*{1.5truecm}
\epsfxsize= 6.5 truecm
\epsfbox{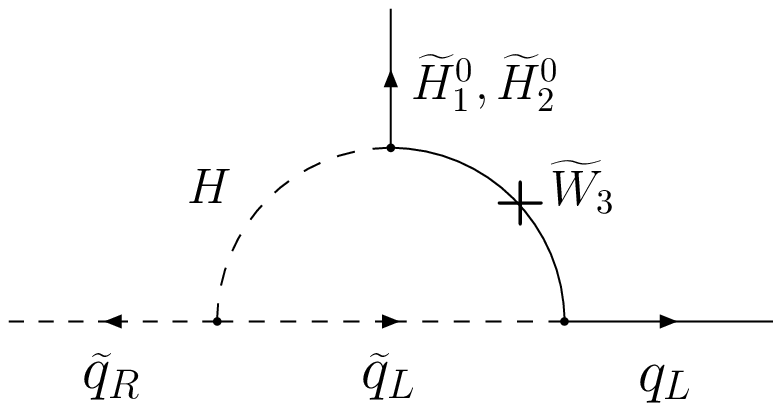}
\end{center}
\caption[f1]{One-loop radiative Wino and Bino contributions
 to the Higgsino--scalar--fermion couplings from soft flavor
 violation. The neutral Higgs $H$ exchanged virtually in the
 loop, can be $h^0$ or $H^0$.}
\label{higgsinodiag}
\end{figure}

The couplings of neutral Higgsinos ${\widetilde H}_1^0$,
${\widetilde H}_2^0$ to fermions $f$ and associated scalar partners
$\tilde f$, arise radiatively in a manner similar to the fermion
masses and Higgs couplings, as shown in fig.~\ref{higgsinodiag}.
The one-loop radiative couplings have to
be calculated separately for the two cases in
which the coupled fermions are both left- or right-handed.

The momentum-dependent coupling $\tilde{h}_{f_R} $
of right-handed fermions to left-handed scalars,
$\tilde{h}_{f_R} \overline{f}_R(q_1) \,
 {\widetilde H}_{\alpha}^0(q) \widetilde{f}_L (q_2)$,
arises, 
for example,
at one loop from pure
$\widetilde{B}$--$\widetilde{B}$ and mixed
$\widetilde{B}$--$\widetilde{W}$ propagators.
With all external particles on-shell, relevant to the decay
$ \widetilde{f}_h (q_2) \to f_R(q_1) \, {\widetilde \chi^0}_i(q) $,
where $ \widetilde{f}_h$ is the scalar partner eigenstate
labelled by $h$ ($h=1,2$), the 
neutralino--neutral Higgs contribution to the 
effective Yukawa coupling is
\begin{equation}
\tilde{h}_{f_R} =
-\frac{\alpha'}{8 \sqrt{2} \pi} \, A \,
 {\widetilde K}_{f_R}^{\,ijlkh} \,
 V_{\,ijlkh}
\label{vertexhinoR}
\end{equation}
with only the indices $ j l k $ summed.
The coupling
${\widetilde K}_{f_R}^{\,ijlkh} $ is given by
\begin{eqnarray}
 {\widetilde K}_{f_R}^{\,ijlkh} & = &
 \frac{Y_{f_R}}{2} N_{jB}
 \left[N_{jB} - \cot \theta_W N_{jW}\right] \times     \nonumber \\
                             &   &
\left[\delta_{l2}(N_{iH_1}\sin \alpha + N_{iH_2} \cos \alpha)
     -\delta_{l1}(N_{iH_1}\cos\alpha - N_{iH_2} \sin \alpha)
\right] \times                                          \nonumber \\
                             &   &
(U_{k1}U_{h2} + U_{k2} U_{h1} ) U_{k1} \,.
\end{eqnarray}
where $N_{iH_1}$, $N_{iH_2}$ are the projection factors for the
external neutralino ${\widetilde \chi}_i^0$ onto the Higgsino states
${\widetilde H}_1^0$ and ${\widetilde H}^0_2$ respectively. Note that
both $N_{iH_1}$ and $N_{iH_2}$ are, in general, non-vanishing,
regardless of which of the two Higgs expectation values induces the
corresponding fermion mass.  The index $j=1,\dots,4$ refers to the
external neutralino mass eigenstate; $l$ to the type of CP-even
neutral Higgs exchanged in the loop, $h^0$, $H^0$. The matrix $U$,
which diagonalizes the scalar mass matrix~(\ref{massmatr}), is defined
in appendix~\ref{sec:sfermion} and the loop function $V_{ijhkl}$ is
given explicitly in appendix~\ref{defC} in terms of $C_0$ and $B_0$
functions.
%
%

In the case of left-handed fermions, the momentum-dependent coupling
$\tilde{h}_{f_L}$ relative to the vertex
$\tilde{h}_{f_L} \overline{f}_L(q_1)
{\widetilde H}_{\alpha}^0(q) \widetilde{f}_R (q_2)$
can be mediated by pure $\widetilde{W}$--$\widetilde{W}$ and
$\widetilde{B}$--$\widetilde{B}$ as well as by mixed
$\widetilde{W}$--$\widetilde{B}$ propagators. For on-shell external
particles, as in the case of the decay
$ \widetilde{f}_h (q_2) \to f_L(q_1) \,{\widetilde \chi^0}_i(q) $,
the 
neutralino--neutral Higgs contribution to the 
effective Yukawa coupling is
\begin{equation}
 \tilde{h}_{f_L} =
 \frac{\alpha_2}{8 \sqrt{2} \pi} \, A \,
 {\widetilde K}_{f_L}^{\,ijlkh} \,
 V_{ijlkh}
\label{vertexhinoL}
\end{equation}
where ${\widetilde K}_{f_L}^{\,ijlkh} $ has the expression
\begin{eqnarray}
 {\widetilde K}_{f_L}^{\,ijlkh} & = &
 \left[T_f N_{jW} + \tan \theta_W \frac{Y_{f_L}}{2} N_{jB} \right]
 \left[N_{jW} - \tan \theta_W N_{jB}\right] \times     \nonumber \\
                             &   &
\left[\delta_{l1}(N_{iH_1}\cos \alpha - N_{iH_2} \sin \alpha)
     -\delta_{l2}(N_{iH_1}\sin\alpha + N_{iH_2} \cos \alpha)
\right] \times                                         \nonumber \\
                              &   &
(U_{k1}U_{h2} + U_{k2} U_{h1} ) U_{k2} \,.
\end{eqnarray}
Also in this case, both projection factors $N_{iH_1}$ and $N_{iH_2}$
for the external neutralino ${\widetilde \chi}_i^0$ onto the two
Higgsino states ${\widetilde H}_1^0$ and ${\widetilde H}^0_2$ are
present.

For both decays, 
$ \widetilde{f}_h (q_2) \to f_R(q_1) \, {\widetilde \chi^0}_i(q) $
and 
$ \widetilde{f}_h (q_2) \to f_L(q_1) \,{\widetilde \chi^0}_i(q) $,
there exist also chargino--charged Higgs loop contributions, which 
are not explicitly given here. Their dependence on gauge couplings 
is the same as that of neutralino--neutral Higgs contributions.

The radiatively generated effective Higgsino Yukawa coupling differs
from the effective mass and Higgs Yukawa couplings in a number of
ways.  Most importantly, radiatively induced quark masses get
contributions from both gluino and neutralino exchange, whereas the
Higgsino couplings only receive contributions from neutralino exchange
in the loop.  This profound difference is due to the absence of a
Higgsino analogue of the flavor-violating scalar tri-linear coupling,
combined with the requirement of gauge invariance.  Thus, radiative
Higgsino couplings differ significantly in magnitude from the
associated radiative mass or Higgs Yukawa couplings.
Furthermore, the fact that the right-handed fermion states couple only
through $U(1)_Y$ interactions, while the left-handed ones couple
through both $U(1)_Y$ and $SU(2)_L$ interactions, yields distinct
Higgsino couplings to the right- and left-handed fermions,
$\tilde{h}_{f_R}$ and $\tilde{h}_{f_L}$, with
$\tilde{h}_{f_R} / \tilde{h}_{f_L} \sim
  {\cal O}(\alpha^{\prime} / \alpha_2)$.
This compares with $\tilde{h}_{f_R} / \tilde{h}_{f_L} =1$ at lowest
order for a fermion mass arising from a hard tree-level Yukawa
coupling.  Finally, the type of Higgsino appearing in the coupling,
$\widetilde{H}_1^0$ or $\widetilde{H}_2^0$, need not coincide with the
type of Higgs expectation value which induces the fermion mass.  In
contrast, for a tree-level Yukawa coupling, only the Higgsino of the
same type appears.

In order to characterize Higgsino couplings to fermions, it is
useful to define their ratio to the effective mass Yukawa coupling,
in analogy with the Higgs coupling ratio~(\ref{higgsratio}).
Dropping momentum dependences, this ratio is:
\begin{equation}
\tilde{r}_{f_{L,R}} \equiv {\tilde{h}_{f_{L,R}} \over
   m_f / | \langle H_{\alpha} \rangle |  } \,.
\end{equation}
The general expression for these ratios is complicated and can be
extracted from~(\ref{vertexhinoR}),~(\ref{vertexhinoL}),
and~(\ref{higgsinovert}).  When superpartner masses are all of the
same order, however, the parametric dependence of
$\tilde{r}_{f_{L,R}}$ is readily determined from the above discussion.
For quarks that obtain a radiative mass from soft flavor breaking,
$\tilde{r}_{f_L}$ is of ${\cal O}(\alpha_2 / \alpha_s)$
and
$\tilde{r}_{f_R}$ of $ {\cal O}(\alpha^{\prime} / \alpha_s)$,
whereas for leptons,
$\tilde{r}_{f_L}$ is of $ {\cal O}(\alpha_2 / \alpha^{\prime})$
and
$\tilde{r}_{f_R}$ of ${\cal O}(1)$.
This is to be compared with a tree-level Yukawa coupling
for which $\tilde{r}_{f_{L,R}} =1$ at the lowest order.

The large differences between the Higgsino and Higgs or mass Yukawa
couplings are a concrete example of apparent hard
supersymmetry breaking from the low-energy point of view.  The large
splittings arise at the leading order and are not small corrections to
a supersymmetric relation.  In contrast, the equality of tree-level
Yukawa couplings in the supersymmetric limit is only slightly modified
by higher-order corrections when supersymmetry is broken.  With
radiative fermion masses, no symmetry exists in the high-energy theory
to enforce equality of the radiatively generated low-energy Higgs and
Higgsino couplings since these couplings vanish in the supersymmetric
limit.  Large splittings for the Higgsino--fermion--scalar couplings
therefore represent a ``smoking gun'' for radiatively generated
fermion masses from soft chiral flavor breaking.  As discussed in
section~\ref{sec:probing}, however, the Higgsino couplings are likely
to be difficult to measure experimentally.

%
%

\section{Phenomenological radiative models}
\label{sec:possible}

The magnitude of fermion masses which arise from soft chiral flavor
violation depends on the soft supersymmetry-breaking parameters
$m_{\lambda}, m_{\tilde{f}_i}$ and $A$ or $A^{\prime}$, as detailed in
section~\ref{sec:mass}.  With the radiative mass given in
eq.~(\ref{mass}), a given quark or lepton mass can be estimated from
the magnitude of the ratio $A m_{\lambda} / \widetilde{m}^{\,2}$ or
$A^{\prime} m_{\lambda} / \widetilde{m}^{\,2}$
\begin{eqnarray}
{\rm Lepton}:~ & &
   \frac{m_{l}}{30\! -\! 100{\mbox{ MeV}}} \sim
\left\{ \sqrt{2} \cos \beta ~
   {A m_{\tilde{B}} \over \widetilde{m}^{\,2}}~,~
        \sqrt{2} \sin \beta ~
   {A^{\prime} m_{\tilde{B}} \over \widetilde{m}^{\,2}}
        \right\}  \nonumber \\
{\rm Down\!-\!type~quark}:~ & &
 \ \ \ \frac{m_{q}}{1\! -\! 3 {\mbox{ GeV}}} \ \ \sim
\left\{  \sqrt{2} \cos \beta ~
   {A m_{\tilde{g}} \over \widetilde{m}^{\,2}}  ~,~
        \sqrt{2} \sin \beta ~
   {A^{\prime} m_{\tilde{g}} \over \widetilde{m}^{\,2}}
        \right\}  \nonumber \\
{\rm Up\!-\!type~quark}:~ & &
 \ \ \ \frac{m_{q}}{1\! - \!3 {\mbox{ GeV}}} \ \ \sim
\left\{ \sqrt{2} \sin \beta ~
   {A m_{\tilde{g}} \over \widetilde{m}^{\,2}}~,~
        \sqrt{2} \cos \beta ~
   {A^{\prime} m_{\tilde{g}} \over \widetilde{m}^{\,2}}
        \right\}  \nonumber \\
\label{a}
\end{eqnarray}
where the first (second) terms in the curly brackets correspond to
holomorphic (non-holomorphic) soft flavor breaking,
$m_{\widetilde{B}}$ and $m_{\tilde{g}}$ are the Bino and gluino mass,
respectively, and $\widetilde{m}
=\max(m_{\tilde{f}_{1}},m_{\tilde{f}_{2}},m_{\lambda})$.  For leptons,
the radiative mass is assumed to be dominated by Bino exchange, which
is the case in the mostly gaugino or Higgsino region of parameter
space.  For the numerical estimates~(\ref{a}) the loop function
$\widetilde{m}^{\,2} I(m^2_{\tilde{f}_1}, m^2_{\tilde{f}_2},
m^2_{\lambda})$ 
is assumed to be in the range ${1 \over 3}$--$1$, and 
$\langle H_{\alpha} \rangle \simeq 175 \cos \beta$ $(\sin \beta)$ GeV
for $\alpha=1$ $(2)$.

The radiative masses are maximized if the scalar partner and gaugino
masses are of the same order, $m_{\lambda} \sim m_{\tilde{f}}$.
Stability of the charge- and color-preserving vacuum, with vanishing
squark and slepton expectation values, places an upper limit on the
scalar tri-linear $A$- or $A^{\prime}$-parameters.  The specific
bounds are model-dependent and are detailed in
section~\ref{sec:stability}.  In general $A/\widetilde{m}$ or
$A^{\prime}/\widetilde{m} \lsim$ few, is required for stability of the
charge- and color-preserving vacuum, where $\widetilde{m}$ is a scalar
mass.  This constraint implies that the ratios 
$A m_{\lambda} / \widetilde{m}^{\,2}$ and 
$A^{\prime} m_{\lambda} / \widetilde{m}^{\,2}$ 
cannot be much larger than a few.  With this upper limit and the
numerical values~(\ref{a}), it is clear that radiative fermion masses
for the first generation can be easily accommodated.
Second-generation masses can also be accommodated, depending on the
size of the ratios $A m_{\lambda} / \widetilde{m}^{\,2}$ or
$A^{\prime} m_{\lambda} / \widetilde{m}^{\,2}$.  For the third
generation, it is (not surprisingly) extremely unlikely that the
$\tau$-lepton or top quark obtain masses from soft chiral flavor
violation, because of the very large tri-linear terms required.
Avoiding the stability constraints discussed in
section~\ref{sec:stability} would require in these cases an extreme
tuning of the potential or loss of perturbativity in the scalar sector
at a very low scale.  However, it is possible in the third generation
to obtain a radiative $b$-quark mass by this mechanism.
\begin{table}
\caption{Magnitude of soft chiral flavor breaking
$A$- or $A^{\prime}$-terms required for soft radiative fermion masses.
For down-type quarks and leptons the 
first term
in brackets corresponds to holomorphic $A$-terms, the
second
to non-holomorphic $A^{\prime}$-terms,
and vice versa for up-type quarks.
Approximate masses are specified at the electroweak scale.
}
\label{table:t2}
\begin{tabular}{ccc}
Fermion & $m_{f}$ (MeV) & $A m_{\lambda} / \widetilde{m}^{\,2}$
                       or $A^{\prime} m_{\lambda} / \widetilde{m}^{\,2}$
  \\
& & $ \times \left\{ \sqrt{2} \cos \beta, \sqrt{2} \sin \beta \right\}$
  \\ \hline
$e$    & $0.5$        & $(0.5$--$1.5) \times 10^{-3}$ \\
$u$    & $1$          & $(0.3$--$1) \times 10^{-3}$   \\
$d$    & $5$          & $(2$--$5) \times 10^{-3}$     \\
$\mu$  & $100$        & $1$--$3$                      \\
$c$    & $700 $       & $0.2$--$0.7 $                 \\
$s$    & $ 100$       & $0.03$--$0.1$                 \\
$b$    & $3000 $      & $ 1$--$3$                     \\
\end{tabular}
\end{table}
The magnitude of the $A$- or $A^{\prime}$-terms required for
soft radiative masses are listed in table~\ref{table:t2}.
(For a recent review of numerical values for running fermion masses,
see~\cite{FK}.)

The magnitude of soft radiative masses depends on the ratio of Higgs
doublet expectation values, $\tan \beta$, as given in
table~\ref{table:t2}.  For holomorphic chiral flavor breaking in
$A$-terms, radiative fermion masses arise from the same Higgs doublet
as for tree-level Yukawa couplings.  However for non-holomorphic
breaking in $A^{\prime}$-terms, radiative fermion masses arise from
the ``wrong Higgs'' doublet, and therefore have the ``wrong''
dependence on $\tan \beta$.  Soft radiative muon or $b$-quark masses
require relatively large scalar tri-linear terms, proportional to
$\cos \beta$ in the holomorphic case and $\sin \beta$ in the
non-holomorphic case.  So if either of these fermions gain a mass from
holomorphic soft chiral flavor breaking in $A$-terms, the upper limit
on the tri-linear terms from stability of the potential,
$A/\widetilde{m} \lsim$ few, discussed in section~\ref{sec:stability},
strongly disfavors large values of $\tan \beta$.  This is simply
because the effective Yukawa coupling required for a given muon or
$b$-quark mass increases with $\tan \beta$.  Alternatively, large
values of $\tan \beta$ can be accommodated with radiative masses for
either the muon or $b$-quark from non-holomorphic flavor breaking in
$A^{\prime}$-terms.  If the charm quark mass, as well as either the
muon or $b$-quark mass, arise from solely holomorphic or solely
non-holomorphic chiral flavor violation, larger values of $\tan \beta$
are disfavored so that none of the required tri-linear terms is too
large.  Note that if all standard model fermions obtained mass
radiatively from chiral flavor violation in purely non-holomorphic
$A^{\prime}$-terms, the role of $H_1$ and $H_2$ in the radiative
Yukawa couplings would be precisely the reverse of what it is in the
minimal case with tree-level masses.  In this case, there would be no
observational signature of ``wrong Higgs'' couplings, because the
physical observables in the bosonic sector of the theory are invariant
under $\tan \beta \leftrightarrow {\rm cot} \beta $.  However, as
noted above, vacuum stability and the absence of fine tuning are not
compatible with purely radiative top-quark and $\tau$-lepton masses.
Therefore in practice with radiative mass arising from
$A^{\prime}$-terms, observable differences from minimal models are
expected, as will be discussed in section~\ref{sec:probing}.

A corollary of a radiative soft fermion mass from chiral flavor
breaking is an enhanced left--right mixing for the associated scalars,
$m_{LR}^2 = A \langle H_{\alpha} \rangle$ or $ A^{\prime} \langle
H_{\alpha} \rangle$.  In a theory with tree-level Yukawa couplings,
the mixing scales as the product of fermion and scalar partner masses
 $m_{LR}^{2} \sim m_{f} m_{\tilde{f}}$.  However with a soft radiative
mass the mixing is effectively a loop factor larger, 
$m_{LR}^{2} \sim (4 \pi / \alpha) m_{f} \widetilde{m}^{\,2}
 / m_{\lambda}$, where $\alpha = \alpha_s$ or $\alpha^{\prime}$ 
for quarks or leptons.  In the case of radiative second-generation
masses, especially for the muon or charm quark, or radiative $b$-quark
mass, the mixing can be near maximal because of the required large
tri-linear term.  This has potentially directly observable
consequences for scalar partner production and decay, as discussed in
section~\ref{sec:probing}.

In the quark sector, the CKM mixing matrix is given by 
$V_{CKM} = V^{\dagger}_u V_d$, where $V_u$ and $V_d$ are the unitary
matrices that diagonalize the left-handed up- and down-type quarks
respectively.  If some of the quark masses arise radiatively, then the
form of $V_u$ and/or $V_d$ follows in part from that of the scalar
tri-linear terms.  In particular, if any quark receives a mass
predominantly from chiral flavor breaking, the source of CP violation
in the CKM matrix and QCD vacuum angle are functions of the relative
phases between the tree-level Yukawa couplings and tri-linear terms.
(The somewhat related idea that tree-level Yukawa couplings could be
purely real with all CP violation arising from $A$-terms was recently
discussed in ref.~\cite{ABEL}.)

The pattern of tree-level and radiative quark masses can in principle
arise in any number of ways.  We do not address specific soft and hard
textures in this paper, but comment on some interesting possibilities.
As discussed above, the down-type quark masses could be purely
radiative.  If this is the case, and all up-type quark masses are
tree-level, then of course $V_d$ follows entirely from the scalar
tri-linear terms, while $V_u$ follows from the superpotential up-type
Yukawa couplings.  In this scenario the breaking of down-type right-handed
quark symmetries solely in soft tri-linear terms can be naturally
implemented by continuous $U(1)_R$ or discrete $R$-symmetries.  An
accidental $U(1)_R$ symmetry of this type would arise automatically
if, for example, down-type quark symmetries are only broken through
auxiliary expectation values, either spontaneously or explicitly in
the supersymmetry-breaking sector, while being respected in the
supersymmetric sector of the theory.  Alternatively, if some of the
up-type quark masses are radiative, or only some of the down-type
quark masses are radiative, with the remainder tree-level, then $V_u$
and/or $V_d$ are block diagonal.  Non-trivial three-generation mixing
requires that $V_u$ and $V_d$ be not simultaneously block diagonal.
This type of texture may be enforced by symmetries.  It is also
possible that in the absence of any specific horizontal
$R$-symmetries, all fermions receive mass both radiatively from soft
breaking, and at tree level from a Yukawa coupling.  This would arise
in a Froggatt--Nielsen mechanism in which flavor symmetries are broken
in both scalar and auxiliary directions.  However, for the radiative
contribution to be significant in this scheme, the breaking in the
auxiliary directions must be stronger than in the scalar directions,
so that they overcome the loop factor.  However, we concentrate on the
cases of purely radiative or dominantly tree-level masses for a given
fermion.

Radiative generation of fermion masses can lead to new sources of
flavor violation.  In the electroweak scale theory, these amount to
misalignments in flavor space between the effective Yukawa couplings,
which determine both the fermion mass eigenstates and Higgs sector
couplings, and the $A$- or $A^{\prime}$-terms and/or left- and
right-handed scalar mass-squared matrices, which determine the squark
and slepton eigenstates.  Misalignment between these terms in general
leads to flavor-changing gaugino couplings, which link the fermion and
scalar partner sectors, and in Higgs boson and longitudinal gauge
couplings to fermions.  The magnitudes of induced flavor violations
are model-dependent functions of the specific flavor textures and
superpartner mass spectrum, and are outside the scope of this work.
However, it is worth while to discuss some general features and
scenarios in which the new sources of flavor violation associated with
radiative masses are suppressed or eliminated.

Dangerous flavor-changing neutral currents in fermionic couplings to
Higgs bosons and longitudinal gauge bosons potentially can be produced
by misalignment in flavor space between the fermionic couplings to the
Higgs doublet expectation values, which give rise to the fermion mass
matrix, and the fermionic couplings to the physical Higgs bosons.
However, this source of flavor violation is automatically avoided, with
no misalignment of the physical Higgs couplings, if the
Glashow--Weinberg~\cite{GWCONDITION} criterion of coupling all
up-type quarks to a single Higgs doublet, all down-type quarks to a
single Higgs doublet, and all leptons to a single Higgs doublet, is
satisfied.  Since the top-quark and $\tau$-lepton masses must be at tree
level, this condition requires, with a minimal Higgs sector, that
any radiative up-type quark or lepton mass arises from holomorphic
chiral flavor breaking, and likewise if only some of the down-type
quark masses arise radiatively.  This potential source of flavor
violation is also avoided if all down-type quark masses arise solely
from non-holomorphic tri-linear terms.  With any of these
conditions, the charged Higgs couplings to fermions are also
proportional to the charged current CKM mixing matrix, and thus
generally safe.

Flavor violation in the superpartner sector is not as readily
eliminated.  Misalignment between the full fermion Yukawa couplings
and $A$- or $A^{\prime}$-terms in general arises if a fermion receives
mass both radiatively and from a tree-level Yukawa, since the flavor
structure of these contributions need not coincide.  This misalignment
between the hard and soft components of effective Yukawa couplings is
eliminated if all fermions receive mass solely from either soft or
hard flavor breaking, but not from both.  For the quark sector, in
which the top quark Yukawa must be at tree level, this source of
misalignment is avoided, for example, in the scenario discussed above,
in which all down-type chiral flavor symmetries are broken in the
supersymmetry-breaking sector, while up-type chiral flavor symmetries
are broken by tree-level Yukawa couplings.  In the lepton sector, in
which the $\tau$-lepton Yukawa must be at tree level, the above
misalignment is analogously eliminated if the muon and electron chiral
flavor symmetries are only broken by auxiliary components in the
supersymmetry-breaking sector.

Even if every fermion receives mass solely from either hard or soft
flavor breaking, precise alignment between the effective Yukawa
couplings and tri-linear $A$- or $A^{\prime}$-terms is not guaranteed,
if the scalar partner mass matrices have a non-trivial flavor
structure.  One possible sufficient condition, which guarantees
complete alignment, is for all the left- and right-handed scalar
superpartners in each sector to be degenerate, with masses squared of
$m_{LL}^2$ and $m_{RR}^2$ respectively.  The radiative Yukawa
couplings are then proportional to the tri-linear $A$- or
$A^{\prime}$-terms.  Consider, for example, the scenario mentioned
above in which up-type quark chiral flavor symmetries are broken only
by tree-level superpotential Yukawa couplings, while down-type quark
symmetries are broken only in the soft tri-linear terms.  Then the
down-type effective Yukawa couplings with degenerate squarks are
$h_{d,ij} =
A_{d,ij} f(m_{\tilde{d_1}}^2,m_{\tilde{d_2}}^2,m_{\lambda}^2)$,
where $i,j$ refer to flavor, and $f$ is a flavor-independent loop
function, which may be obtained from~(\ref{mass}).  The
proportionality implies that the fermion and scalar mass matrices are
simultaneously diagonalized, with the result that neutralino couplings
are flavor-conserving.  Chargino couplings are also proportional to
the charged current CKM mixing matrix.  Precise alignment in this case
may also be understood by promoting the flavor symmetries to
background symmetries and treat the down-type quark tri-linear terms
and up-type quark Yukawa couplings as spurions that spontaneously
break the symmetries.  Invariance with respect to the background
symmetries then implies that any chirality-violating amplitude between
external quarks is exactly proportional to a single power of either of
these spurions, and therefore has precisely the same flavor structure
as the quark mass matrices.  This particular alignment is of course in
general spoiled by non-degeneracy among the scalar partners.  Also,
the required degeneracy does not follow {\it a priori} from any
symmetry since the flavor symmetries are broken in the tri-linear
terms or Yukawa couplings.  Near degeneracy might arise dynamically,
however, if there is a relatively large flavor neutral contribution to
the soft scalar masses.

There are other scenarios that avoid dangerous levels of quark-sector
flavor violation associated with soft radiative masses.  If all
down-type quarks receive mass from soft flavor breaking as outlined
above, but the individual down-type quark number is conserved, the
down-type squark mass-squared matrices and $A$- or $A^{\prime}$-terms
are diagonal (though not necessarily degenerate), as are the down-type
effective Yukawa couplings.  In this case $V_{CKM} = V_u^{\dagger}$,
and all quark mixing and CP violation in the CKM matrix arise from
tree-level Yukawa couplings in the up sector.  In addition, the
chargino couplings are proportional to the charged current CKM mixing
matrix.  Then all supersymmetric flavor violation occurs in the
up-type quark sector, from possible mismatch between the quark and
squark mass matrices.  At present, experimental probes in this sector
are not very sensitive to such flavor violation.  An analogous
scenario may also be extended to the leptons, with individual lepton
number conserved~\cite{MaNgWong}.

A very general scenario in which low-energy flavor violations arising
from supersymmetric effects are suppressed results if some of the
superpartners are much heavier than the electroweak scale.  Requiring
that two-loop quadratically divergent gauge contributions to the Higgs
potential from heavy scalars lead to not excessive tuning of
electroweak symmetry breaking implies that first- and
second-generation scalar partners and the gluino are lighter than
${\cal O}(20 ~{\rm TeV})$~\cite{CKN}.  The analogous naturalness
requirement from one-loop quadratically divergent contributions from
heavy electroweak gauginos implies that these gauginos are lighter
than ${\cal O}(2~{\rm TeV})$.  Finally, for the third generation, a
similar naturalness requirement from one-loop quadratically divergent
contribution through the tree-level top Yukawa implies that the left-
and right-handed top squarks, and the left-handed $b$-squark are
lighter than ${\cal O}(1~{\rm TeV})$.  In the scenario for radiative
fermion masses, consider the case where the first-generation scalars
are as heavy as the naturalness bounds allow.  If all gauginos have
masses at the electroweak scale, the hierarchy 
$m_{\lambda} \ll m_{\tilde{f}}$ provides an attractive explanation for
the smallness of first-generation radiative masses.  As discussed
above, radiative masses for second-generation quarks and the $b$-quark
require $m_{\lambda} \sim m_{\tilde{f}}$.  In this case, the
associated scalars cannot be as heavy as allowed by the naturalness
bounds unless the gluino is also very heavy.  Even if only the
first-generation scalars are heavy, flavor violation involving
first-generation fermions arising from scalar partner mixing is
suppressed by the large mass splitting between first- and
second-generation scalars.

%
%

\section{Stability analysis of the scalar potential}
\label{sec:stability}

Radiatively generated fermion masses arising from soft chiral flavor
violation require sizeable scalar tri-linear couplings, in particular
$A \gg h {m}_{\tilde{f}}$, where $h$ is the effective Yukawa coupling,
and ${m}_{\tilde{f}}$ represents the scalar masses.  For
second-generation and $b$-quark radiative masses, 
$A \sim {m}_{\tilde{f}}$ are required, as discussed in the previous
section and shown in table~\ref {table:t2}.  Such large tri-linear
terms give negative contributions to the scalar potential along
certain directions in field space.  If large enough, these can lead to
color- and/or charge-breaking minima with non-vanishing squark or
slepton expectation values.  The possible existence of these vacua
gives an upper limit on the magnitude of the tri-linear couplings, and
consequently on the radiatively induced fermion mass.

In the vacuum with electroweak symmetry broken by Higgs field
expectation values, the scalar tri-linear terms~({\ref{ophol})
and~(\ref{opnonhol}) contribute a left--right mixing term to the
squark or slepton mass-squared matrices, 
$m_{LR}^{2} = A \langle H_{\alpha} \rangle$ or 
$A^{\prime} \langle H_{\alpha}^{*} \rangle$, which induces the
radiative fermion masses.  These mixings cause a level repulsion
between the scalar mass eigenvalues.  Stable electroweak symmetry
breaking without color or charge breaking requires that the lightest
squark or slepton eigenstate does not become tachyonic.  This puts an
upper limit on the $A$-parameters of 
$|A \langle H_{\alpha} \rangle| < m_{LL} m_{RR}$ or 
$|A^{\prime} \langle H_{\alpha} \rangle | < m_{LL} m_{RR}$.  For
second-generation and $b$-quark radiative masses, for which 
$A \gsim {m}_{LL,RR}$, then 
$m_{LL,RR} \gsim \langle H_{\alpha} \rangle$ is required in order to
achieve stable electroweak symmetry breaking.  This is easily
satisfied in the scenarios discussed in section~\ref{sec:possible}
with sufficiently massive squark or sleptons.

The negative contribution of the tri-linear terms to the total
potential is maximized in a direction along which all scalar fields
that appear in the tri-linear term have equal expectation values:
\begin{equation}
|H_{\alpha}| = |\phi_L| = |\phi_R| = {1 \over \sqrt{6}} \phi \,,
\label{direction}
\end{equation}
where gauge indices and an over-all phase have been suppressed, and
$\phi$ is a real valued collective coordinate~\cite{GHS,LP}.  Even if
the constraint given above for locally stable electroweak symmetry
breaking is satisfied, color- and/or charge-breaking vacua can appear
along this direction at larger field values. With the
normalization~(\ref{direction}), $\phi$ is a canonically normalized
real scalar field with general renormalizable tree-level potential
\begin{equation}
V(\phi) = {1 \over 2} \widetilde{m}^{\,2} \phi^2
    - {1 \over 3 \sqrt{6}} A \phi^3
        + {1 \over 36} \lambda^2 \phi^4 \,,
\label{phipot}
\end{equation}
where
$\widetilde{m}^{\,2} \equiv {1 \over 3} \left( m_{LL}^2
 + m_{RR}^2 + m_{H_{\alpha}}^2 \right)$.
The Higgs mass term $ m_{H_{\alpha}}^2$ receives a contribution from
the Higgs Dirac mass in the superpotential $W \supset \mu H_1 H_2$.
Hence, $\widetilde{m}^{\,2} \gg m_{LL,RR}^2$ is possible in principle
with a large Dirac mass, $\mu^2 \gg m_{LL,RR}^2$.  However, minimal
tuning of electroweak symmetry breaking usually implies
$|m_{H_{\alpha}}^2| \sim m_Z^2 \lesssim m_{LL,RR}^2$.  In what follows
$\widetilde{m} \sim m_{LL,RR}$ is implicitly assumed.  The quartic
coupling $(1 / 36) \lambda^2 \phi^4$ is crucial for global stability
and can arise in a number of ways, as discussed below.

The potential~(\ref{phipot}) in general has a local charge- and/or
color-breaking minimum at non-vanishing $\phi$, in addition to the
charge and color preserving minimum $\phi=0$.  A sufficient condition
to avoid color or charge breaking is to require that the deepest
minimum along~(\ref{direction}) is at the origin.  The global minimum
of the theory then conserves color and charge and is absolutely
stable.  For the potential~(\ref{phipot}) this stability constraint
gives an upper limit $A/ \widetilde{m} < \sqrt{3}
\lambda$~\cite{GHS,LP}, and is plotted in fig.~\ref{bouncefig}.
%
%
\begin{figure}[ht]
\begin{center}
\epsfxsize= 11.8 cm
\leavevmode
\epsfbox{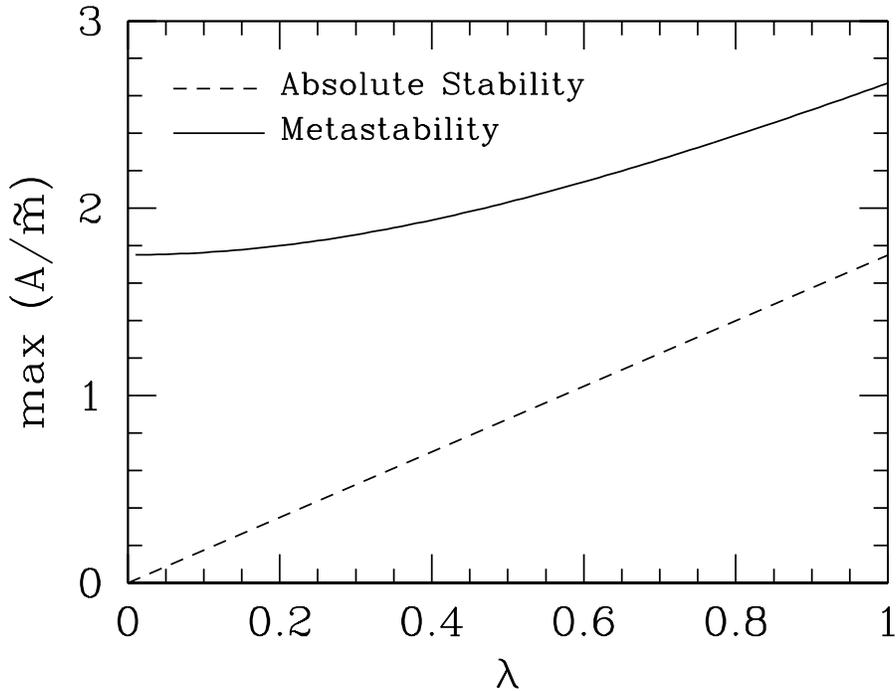}
\end{center}
\caption[f1]{Maximal value of $A/\widetilde{m}$ for absolute
 stability and metastability of tree-level scalar
 potential as a function of the quartic coupling $\lambda$,
 where $\widetilde{m}^{\,2} \equiv {1 \over 3}
 \left( m_{LL}^2\!+\! m_{RR}^2 \!+\! m_{H_{\alpha}}^2 \right)$.
 Absolute stability refers to the absence of global
 charge-breaking minima, while metastability refers to a
 lifetime of the charge preserving vacuum greater than the
 age of the Universe, corresponding to a bounce action
 $S > 400$.}
\label{bouncefig}
\end{figure}

If the absolute stability constraint is not satisfied, and the color-
and charge-breaking vacuum is the global minimum, a weaker necessary
condition is that the time to tunnel from the color and charge
preserving metastable vacuum be greater than the present age of the
Universe.  Cosmological selection of the color and charge preserving
vacuum at the origin is natural, since this is a point of enhanced
symmetry and always a local minimum of the free energy at high enough
temperature.  For the potential~(\ref{phipot}) with electroweak scale
mass $\widetilde{m}$, a lifetime greater than the present age of the
Universe corresponds to a bounce action out of the metastable vacuum
along the direction~(\ref{direction}) of 
$S \gsim 400$~\cite{STABILITY}.  
The bounce action for the potential~(\ref{phipot}) may be
calculated numerically~\cite{STABILITY,SARID}.  The action
interpolates between the thin-wall limit, 
$\lambda^2 {\widetilde{m}^2}/A^2\rightarrow 1/3^-$, in which the two 
vacua are nearly degenerate~\cite{THINWALL}:
$$
S_{\rm thin} \simeq \frac{9 \pi^2}{2}
 \left(\frac{\widetilde{m}^{\,2}}{A^2}\right)
 \left( 1- 3 \lambda^2 \frac{\widetilde{m}^{\,2}}{A^2}\right)^{-3}\,,
$$
and the thick-wall limit,
$\left\vert\lambda^2 {\widetilde{m}^{2}}/{A^2}\right\vert \ll 1$,
in which the quartic term is unimportant~\cite{THICKWALL}
$$
S_{\rm thick} \simeq 1225
\left(\frac{\widetilde{m}^{\,2}}{A^2}\right) \,.
$$
The maximum value of $A/\widetilde{m}$ for which $S > 400$ is shown in
fig.~\ref{bouncefig} as a function of the quartic coupling coefficient
$\lambda$.  This figure employs the empirical fit to the numerically
calculated bounce action given in ref.~\cite{SARID}.  The maximum
allowed $A/\widetilde{m}$ continues as a smooth monotonically
decreasing function for $\lambda^2 <0$, even though the renormalizable
potential is unbounded from below in this case.  For 
$|\lambda^2| \ll 1$ the weak requirement of metastability on a
cosmological time scale is met for $A/\widetilde{m} \lesssim 1.75$.

The magnitude of the quartic coupling for the
direction~(\ref{direction}) is clearly crucial for determining the
maximum allowed value of $A/\widetilde{m}$ implied by the absolute or
metastability constraints given above.  This in turn determines the
maximum allowed radiatively induced fermion masses.  Quartic couplings
can arise from the $D$-term gauge potential or tree-level Yukawa
coupling $F$-terms with standard model fields or mirror matter.  These
contributions distinguish between different classes of models.

$\bullet$
{\em Minimal holomorphic models}

A minimal class of models with soft radiative fermion masses are those
with the holomorphic tri-linear terms $AH_{\alpha} \phi_L \phi_R$, and
no additional significant tree-level Yukawa couplings to the
associated fermions.  With holomorphic tri-linear terms, gauge
invariance implies that the $D$-term gauge potential vanishes along
the direction~(\ref{direction}).  This is guaranteed since holomorphic
directions are invariant with respect to complexified gauge
transformations and are therefore $D$-flat.  Without additional
superpotential couplings, the quartic coupling along the
direction~(\ref{direction}) vanishes and the tree-level
potential~(\ref{phipot}) is then unbounded from below.  This persists
at one loop, where a small negative quartic coupling is generated:
$\lambda^2 \simeq - (A/\widetilde{m})^4/96 \pi^2$.  Radiative fermion
masses in this class of models therefore have a metastable vacuum, and
absolute stability of the charge and color preserving vacuum is not
possible.
In this case, however, the cosmological
metastability bound is easily satisfied for first-generation masses
for which $A /\widetilde{m} \sim 10^{-3}$ is required.
Second-generation masses for charm and strange quarks, which require
$A/\widetilde{m} \sim 10^{-1}$--$1$, can also be accommodated, while
for the muon and $b$-quark $A/\widetilde{m} \sim 1$--$3$ is very close
to the maximum value allowed by metastability of 
$A/\widetilde{m} \lesssim 1.75$.


$\bullet$
{\em Minimal non-holomorphic models}

A non-holomorphic tri-linear term 
$A^{\prime} H^*_{\alpha} \phi_L \phi_R$ can also give rise to
radiative fermion masses.  In this case the operator does not
correspond to a holomorphic direction and is not necessarily $D$-flat.
The $SU(3)_C$ and $SU(2)_L$ $D$-term gauge potentials do in fact
vanish along the direction~(\ref{direction}), but the $U(1)_Y$ gauge
potential does not.  The $D$-term potential in this case is then
\begin{equation}
V_D = {{g^{\prime}}^2 \over 2} \left| 
       {\rm Tr}~{Y_i \over 2}
          \phi_i^* \phi_i \right|^2
    = {{g^{\prime}}^2 \over 288} \left( 
           {\rm Tr}~ Y_i \right)^2 \phi^4 \,,
\end{equation}
where the trace is over all fields along the collective direction.
{}For either the quark or lepton non-holomorphic tri-linear terms,
this trace is $|{\rm Tr} ~Y_i | = 2$.  The $U(1)_Y$ $D$-term 
contribution to the quartic coupling along these directions is then 
$\lambda = g^{\prime}/\sqrt{2} \simeq 0.25$.

With this tree-level quartic coupling, global stability of the charge
and color preserving vacuum is easily achieved for first-generation
radiative masses.  In addition, second-generation strange and charm
quark and muon radiative masses do not necessarily lead to a deeper
charge- and color-obreaking minimum.  A radiative $b$-quark mass does
lead to a metastable vacuum in this class of models. From
fig.~\ref{bouncefig} it is apparent that the value of the quartic
coupling arising from the $U(1)_Y$ gauge potential is not large enough
to significantly modify the metastability constraints with respect to
the case of vanishing quartic coupling.


$\bullet$
{\em Tree--level top Yukawa}

Quartic couplings for the direction~(\ref{direction}) can also arise
from superpotential $F$-terms.  Without additional matter, this
requires that some of the standard model fermions have tree-level
superpotential Yukawa couplings.  In the case of a tree-level top
quark Yukawa coupling, $W = h_{t}H_{2}QU$, an effective quartic
coupling is generated for the direction~(\ref{direction}) relevant to
a radiative $b$-quark mass from a non-holomorphic $A_b^\prime$ term:
\begin{equation}
V_F = \left| {\partial W  \over \partial U} \right|^{2}
  = h_{t}^{2}  \left| H_{2} Q \right|^2
  = {1 \over 36} \lambda^2 \phi^4 \,.
\end{equation}
The tree-level top Yukawa contribution to the quartic coupling along
this direction is then $\lambda=h_t$.  With such a large quartic
coupling, both the metastability and absolute stability bounds on
$A^{\prime}/\widetilde{m}$ are increased and a radiative $b$-quark
mass can be obtained with 
$A_{b}^{\prime}/\widetilde{m} \sim 1$--$2.5$, while preserving the
absolute stability of the color and charge preserving global minimum.
In a hybrid model with non-holomorphic tri-linear scalar terms, and a
tree-level top Yukawa, it is therefore possible to obtain all fermion
masses radiatively, aside from the top quark and $\tau$-lepton, while
preserving the global charge and color preserving minimum.


$\bullet$
{\em Models with mirror matter}

Effective quartic scalar couplings can also arise from tree-level
superpotential couplings between standard model and non-standard model
mirror matter.  Vector-like Dirac pairs of mirror matter with the same
gauge quantum numbers as some standard model fields exist in many
extensions of the standard model.  Mirror matter may however transform
differently from standard model matter under flavor or discrete
$R$-symmetries.  It is therefore possible that such symmetries, which
forbid or highly suppress Yukawa couplings for standard model fields,
allow large couplings to mirror matter.  For example, with massive
mirror matter $\Psi$ and $\overline{\Psi}$, mixing with quark or
lepton fields can arise from the superpotential coupling
\begin{equation}
W = h H_{\alpha} \Phi \Psi + M_{\Psi} \Psi \overline{\Psi}\,,
\end{equation}
where $\Phi$ represents $\Phi_L$ or $\Phi_R$ standard model fields.
Integrating out the mirror matter with supersymmetry-breaking
soft mass potential
\begin{equation}
V  =  m_{\Psi}^2 \left( \Psi^* \Psi +
                  \overline{\Psi}^* \overline{\Psi} \right) \\
\end{equation}
gives the quartic potential
$V \supset \lambda^2 |H_{\alpha} \Phi |^2$, with
\begin{equation}
\lambda^2  = h^{2}\left( { m_{\Psi}^2 \over  M_{\Psi}^2 + m_{\Psi}^2 }
     \right) \,.
\label{lambda3}
\end{equation}
In the supersymmetric limit, $m_{\Psi} \rightarrow 0$, the quartic
coupling vanishes as required by supersymmetric decoupling and
exactness of the superpotential.  Obtaining a sizeable effective
scalar quartic coupling for the standard model fields therefore
requires both a large mirror matter Yukawa coupling, $h$, and that the
mirror matter soft supersymmetry-breaking mass, $m_\Psi$, be of the
order of the supersymmetric Dirac mass, $M_\Psi$.  The quartic
couplings in this case therefore represent an apparent large violation
of supersymmetry in the low-energy theory with the mirror matter
integrated out. Mirror matter mixing with the Higgs fields can also
give rise to analogous effective quartic couplings.

Mirror matter with the properties discussed above arises in many
extensions of the standard model.  In a grand unified theory, for
example, a ${\bf 27}$ of $E_6$ contains, in addition to an entire
generation, Dirac pairs of lepton doublets and down-type right-handed
quarks.  Entire massive pairs of generations and antigenerations also
often occur in string compactifications.  With Yukawa couplings to
standard model fields, and electroweak scale Dirac and
supersymmetry-breaking masses, these fields yield effective quartic
scalar couplings.  As another example, theories of gauge mediated
supersymmetry breaking often have messenger matter, which has
identical gauge quantum numbers as some of the standard model fields.
If allowed by flavor or discrete symmetries, the messengers can
therefore mix with standard model fields through superpotential Yukawa
couplings~\cite{MESSENGER}.  In addition, for one-scale theories, the
Dirac mass messenger scale and supersymmetry-breaking scale are
roughly equal, ${\cal O}(10$--$100)$ TeV, as required to obtain a
large effective quartic coupling, as discussed above.  Finally,
theories with additional compact dimensions also often contain mirror
matter.  For example, massive $N=2$ hypermultiplet excitations, which
exist in the bulk of the compact space, may mix with standard model
matter.  If the scale of supersymmetry breaking is related to the size
of the compact manifold, for example by twisted boundary
conditions~\cite{PQ}, both the supersymmetry-breaking, and
supersymmetry-preserving $N=2$ masses can be of the same order,
thereby leading to a large effective quartic coupling.  For any type
of mirror matter, effective quartic couplings can arise from mixing
with either quark or lepton fields, or the Higgs fields.

With large effective quartic couplings arising from mirror matter
mixing, the upper limits on soft chiral flavor breaking implied by
both metastability and absolute stability can be increased to
$A/\widetilde{m} \lesssim$ few.  This could in principle allow all
fermions except the top quark and $\tau$-lepton to obtain masses
radiatively from holomorphic soft chiral flavor breaking, while
keeping the charge and color preserving vacuum as the global minimum.
By appropriate choice of the representation, mirror matter can also
stabilize non-holomorphic $A^\prime$ terms.

\begin{table}
\caption{The quartic coupling, $\lambda^2$,
along the equal field direction for different classes of models.
For the tree-level top Yukawa models $\tan \beta >1$ is assumed
for the stability limits.}
\label{table:t1}
\begin{tabular}{lcllc}
Model &
$\lambda^2$ &
Absolute stability &
Metastability  &
Comments \\
       &             &  limits & limits & \\
 \hline
Minimal holomorphic &
$ \sim \frac{-1}{96\pi^2}\left(\frac{A}{\widetilde{m}}\right)^4$ &
$ \frac{A}{\widetilde{m}}  \ll 1$ &
$ \frac{A}{\widetilde{m}} \lesssim 1.75$ &
Metastable vacuum
 \\[1.3ex]
Minimal non-holomorphic &
$ \frac{g^{\prime 2}}{2}$ &
$ \frac{A^{\prime}}{\widetilde{m}} \lesssim 0.4$ &
$ \frac{A^{\prime}}{\widetilde{m}} \lesssim 1.8$ &
 \\[1.3ex]
Tree-level top Yukawa &
$ h_{t}^{2}$ &
$ \frac{A_{b}^{\prime}}{\widetilde{m}}
          \lesssim 1.7-2.5 $ &
$ \frac{A_{b}^{\prime}}{\widetilde{m}}
          \lesssim 2.7-3.2 $ &
Relevant to $b$
 \\[1.3ex]
Mirror matter &
$ \frac{h^2\,m^2_{\Psi}}{(M^2_{\Psi} + m^2_{\Psi})}$ &
$ \frac{A}{\widetilde{m}}
       \lesssim$  few &
$ \frac{A}{\widetilde{m}}
           \lesssim$ few &
\end{tabular}
\end{table}

The different classes of models for the origin of the effective
scalar quartic couplings along with the associated absolute
or metastability constraints are summarized in table~{\ref{table:t1}}.

%
%

\section{Classification of operators}
\label{sec:class}

Soft chiral flavor breaking in the scalar tri-linear operators
$AH_{\alpha}\phi_{L}\phi_{R}$ and
$A^{\prime}H_{\alpha}^{*}\phi_{L}\phi_{R}$, along with gluino or
neutralino masses, are sufficient to generate radiative fermion masses
and Higgs and Higgsino couplings at one loop.  Both these operators
are gauge-invariant and may be included in the most general low-energy
soft supersymmetry-breaking scalar potential~\cite{HR}.  It is
instructive to consider the effective operators that couple the
visible and supersymmetry-breaking sectors and give rise to these soft
terms.

Supersymmetry-breaking is associated with non-zero auxiliary-component
expectation values in the supersymmetry-breaking sector.  Soft
supersymmetry-breaking terms in the visible sector originate in
operators that couple these auxiliary components to visible sector
fields.  Below the messenger scale for transmitting
supersymmetry breaking, these give rise to effective operators
suppressed by appropriate powers of the messenger scale.  The
lowest-order operator, which produces holomorphic tri-linear soft
terms, is a superpotential term of the form
\begin{equation}
{1 \over M}
\int d^{2}\theta ~{\cal{Z}} H_{\alpha} \Phi_{L} \Phi_{R}\,,
\label{integral1}
\end{equation}
where, as throughout, $M$ represents the messenger scale, the explicit
flavor structure is suppressed, and 
${\cal{Z}} = {\cal{A}} +\theta^{2}{\cal{F}}$ represent chiral
superfields in the supersymmetry-breaking sector.  A non-vanishing
auxiliary component gives rise to holomorphic soft tri-linear terms,
$A H_{\alpha} \phi_L \phi_R$, with $A \sim {\cal F} /M$.  The scale of
soft supersymmetry breaking in the visible sector, in particular the
scale for the superpartner and gaugino masses, is 
$\widetilde{m} \sim {\cal F} / M$.  The holomorphic $A$-parameters are
then parametrically of the same order (up to model-dependent couplings
and flavor suppressions), $A \sim \widetilde{m}$, independent of the
messenger scale.  In high-scale gravity mediated
supersymmetry breaking, the messenger scale is the Planck scale, 
$M \sim M_{P}$, and $\widetilde{m} \sim m_{3/2} \sim {\cal F} / M_P$.

The existence of the coupling~(\ref{integral1}) requires non-trivial
flavor interactions at the messenger scale, or equivalently that the
supersymmetry-breaking sector field(s) ${\cal Z}$ transform
non-trivially under flavor.  Flavor must therefore be intimately
linked to the transmission and/or breaking of supersymmetry.
Specifically, the flavor symmetries (for fermions with soft radiative
masses) are either broken explicitly by the messenger sector
interactions and/or spontaneously in the supersymmetry-breaking
sector.  As discussed in section~\ref{sec:possible}, second- and
third-generation radiative masses require $A \sim \widetilde{m}$.  So
in this case the flavor scale cannot be greater than the messenger
scale.  Alternatively, for first-generation radiative masses, which
require somewhat smaller tri-linear $A$-terms, there could be a small
hierarchy between the messenger and flavor scales, so that the
operators~(\ref{integral1}) possess an additional suppression
$M/M_{flav}$ in this case, where $M_{flav}$ is the flavor scale.

In addition to the soft tri-linear couplings arising
from~(\ref{integral1}), with auxiliary expectation values ${\cal F}$,
superpotential Yukawa couplings in general arise with scalar
expectation values ${\cal A}$.  The latter case of scalar expectation
values, or spurions in the low-energy theory, which give rise to
Yukawa couplings, is nothing but the well-known Froggatt--Nielsen
mechanism~\cite{FN}, with the hierarchies in the Yukawa coupling
matrix arising from the ratio ${\cal A}/M$.  The soft breaking of
chiral flavor symmetries considered here may therefore be thought of
as a version of the Froggatt--Nielsen mechanism in which auxiliary
spurions break chiral flavor symmetries rather than scalar spurions.
In order for the induced Yukawa couplings to be unimportant the scalar
expectation values must be insignificant, ${\cal A} \ll M$, or vanish.
This is the case if all scalar expectation values in the
supersymmetry-breaking sector are small respect to the messenger
scale.  This could occur, for example, in a renormalizable hidden
sector with high-scale mediation for which 
${\cal A} / M \sim m_{3/2} / \sqrt{\cal F} \ll 1$.

Since the breaking of chiral flavor symmetries must be linked with the
messenger and/or supersymmetry-breaking sectors, there are additional
operators beyond~(\ref{integral1}) which have non-trivial flavor
transformation properties in the low-energy theory.  In particular, no
symmetry can forbid operators of the form
\begin{equation}
{1 \over M^2} \int d^2 \theta d^{2}\bar{\theta}~
 {\cal Z}^{\dagger} {\cal Z}   \Phi^{\dagger} \Phi \,,
\label{phiphiop}
\end{equation}
where $\Phi=\Phi_L$ or $\Phi_R$.  With auxiliary expectation values,
these operators give rise to soft scalar masses in the visible sector,
$\widetilde{m}^{\,2} \phi^* \phi$, where $\widetilde{m} \sim {\cal
F}/M$.  Arbitrary breaking of flavor symmetries in the scalar masses
of course leads in general to unacceptable flavor-changing neutral
currents.  The specific magnitude of the flavor breaking depends on
the precise flavor structure of the operators~(\ref{phiphiop}) and/or
on how the flavor symmetries are broken by the auxiliary expectation
values, and may be minimized in particular models.  In addition, the
problem of excessive flavor-changing can be largely ameliorated if the
scalar partners of the first generation, and to a lesser extent those
of the second generation, are much heavier than the electroweak
scale~\cite{CKN}, as discussed in section~\ref{sec:possible}.

The form of the coupling~(\ref{integral1}) between the visible and
supersymmetry-breaking sectors, and the subsequent absence of
tree-level superpotential Yukawa coupling, can be enforced by discrete
or continuous flavor or $R$-symmetries.  For example, a $Z_2$
$R$-symmetry~\cite{FF} under which ${\cal Z} \to - {\cal Z}$ and
$H_{\alpha} \Phi_L \Phi_R \to - H_{\alpha} \Phi_L \Phi_R$ allows the
coupling~(\ref{integral1}) but forbids a superpotential Yukawa
coupling.  This is easily extended to continuous $U(1)_R$ symmetries,
or specific flavor symmetries under which ${\cal Z}$ transforms.

Non-holomorphic tri-linear soft terms can only arise from K\"ahler
potential terms rather than from the superpotential.  The lowest-order
operator that gives rise to such terms is of the form
\begin{equation}
{1 \over M^3}
\int d^2 \theta d^{2}\bar{\theta} ~{\cal{Z}}{\cal{Z}}^{\dagger}
H^{\dagger}_{\alpha} \Phi_{L} \Phi_{R} \,.
\label{integral2}
\end{equation}
Non-vanishing auxiliary terms give rise to soft tri-linear terms,
$A^{\prime} H_{\alpha}^* \phi_L \phi_R$, with
$A^{\prime} \sim {\cal F}^2 / M^3 \sim \widetilde{m}^{\,2} / M$.
Because of the dimensionality of the K\"ahler potential, these
non-holomorphic tri-linear terms are suppressed for a messenger scale
well above the supersymmetry breaking by 
${\cal O}(\widetilde{m} /M)$.  This non-holomorphic source of chiral
flavor breaking can therefore be relevant only with a low scale for
both flavor and supersymmetry breaking.  In this case the small ratio
$A /\widetilde{m} \sim 10^{-3}$ required for first-generation
radiative masses could arise wholly or in part from the ratio of the
electroweak to messenger/flavor scale(s), $\widetilde{m} / M$.

%
%

\section{Anomalous Magnetic Moments}
\label{musection}

With soft chiral flavor breaking, radiative fermion masses arise
from effective operators generated at the superpartner mass scale.
The same virtual processes which give rise to the fermion mass
also generate other chirality violating operators below the
superpartner scale.  The experimentally most important of these
operators is the fermion anomalous magnetic moment.
As shown in fig.~\ref{gmtwoadiag}, the anomalous magnetic
moment arises from the same one-loop diagram which gives rise
to the fermion mass shown in fig.~\ref{massdiag}, with an
external photon coupling to the virtual scalar partner.
%
%
\begin{figure}[ht]
\begin{center}
\epsfxsize= 6.5 cm
\leavevmode
\epsfbox{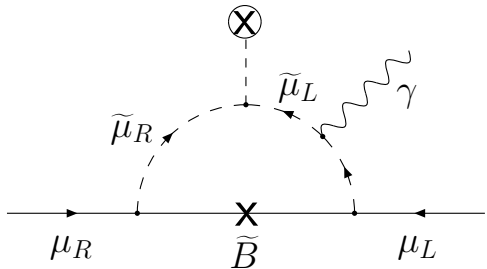}
\end{center}
\caption[f1]{One-loop radiative contribution to the muon anomalous
 magnetic moment from soft chiral flavor violation and Bino mass.}
\label{gmtwoadiag}
\end{figure}
Since the mass and magnetic moment both arise at one loop with the
same chiral structure, the supersymmetric contribution to the
dimension-six operator, which gives a fermion anomalous magnetic
moment, $a_f \equiv {1 \over 2}(g-2)$, is given parametrically by
$a_{f}^{\rm SUSY} \sim m_f^2 / \widetilde{m}^{\,2}$, where
$\widetilde{m} =
 {\rm max}(m_{\tilde{f}_1}, m_{\tilde{f}_2}, m_{\lambda})$.  Notice
that this is {\it not} suppressed by a loop factor even though it
arises perturbatively.  Anomalous magnetic moments suppressed by the
relevant heavy mass scale without a loop factor suppression are a
generic feature of any theory of radiative fermion
masses~\cite{NOLOOP}.  The relatively large supersymmetric
contribution to the magnetic moments is in analogy to the relatively
large Higgs Yukawa radii discussed in section~\ref{sec:higgs}, which
also are effectively not suppressed by a loop factor.  In fact, in the
present scenario for radiative fermion masses, any chirality-violating
operator generated at the superpartner mass scale is 
effectively a loop factor larger than with tree-level Yukawa
couplings.

The best measured fermion magnetic moment in relation
to its mass squared is by far that of the muon.
If the muon mass arises radiatively from soft chiral flavor
violation, the loop function for the anomalous magnetic moment
can be related to that for the radiative mass
\begin{equation}
  a_\mu^{\rm SUSY} = +2
  m_\mu^2
 \frac{\quad \sum_{j}\, K_\mu^j \, m_{\tilde{\chi}_j^{0}} \,
  I_{g-2} ( m^2_{\tilde{\mu}_1}, m^2_{\tilde{\mu}_2},
            m^2_{\tilde{\chi}_j^0} ) }
      {\sum_{j}\, K_\mu^{j}  \, m_{\tilde{\chi}_{j}^{0}} \,
  I ( m^{2}_{\tilde{\mu}_{1}}, m^{2}_{\tilde{\mu}_{2}},
      m^2_{\tilde{\chi}_{j}^{0}}) }  \,.
\label{mammform}
\end{equation}
The loop function
$ I_{g-2} (m^2_{\tilde{\mu}_1}, m^2_{\tilde{\mu}_2}
           m^2_{\tilde{\chi}_j^0} )$ 
and details of the calculation are given in appendix~\ref{sec:muon}.
{}From the general expressions given in the appendix, one has 
\begin{equation}
\hspace*{0.5truecm}
\frac{
 I_{g-2} (m^2_{\tilde{\mu}_1}, m^2_{\tilde{\mu}_2}, m^2_{\lambda})}
{I(m^2_{\tilde{\mu}_1}, m^2_{\tilde{\mu}_2}, m^2_{\lambda})}
 \sim \frac{1}{6 \widetilde{m}^{\,2}} \,;
\hspace*{1.0truecm}
\widetilde{m} =
 {\rm max}(m_{\tilde{f}_1}, m_{\tilde{f}_2}, m_{\lambda})\,,
\label{estimate}
\end{equation}
where the prefactor $1/6$ arises for degenerate superpartners.  The
supersymmetric contribution to the muon anomalous moment is then
$a_{\mu}^{\rm SUSY} \sim m_{\mu}^2 / (3 \widetilde{m}^{\,2})$.
This result differs in a number of ways from the supersymmetric
contribution with tree-level Yukawa couplings, which is dominated by
chargino exchange over much of the parameter space and is
$a_{\mu}^{\rm SUSY} \sim (g^2 / 16 \pi^2) (m_{\mu}^2 m_Z \tan \beta
  / \widetilde{m}^{\,3}) $~\cite{AMU}.
Besides not being suppressed by a loop factor, the anomalous magnetic
moment obtained in the radiative mass scenario is dominated by Bino
exchange.  Because of the absence of a chargino contribution, which
vanishes without a tree-level Yukawa, it is largely independent of
$\tan \beta$.  Finally, it is {\it necessarily} positive.  The
positivity may be understood from the real space form of the radiative
magnetic moment amplitude.  The external low frequency photon field
couples to the scalar partner virtual cloud, which necessarily has the
same charge as the fermion, and in real space necessarily has positive
extent.  In addition, the radiative mass and contribution to the
magnetic moment have the same fermionic chiral structure as that
arising from the virtual neutralino.  It then follows that the
supersymmetric contribution to the magnetic moment has the same sign
as the minimal Dirac term.  The positivity of the supersymmetric
contribution to the anomalous magnetic moment is an interesting
feature and a definite prediction of radiative fermion masses from
soft chiral flavor breaking.  With a tree-level Yukawa coupling, the
supersymmetric contribution is correlated with ${\rm sgn}(\mu)$ over
much of the parameter space, and can have either sign~\cite{AMU}.

%
%
%
\begin{figure}[ht]
\begin{center}
\epsfxsize= 12.5cm
\leavevmode
\epsfbox{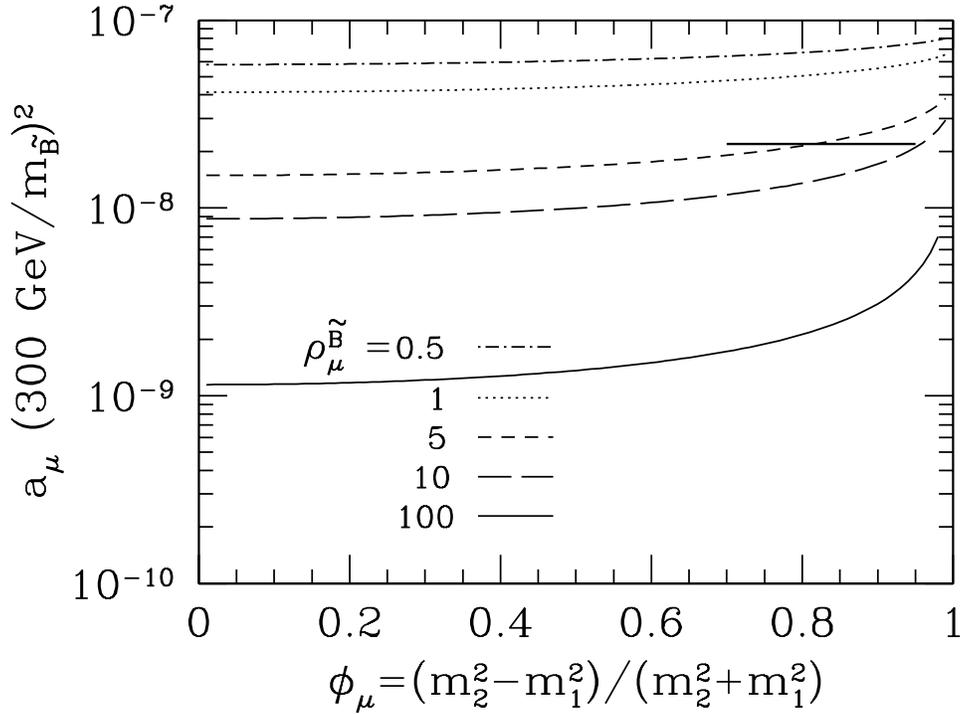}
\end{center}
\caption[f1]{
 Supersymmetric contribution
 to the muon anomalous magnetic moment,
 $a_{\mu} \equiv (g_{\mu}\! -\! 2)/2$, for a radiative muon mass,
 as a function of the fractional smuon mass splitting
 $\phi_\mu$.  The anomalous moment is given in units of
 $ (m_{\tilde B} / 300\  {\rm GeV})^{-2}$
 for various values of the ratio
 $\rho_{\mu}^{\tilde B}$,
 neglecting neutralino mixing effects.
 The current experimental bound of
 $a_{\mu}^{\rm new} < 220 \times 10^{-10}$
 is indicated for comparison, and already constrains the parameter
 space for low values of $m_{\tilde{B}}$ and $\rho_{\mu}$.}
\label{gmtwofig}
\end{figure}
The anomalous magnetic moment~(\ref{mammform}) is plotted in
fig.~\ref{gmtwofig} as a function of the fractional smuon mass-squared
splitting, $\phi_\mu$, for various values of the ratio
$\rho_{\mu}^{\tilde{B}}$ in the pure gaugino limit.  Note that the
dependence on the $A$- or $A^{\prime}$-parameter, or equivalently the
smuon left--right mixing angle, is implicitly contained in the muon
mass $m_{\mu}$, so that $a_{\mu}^{\rm SUSY}$ is only a function of the
superpartner mass spectrum through the loop functions.  The current
experimental measurement of $a_{\mu}^{\rm exp}$~\cite{MUONBOUND} is in
good agreement with the theoretical calculations, which include 
${\cal O}(\alpha^5)$ QED corrections and hadronic vacuum polarization
to ${\cal O}(\alpha^3)$~\cite{MUONCALC}.  This agreement allows to put 
a bound on positive, non-standard model contributions of 
$a_{\mu}^{\rm new} < 220 \times 10^{-10}$ at the 95\%
CL~\cite{MUONBOUND,MUONCALC}, which is indicated in
fig.~\ref{gmtwofig} for comparison.  If the muon mass arises
radiatively, the current bound already rules out
$m_{\tilde{\mu}_{1,2}} \lsim 400\,$GeV for $m_{\tilde{B}} \lsim
m_{\tilde{\mu}_{1,2}}$.  The Brookhaven E821 muon $g-2$ experiment is
expected to reach a level of sensitivity of 
$\delta a_{\mu}^{\rm exp} \sim 4 \times 10^{-10}$~\cite{BROOKHAVEN}.
If all the standard model contributions can be calculated to this
precision, smuon masses up to $m_{\tilde{\mu}_{1,2}} \sim 3$ TeV can
be probed for $m_{\tilde{B}} \lsim m_{\tilde{\mu}_{1,2}}$ if the muon
mass is radiative.  The muon anomalous magnetic moment is by far the
best experimental probe of a radiative muon mass arising from soft
chiral flavor violation.

%
%
\begin{figure}[ht]
\begin{center}
\epsfxsize= 12.5 cm
\leavevmode
\epsfbox{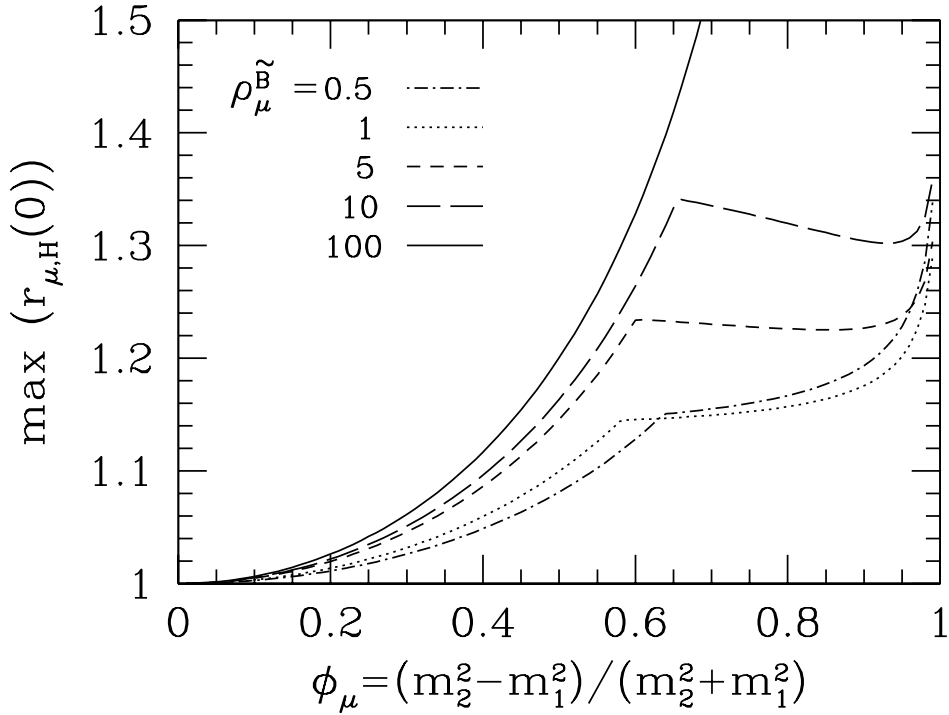}
\end{center}
\caption[f1]{Upper limit on the Higgs--Yukawa enhancement parameter
 $r_{\mu,H}(0)$ at zero momentum transfer for a radiative muon mass
 as a function of the smuon mass squared splitting $\phi_\mu $,
 for various values of the ratio $\rho_{\mu}^{\tilde{B}}$.
 The upper limit follows indirectly from the current experimental
 constraint on non-standard model contributions to the muon
 anomalous magnetic moment of $a_{\mu}^{\rm new} < 220 \times 10^{-10}$.}
\label{addfig1}
\end{figure}
If the muon mass does arise radiatively, the indirect constraints on
the superpartner masses from the experimental bound on $a_{\mu}^{\rm
new}$ discussed above may be used to give an upper limit on the
enhancement of the effective Higgs Yukawa coupling over the mass
Yukawa,
$r_{\mu} = \bar{h}_{f}(\langle H_{\alpha} \rangle / m_{\mu})$.  

For fixed $\phi_\mu$ and $\rho_{\mu}^{\tilde{B}}$ this 
bound, $m_{\tilde{\mu}_{1,2}} \gsim 400\,$GeV for 
$m_{\tilde{B}} \lsim m_{\tilde{\mu}_{1,2}}$,  
may be translated into a lower limit on the combination 
$m_{\widetilde B}^3 \, \rho_\mu^{\tilde B}\,
 I(m^2_{\tilde{\mu}_1}, m^2_{\tilde{\mu}_2},m^2_{\tilde{B}}) $, 
which is related to the radiative muon mass~(\ref{massnew}).  Given
the known muon mass, this may then be used to place an upper limit on
the smuon mixing angle $\sin 2 \theta_\mu$ using~(\ref{massnew}).
Finally, these bounds may all be put together to bound the $q^2=0$
effective Higgs Yukawa coupling ratio $r_{\mu,H}(0)$,
using~(\ref{vertex2}) and~(\ref{higgsratio}).  This upper limit on
$r_{\mu,H}(0)$ from the current experimental bound on 
$a_{\mu}^{\rm new}$ is given in fig.~\ref{addfig1}.
For $ \phi_\mu \lsim 0.6$--$0.7$, the current experimental bound on
$a_{\mu}^{\rm new}$ provides no limit on $\sin 2 \theta_\mu$, and the
maximum value of $r_{\mu,H}(0)$ just follows from~(\ref{vertex2})
and~(\ref{higgsratio}) with $\sin 2 \theta_\mu =1$ (compare
fig.~\ref{higgscoupling}).  This is because, in this region of
fractional smuon mass squared splitting $\phi_\mu$, the correct value
of the muon mass is obtained only with relatively massive
superpartners, which are not bounded by $a_{\mu}^{\rm new}$. However,
for $ \phi_\mu \gsim 0.6$--$0.7$ the current bound on $a_{\mu}^{\rm
new}$ does provide a non-trivial limit on $\sin 2 \theta_\mu$.  This
is reflected in the maximum value of $r_{\mu,H}(0)$ shown in
fig.~\ref{addfig1}. As discussed in detail in
section~\ref{sec:probing}, the ratios $r_{\mu,H}(m_h^2)$ and
$r_{\mu,H}(m_H^2)$ could also be probed at a $\mu^+ \mu^-$ collider by
direct measurement of the Higgs--muon coupling.  This would give
additional sensitivity to possible momentum dependence of the
Higgs--muon couplings.  However, the Brookhaven E821 muon $g-2$
experiment will in general be much more sensitive to a radiative muon
mass.

%
%
\begin{figure}[ht]
\begin{center}
\epsfxsize= 12.5 cm
\leavevmode
\epsfbox{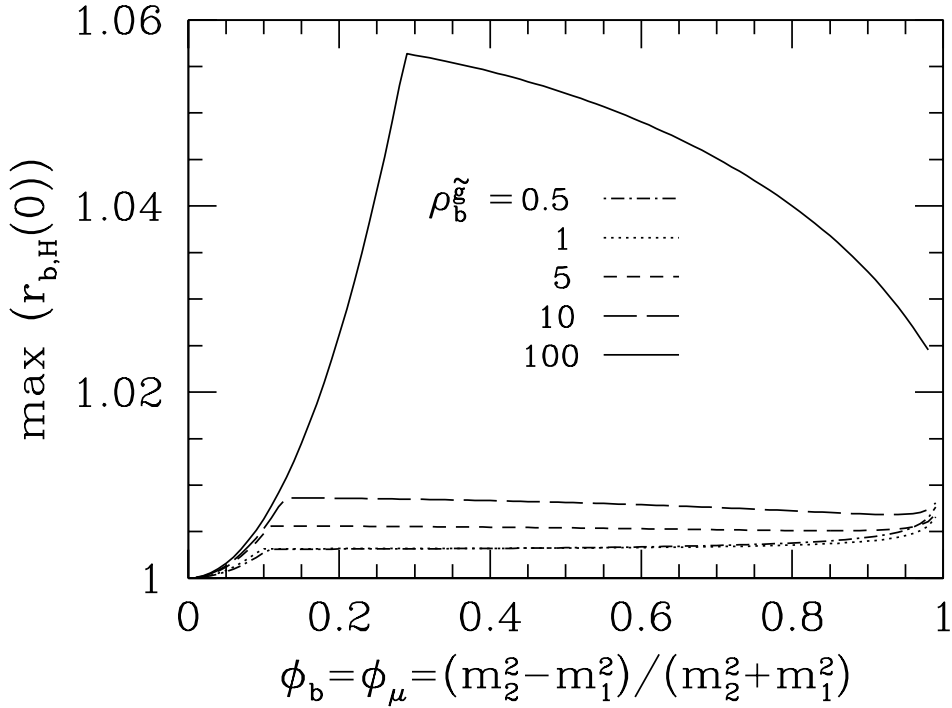}
\end{center}
\caption[f1]{Upper limit on the Higgs--Yukawa enhancement parameter
 $r_{b,H}(0)$ at zero momentum transfer for a radiative $b$-quark
 mass as a function of the sbottom mass splitting $\phi_b$,
 for various values of the ratio
 $\rho_b^{\tilde{g}} \!=\! \rho_{\mu}^{\tilde{B}}$, and assuming
$\phi_{b} = \phi_{\mu}$.
 The upper limit follows indirectly from the current experimental
 constraint on non-standard model contributions to the muon
 anomalous magnetic moment of $a_{\mu}^{\rm new} < 220 \times 10^{-10}$
 under the assumption of gaugino unification mass relations.
 The bound is only obtained if {\it both} the muon and $b$-quark
 obtain mass radiatively from soft chiral flavor breaking.}
\label{addfig2}
\end{figure}
The limits on Higgs coupling enhancements for a radiative muon mass
implied by the current experimental bound on the muon anomalous
magnetic moment may be extended to other fermions with radiative
masses, if assumptions about relations within the superpartner mass
spectrum are postulated.  The Higgs--$b$-quark coupling may be well
measured at future colliders, as detailed in
section~\ref{sec:probing}.  If {\it both} the muon and $b$-quark
masses arise radiatively, an upper limit on $r_{b,H}(0)$ can be
obtained indirectly from the $a_{\mu}^{\rm new}$ experimental bound.
For example, assuming {\it i)} that there is a gaugino unification
relation between the Bino and gluino masses, 
($m_{\tilde{g}} = {3 \over 5} (\alpha_s / \alpha^{\prime})
  m_{\tilde{B}} \simeq 6~ m_{\tilde{B}}$),
{\it ii} that $ \rho_{\mu}^{\tilde{B}} = \rho_b^{\tilde{g}}$, 
{\it iii} that fractional smuon and sbottom mass squared splittings
are the same ($\phi_\mu = \phi_b$), {\it iv} that the $b$-quark mass
has the known value, the bound on $a_{\mu}^{\rm new}$ may be
translated into the analogous indirect bound on $r_{b,H}(0)$ given in
fig.~\ref{addfig2}.  The constraint implied on the Higgs--$b$-quark
coupling is more severe than that for the Higgs--muon coupling simply
because, under the assumption of gaugino unification, the strongly
interacting superpartners, which radiatively generate the $b$-quark
mass, are somewhat heavier than the weakly interacting superpartners,
which generate the muon mass; and $r_{f,H}(0) \to 1$ in the
superpartner decoupling limit.  Other assumptions for the relations
between the $b$-squark and smuon masses and the gaugino masses would
yield different bounds on $r_{b,H}(0)$.  If the muon mass is
tree-level, or if no relations are assumed, then of course no indirect
bound on the Higgs--$b$-quark coupling for a radiative $b$-quark mass
can be gleaned from the muon anomalous magnetic moment.

It is also possible, in general, for flavor-changing electromagnetic-
and chromo-dipole operators to be generated, depending on the precise
textures that appear in the scalar partner mass matrices.  Within any
specific model for the soft and hard textures in the Yukawa couplings,
such flavor-changing dipole operators may give the most interesting
low-energy probe of supersymmetric flavor violation.  These are,
however, very model-dependent and outside the scope of this work.  In
the scenario discussed in section~\ref{sec:possible}, in which all the
down-type quarks gain a mass radiatively, in the limit of degenerate
squarks in each sector, all the transition quark dipole moments
vanish.  In the same scenario, but with non-degenerate squarks and
down-type quark number conserved, all supersymmetric contributions to
the flavor-changing down-type dipoles associated directly with the
radiative masses also vanish, and likewise for the analogous scenario
with radiative electron and/or muon masses.


\section{CP Violation}
\label{sec:cpviol}

In generic supersymmetric theories, virtual contributions to
low-energy CP-violating operators, such as electric dipole
moments (EDMs) and CP-odd Higgs--fermion couplings, appear due to the
presence of intrinsic supersymmetric CP-violating phases.
The flavor-conserving phases reside in the complex Lagrangian parameters:
$A$- or $A^{\prime}$-terms,
the gaugino masses, $m_{\lambda}$,
the Higgsino mass term $\mu$,
and the scalar Higgs mass coupling $m_{12}^2$.
In the case of hard tree-level Yukawa couplings,
contributions to the EDMs of the neutron and atoms are generated
at one loop.
For superpartners at the electroweak scale,
these contributions are relatively large
with respect to current experimental bounds.
The apparent requisite smallness of the intrinsic phases,
typically $10^{-2}$--$10^{-3}$ \cite{NOLOOP,oldneutron,phasebounds},
is generally referred to as the SUSY CP problem.

With radiative fermion masses arising from soft chiral flavor
breaking, the masses and possible CP-violating operators both arise at
the one-loop level.  At first sight, the lack of suppression of the
CP-violating operators by a relative loop factor would appear to
result in an even worse SUSY CP problem with the scenario for
radiative masses presented here.  In this scenario, however, the
phases of chirality-violating operators can be aligned to a very high
precision with the phase of the relevant fermion mass.  This occurs
because the radiative fermion mass and chirality-violating operators
arise from very similar diagrams with similar phase structure, as
shown in fig.~\ref{higgsdiag} for the Higgs couplings, and
fig.~\ref{gmtwoadiag} for the electromagnetic dipole moment.  The
natural alignment can significantly suppress the magnitude of such
CP-violating operators and may in fact provide an effective solution
to the SUSY CP problem within the context of radiatively generated
fermion masses.  If the light-quark masses arise radiatively, the
neutron EDM is highly suppressed, as detailed below, and is compatible
with the current experimental bounds~\cite{MaNg}.  
The electron EDM is in general
not as strongly suppressed.  However, as described below, if the
electron mass arises radiatively, there are interesting regions of
superpartner parameter space in which it is also sufficiently
suppressed.

A CP-odd observable associated with a chirality and CP-violating
effective operator such as an EDM is proportional in a general basis
to the sine of the relative phase between the mass $m$ and the
coefficient of the chirality-violating operator $d$:
$$
\sin \left({\rm Arg}(dm^*) \right).
\label{none}
$$
In terms of field redefinitions, this implies that, the CP-odd
observable is proportional to the imaginary part of the coefficient of
the effective operator in a basis in which the fermion mass is taken
to be real.  In the absence of tree-level Yukawa couplings, gauge
invariance implies that only the neutralinos, and for quarks the
gluino, contribute to the radiative mass at one loop, as noted in
section~\ref{sec:soft}. The same is true of radiative contributions to
the chirality-violating dipole moments and Higgs--fermion couplings.
The full one-loop contributions to the (in general complex) mass $m$
and to the dipole coefficient or Higgs-fermion coupling $d$ then are:
\begin{equation}
m = \sum_j m_j\,, \quad \quad  d = \sum_j d_j,
\label{sumj}
\end{equation}
where the sums are over the individual contributions of each
neutralino, and gluino in the case of quarks.  In general, the phases
of each individual contribution are not related, ${\rm Arg}(m_i m_j^*)
\neq 0$ and ${\rm Arg}(d_i d_j^*) \neq 0$.  However, the one-loop
contribution of any given neutralino or gluino to the mass does have
{\it exactly} the same phase as the corresponding contribution to the
dipole coefficient or Higgs--fermion coupling,
\begin{equation}
{\rm Arg}(d_j m_j^*) =0.
\label{alignj}
\end{equation}
This is because the individual one-loop contributions to the mass and
dipole moment only differ diagrammatically by the coupling to an
external electromagnetic field, so that they necessarily have the same
over-all phase up to a possible sign. Neglecting any
scalar--pseudoscalar Higgs mixing induced by CP violation in the Higgs
potential (which vanishes at tree-level in the MSSM), the same is true
of the Higgs--fermion couplings.  For the dipole coefficient, this can
be seen explicitly by comparing eqs.~(\ref{mass}) and~(\ref{rawgmtwo})
and for the Higgs--fermion coupling directly from eq.~(\ref{vertex1}).
The arguments and results given in section~\ref{musection} for the
positivity of the anomalous magnetic moment further imply that the
contribution of any given neutralino or gluino to the mass and
electromagnetic dipole are in fact aligned rather than anti-aligned
with opposite sign.

The automatic alignment of the individual one-loop
contributions~(\ref{alignj}) has some interesting consequences in the
case that a single neutralino or gluino dominates the total radiative
effects.  This is in fact expected to be the case for radiative quark
masses, and in certain regions of parameter space for radiative lepton
masses, as described in detail below.  If the $i$-th contribution
dominates the sums~(\ref{sumj}), so that 
$|m_j /m_i| \ll 1$ and $|d_j /d_i| \ll 1$ for $j \neq i$, the physical
CP-violating phase of the effective operator is approximately
\begin{equation}
\sin \left( {\rm Arg} (d m^*) \right) \simeq
\sum_{j \neq i} \left(  \left| d_j \over d_i \right|  -
                        \left| m_j \over m_i \right| \right)
   \sin \left( {\rm Arg} (m_j m_i^*) \right) \,,
   \label{dm}
\end{equation}
where~(\ref{alignj}) has been used.  The leading contribution alone
does not give rise to a physical phase because of the automatic
alignment of individual contributions noted above in~(\ref{alignj}).
A non-zero physical phase only arises from {\it interference} between
the leading and subdominant contributions.  This leads to the
suppression factor in parenthesis, multiplying the intrinsic phase
contributions to the physical phase.  {}From expression~(\ref{dm}) it
is apparent that the physical CP-violating phase may be suppressed by
any of the following conditions:
\begin{enumerate}
\item
The subdominant contributions are much smaller than that of the
leading diagram: $|m_j /m_i| \ll 1$ and  $|d_j /d_i| \ll 1$.
\item
The {\it relative} magnitudes of the subdominant contributions to both
the mass and dipole moment operator are nearly equal: 
$|m_j /m_i| \simeq |d_j /d_i|$.
\item
The phase of the dominant and subdominant contributions are nearly
equal:
$\sin \left( {\rm Arg} (m_j m_i^*) \right) \ll 1$.
\end{enumerate}
As detailed below, the first effect significantly suppresses both
CP-violating Higgs--quark couplings and quark EDMs; the second effect
strongly suppresses CP-violating Higgs--fermion couplings and EDMs in
some regions of parameter space; and the third effect further
suppresses CP-violating Higgs-quark couplings and quark EDMs under the
assumption of strict gaugino mass unification.

Consider first the possible CP-violating couplings of the Higgs bosons
to fermions.  The CP-conserving and -violating couplings of the
lightest Higgs boson to fermions, through scalar and pseudoscalar
fermion bilinears respectively, are
\begin{equation}
{\cal L} \supset |\lambda_{h^0}| \cos \left(
   {\rm Arg} (m_f^* h_{f,H} ) \right)
   \overline{f} f h^0
 +  |\lambda_{h^0}| \sin \left(
   {\rm Arg} (m_f^* h_{f,H} ) \right)
   \overline{f} i \gamma_5 f h^0     \,,
\label{hzerocoup}
\end{equation}
where the magnitude of the Higgs coupling is
$|\lambda_{h^0}|= \Theta \{ \cos \beta, \sin \beta \}
   |h_{f,H}| / \sqrt{2}$, where the first (second) term in
brackets is for $\alpha=1$ $(2)$, and $\Theta$ is the $h^0$ mixing
coefficient given in section~\ref{sec:probing}.  Couplings to other
Higgs bosons have similar forms, with scalar and pseudoscalar fermion
bilinears exchanged for the pseudoscalar Higgs $A^0$.  The full
expression for the (in general complex) radiatively generated Yukawa
coupling, $h_{f,H}$, is a sum over the neutralino contributions, and
for quarks, the gluino contribution.  Neglecting
$\widetilde{B}$--$\widetilde{W}_3$ mixing, only the Bino, and for
quarks the gluino, contributes at one loop.  For a radiative quark
mass, in this limit the ratio of Bino contribution to the quark mass,
$m_q^{\tilde{B}}$, to that of the gluino $m_q^{\tilde{g}}$, from the
general expression for a radiative fermion mass~(\ref{mass}) is
\begin{equation}
\left| m_q^{\tilde{B}} \over m_q^{\tilde{g}} \right| =
{3 \alpha^{\prime} \over 16 \alpha_s}
 \left| { m_{\tilde{B}} \over m_{\tilde{g}}} \right|
   Y_{q_L} Y_{q_R} ~
       {I(m_{\tilde{q}_1},m_{\tilde{q}_2},m_{\tilde{B}}) \over
        I(m_{\tilde{q}_1},m_{\tilde{q}_2},m_{\tilde{g}}) } \,, 
\label{Bgmass}
\end{equation}
where $m_{\tilde{B}} $ and $ m_{\tilde{g}}$ are in general complex,
but throughout the masses appearing in the loop functions are
understood to be the real positive mass eigenvalues.  For up- and
down-type quarks the product of hypercharges are
$Y_{u_L} Y_{u_R}={4 \over 9}$ and $Y_{d_L} Y_{d_R}=-{2 \over 9}$
respectively.  The Bino contribution to the radiative mass is
naturally suppressed for a number of reasons.  The individual
contributions are proportional to the gauge couplings squared times
the chirality-violating gaugino mass.  Under the assumption of
gaugino-mass unification for the magnitudes of the gaugino masses,
$|m_{\tilde{B}}/m_{\tilde{g}}|={5 \over 3}\alpha^{\prime}/\alpha_s$,
this results in a suppression of
$m_{\tilde{B}} \alpha^{\prime} /(m_{\tilde{g}} \alpha_s) =
 {5 \over 3}\alpha^{\prime 2}/\alpha_s^2 \sim 10^{-2}$.
The ratio of loop functions is not particularly small in this limit:
$I(m_{\tilde{q}_1},m_{\tilde{q}_2},m_{\tilde{B}}) /
 I(m_{\tilde{q}_1},m_{\tilde{q}_2},m_{\tilde{g}}) \sim 2$, 
and equals 2 for
$m_{\tilde{q}_1}=m_{\tilde{q}_2}=m_{\tilde{g}} \gg m_{\tilde{B}}$.
In addition, the color factor and the product of hypercharges give a
suppression ${3 \over 16} Y_{q_L} Y_{q_R} \sim 10^{-1}$.  Numerically,
then, $|m_q^{\tilde{B}} / m_q^{\tilde{g}}| \lsim 10^{-3}$, so the
situation discussed above in which a single contribution dominates is
achieved.

The ratio of Bino to gluino contributions to the Higgs--quark
coupling~(\ref{vertex1}) is very similar to the ratio for the
radiative mass~(\ref{Bgmass}), modified only by the ratios of Higgs to
mass effective Yukawa couplings defined in~(\ref{higgsratio}).
Altogether then, the coefficient of the CP-violating Higgs--quark
couplings in~(\ref{hzerocoup}) in the approximation~(\ref{dm}) is
$$
\sin \left(
   {\rm Arg} (m_q^* h_{q,H} ) \right) \simeq
  {3 \alpha^{\prime} \over 16 \alpha_s}
 \left| { m_{\tilde{B}} \over m_{\tilde{g}}} \right|
   Y_{q_L} Y_{q_R}
   \left( 1 -  { r^{\tilde{B}}_{q,H}(m_{h^0}^2) \over
                 r^{\tilde{g}}_{q,H}(m_{h^0}^2) } \right)
 ~ {I(m_{\tilde{q}_1},m_{\tilde{q}_2},m_{\tilde{B}}) \over
        I(m_{\tilde{q}_1},m_{\tilde{q}_2},m_{\tilde{g}}) }
$$
\begin{equation}
  ~~~~~~~~~~\times
                \sin   \left( {\rm Arg}(m_{\tilde{B}}
      m_{\tilde{g}}^*) \right),
\label{hphase}
\end{equation}
where $r^{\tilde{B}}_{q,H}(m_{h^0}^2)$ and
$r^{\tilde{g}}_{q,H}(m_{h^0}^2)$
are the ratios of Higgs Yukawa coupling at $q^2=m_{h^0}^2$ to mass
Yukawa coupling for the Bino and gluino contributions respectively.
The physical CP-violating phase arises from the interference between
the Bino and gluino contributions.  It is proportional to the relative
phase between the Bino and gluino masses, and is suppressed because of
the small magnitude of the subdominant Bino contribution.  In
addition, the factor
 $(1 -  { r_{q,H}^{\tilde{B}}(m_{h^0}^2)/
          r_{q,H}^{\tilde{g}}(m_{h^0}^2) } )$
reflects the second of the general suppressions mentioned above,
leading to the alignment of the phases when the {\it relative
contributions} of the Bino to both the mass and Higgs couplings are
similar to those of the gluino.  Since the loop functions for the mass
and Higgs couplings are so similar, this is generally a non-trivial
suppression, especially if the superpartners are much heavier than the
Higgs boson, for which $r_{f,H}(m_{h^0}^2) \rightarrow 1$.  As
discussed in section~\ref{musection}, if both a quark mass and the
muon mass arise radiatively, the current limit on the muon anomalous
magnetic moment can be used to bound $r_{q,H}(0)$ if certain relations
are assumed within the superpartner mass spectrum.  For
$m_{\tilde{q}_i} \sim m_{\tilde{g}}$ the bounds on
$r_{q,H}^{\tilde{g}}(0)$ shown in fig. \ref{addfig2}, obtained under
the assumption of gaugino unification for the magnitude of the gaugino
masses, also approximately apply to $r_{q,H}^{\tilde{B}}(0)$.  In this
case the $q^2= m_{h^0}^2$ corrections should be quite small.  With all
these assumptions, this additional suppression from the bounds in
fig.~\ref{addfig2}, is conservatively
$(1 -  { r_{q,H}^{\tilde{B}}(m_{h^0}^2)/
         r_{q,H}^{\tilde{g}}(m_{h^0}^2) } ) \lsim 10^{-1}$.
Taken together all these suppressions imply that the CP-violating
Higgs--quark couplings for a quark with a radiative mass are
naturally small, i.e.
 $\sin (
   {\rm Arg} (m_q^* h_{q,H} ) )
 \lsim 10^{-(3-4)}
 \sin ( {\rm Arg}(m_{\tilde{B}}
      m_{\tilde{g}}^*) )$.
In addition, under the assumption of strict gaugino unification,
$m_{\tilde{g}}=m_{\tilde{W}}=m_{\tilde{B}}$ at the unification scale,
the phases of the gaugino masses are also correlated.  This is the
final source of possible suppressions of the physical phases mentioned
above.  In this case 
${\rm Arg}(m_{\tilde{B}} m_{\tilde{g}}^*) =0$ at lowest order, 
and the CP-violating Higgs--quarks couplings vanish at one loop in the
pure Bino limit.  The only contributions at this order arise from
gaugino-Higgsino mixing effects.

For radiative lepton masses, in the pure gaugino limit, the 
$\widetilde{B}$--$\widetilde{W}_3$ mixing vanishes, so only the Bino
contributes at one loop to both the radiative mass and Higgs 
Yukawa couplings
of leptons.  The phases are then precisely aligned at one loop and the
CP-violating Higgs couplings vanish in this limit.  Neutralino mixing
through electroweak symmetry breaking can however introduce
non-trivial phases.  In the mostly gaugino or Higgsino region of
parameter space, the dominant radiative contributions come from the
mostly Bino state, denoted $\tilde{\chi}_1^0 \simeq \tilde{B}$, as
discussed in section~\ref{sec:soft}.  Thus the
approximation~(\ref{dm}) for the CP-violating Higgs--lepton couplings
may be employed.  Summing over the neutralino contributions to the
radiative fermion mass~(\ref{mass}) and to the 
effective Higgs Yukawa coupling parametrized by the
ratios~(\ref{higgsratio}), yields:
$$
\sin \left(
   {\rm Arg} (m_l^* h_{l,H} ) \right) \simeq
\sum_{j=2}^4  \left|
    { K^j_l m_{\tilde{\chi}_j^0} \over
          K^1_l m_{\tilde{\chi}_1^0} }  \right|
  \left( 1 -  { r^{\tilde{\chi}_j^0}_{l,H}(m_{h^0}^2) \over
                 r^{\tilde{\chi}_1^0}_{l,H}(m_{h^0}^2)   } \right)
 {I(m_{\tilde{l}_1},m_{\tilde{l}_2},m_{\tilde{\chi}_j^0}) \over
        I(m_{\tilde{l}_1},m_{\tilde{l}_2},m_{\tilde{\chi}_1^0}) }
$$
\begin{equation}
   ~~~~~~~~~~\times ~
 \sin \left( {\rm Arg}
    \left( K_l^j m_{\tilde{\chi}_j^0}
           K_l^{1^*} m_{\tilde{\chi}_1^{0*}}
                \right)
        \right),
\label{lephiggs}
\end{equation}
where $K_l^j$ are the neutralino coupling coefficients defined in
eq.~(\ref{Kfactor}).  In the mostly gaugino or Higgsino region of
parameter space, the neutralino mixing may be treated perturbatively.
To first order in mixing, the product of neutralino eigenvectors
in~(\ref{Kfactor}) is unmodified, $K_l^j = {1 \over 2} \delta_{j1}$,
and the physical phase vanishes at this order.  To second order,
however, a non-trivial phase dependence in the interference terms is
introduced by mixing.  At this order the Bino mixes with the Higgsino
states, giving a small coupling 
$K^j_l \sim {\cal O}(m_Z^2/\tilde{m}^2)$ for $j$ the mostly Higgsino
states, where $\tilde{m}={\rm max}(m_{\tilde{B}}, \mu)$.  The physical
phase then arises as an interference between the mostly Bino and
Higgsino states.  The intrinsic phase that appears on the right-hand
side of~(\ref{lephiggs}) is therefore the phase between the masses of
the mostly Bino and Higgsino states, ${\rm Arg}(m_{\tilde{B}} \mu
(m_{12}^2)^*)$, where $V \supset - m_{12}^2 H_1 H_2 + {\rm h.c.}$
determines the phase of the Higgs condensate in a general basis.

In the mostly gaugino or Higgsino region of parameter space, the
contribution of the intrinsic phase to the physical CP-violating phase
is further suppressed beyond the small $K^j_l$ couplings of the mostly
Higgsino states.  The ratio of the loop functions in~(\ref{lephiggs})
can be small.  For $m_{\tilde{B}},\mu \ll m_{\tilde{l}}$ this
 ratio is of ${\cal O}(1)$. It is, however, small for 
 $m_{\tilde{B}} \lsim m_{\tilde{l}} \ll \mu$, i.e. 
 $I(m_{\tilde{l}_1},m_{\tilde{l}_2},m_{\tilde{\chi}_j^0}) /
        I(m_{\tilde{l}_1},m_{\tilde{l}_2},m_{\tilde{\chi}_1^0})
                \sim m_{\tilde{l}}^2 / \mu^2$.
In addition, the relative contribution of the mostly Higgsino states
to the mass and Higgs-lepton couplings can be very similar, causing
the factor
$( 1 -  { r^{{\tilde{\chi}}_j^0}_{l,H}(m_{h^0}^2) /
   r^{{\tilde{\chi}}_1^0}_{l,H} (m_{h^0}^2) } )$
to give a significant suppression.  This relative suppression is
dominated by the largest individual deviation from unity of the ratios
$r^{{\tilde{\chi}}_i^0}_{l,H}$, $i=1,\dots,4$.  {}From the discussion
in section~\ref{sec:higgs}, the $D$-term and finite-momentum
contributions to $r^{{\tilde{\chi}}_j^0}$ decouple most slowly.
For $m_{h^0}^2 \sim m_Z^2$, these are both parametrically
$r^{{\tilde{\chi}}_i^0}_{l,H} \sim 1 + 
  {\cal O}(m_Z^2/ \tilde{m}^{\prime 2})$,
where 
$\tilde{m}^{\prime} = {\rm max}(m_{\tilde{\chi}_i^0},\tilde{m}_l)$.
This suppression, due to the relative similarity of the mostly
Higgsino contributions to both the mass and Higgs couplings, can be
understood in terms of the effective operator discussion of the Yukawa
couplings given in section~\ref{sec:higgs}.  At the renormalizable
level a single operator, namely the effective Yukawa coupling,
contributes to both the mass and Higgs couplings.  This operator
has a definite phase, even if more than one neutralino contributes.
The deviations
$( 1 -  { r^{{\tilde{\chi}}_j^0}_{l,H}(m_{h^0}^2) /
          r^{{\tilde{\chi}}_1^0}_{l,H} (m_{h^0}^2) } )$ represent
non-renormalizable operators, which give different contributions to
the mass and Higgs effective Yukawa couplings.  Thus, it is only the
interference between the renormalizable Yukawa coupling, and the
non-renormalizable operators that gives rise to a physical phase.

The total suppression of the CP-violating Higgs--lepton coupling can
be substantial in the mostly gaugino or Higgsino region of parameter
space.  
Alltogether, the following parametric suppression of the physical 
phase is obtained:
$$
\sin \left(
   {\rm Arg} (m_l^* h_{l,H} ) \right) \sim
   {\cal O}(m_Z^4 / (\mu {m}_{\tilde{B}} m_{\tilde{l}}^2))
   \sin ({\rm Arg}(m_{\tilde{B}} \mu (m_{12}^2)^*))\,,
$$
for $m_{\tilde{B}} \lsim \mu \ll m_{\tilde{l}}$, and 
$$
\sin \left(
   {\rm Arg} (m_l^* h_{l,H} ) \right) \sim
   {\cal O}(m_Z^4 m_{\tilde{B}}/ (\mu^3 m_{\tilde{l}}^2))
   \sin ({\rm Arg}(m_{\tilde{B}} \mu (m_{12}^2)^*)) \,,
$$
for $m_{\tilde{B}} \lsim m_{\tilde{l}} \ll \mu $. 
If the superpartners are somewhat heavier than the $Z$ boson, the
suppression can be significant, yielding very small CP-violating
Higgs--lepton couplings.

Now consider the EDM of the neutron and electron in the case where the
light-quark and electron masses are radiative.  The discussion
parallels that of the Higgs Yukawa couplings in many respects.  For
the neutron EDM, the dominance of the gluino contribution to
light-quark radiative masses turns out to be sufficient, on its own,
to render the supersymmetric contribution compatible with current
experimental bounds.  For the electron EDM, the suppression depends
sensitively on the neutralino masses and mixings as described below.

{}From the definitions in appendix~\ref{sec:muon}, a fermion EDM,
$d_f^e$, is related to the complex coefficient of the electromagnetic
dipole operator, $d_f$, by eq.~(\ref{dfsin})
$$
d_f^e = |d_f| \sin  \left( {\rm Arg}( d_f m_f^*) \right)\,.
$$
For a radiative quark mass, neglecting
$\widetilde{B}$--$\widetilde{W}_3$ mixing, only the gluino and Bino
contribute to chirality-violating operators at one loop.  The gluino
contribution is likely to dominate the Bino contribution, as
illustrated in~(\ref{Bgmass}) and discussed above.  Given the
dominance of a single contribution, the approximation~(\ref{dm}) may
be employed for the relative phase of the mass and dipole-moment
coefficient.  The magnitude of the Bino and gluino contributions to
the dipole-moment coefficient may be obtained from
eqs.~(\ref{rawgmtwo}) and~(\ref{gluedip}) respectively.  With this,
the leading contribution to the EDM of a quark with radiative mass is
$$
d_q^e \simeq {3 \alpha^{\prime} \over 8 \alpha_s}
 \left| { m_{\tilde{B}} \over m_{\tilde{g}}} \right|
 e Q_q Y_{q_L} Y_{q_R} m_q
 ~{I_{g-2}(m_{\tilde{q}_1},m_{\tilde{q}_2},m_{\tilde{g}}) \over
        I(m_{\tilde{q}_1},m_{\tilde{q}_2},m_{\tilde{g}}) }
 ~ \sin   \left( {\rm Arg}(m_{\tilde{B}} m_{\tilde{g}}^*) \right)
$$
\begin{equation}
  ~~~~~~~~~\times
   \left({I_{g-2}(m_{\tilde{q}_1},m_{\tilde{q}_2},m_{\tilde{B}}) \over
        I_{g-2}(m_{\tilde{q}_1},m_{\tilde{q}_2},m_{\tilde{g}}) }
 -     {I(m_{\tilde{q}_1},m_{\tilde{q}_2},m_{\tilde{B}}) \over
        I(m_{\tilde{q}_1},m_{\tilde{q}_2},m_{\tilde{g}}) }
        \right).
        \label{quarkedm}
\end{equation}
The EDM arises from the interference between the subdominant Bino
contribution and leading gluino contribution, and is proportional to
the relative phase between the gluino and Bino masses.  It is
suppressed not by a relative loop factor, but by the small magnitude
of the Bino coupling, in analogy to the suppression of the
CP-violating Higgs--quark couplings.  Under the assumption of
unification of gaugino masses,
$|m_{\tilde{B}}/m_{\tilde{g}}|={5 \over 3}\alpha^{\prime}/\alpha_s$, 
the factor 
$m_{\tilde{B}} \alpha^{\prime} / ( m_{\tilde{g}} \alpha_s) =
 {5 \over 3} \alpha^{\prime 2} / \alpha_s^2$ 
results in a suppression $\sim 10^{-2}$. In addition, the product of
color factor and gauge couplings gives a further suppression
${3 \over 8} Q_q Y_{q_L} Y_{q_R} \sim 10^{-1}$.  The difference of the
ratio of loop functions in the parenthesis in~(\ref{quarkedm}) need
not be particularly small, since the mass and dipole loop functions
are very different.  In the limit
$m_{\tilde{g}}=m_{\tilde{q}_1}=m_{\tilde{q}_2}= \tilde{m}
           \gg m_{\tilde{B}}$, 
the difference approaches $(\cdots) \rightarrow 4$.  So unlike the
case of the Higgs--quark CP-violating couplings, there is generally no
suppression coming from a relative similarity of Bino contributions to
the mass and dipole moment.  {}From the general expressions in
appendices~\ref{sec:defI} and~\ref{sec:muon}, the over-all ratio of
loop functions appearing in~(\ref{quarkedm}) is
$I_{g-2}(m_{\tilde{q}_1},m_{\tilde{q}_2},m_{\tilde{g}})/
 I(m_{\tilde{q}_1},m_{\tilde{q}_2},m_{\tilde{g}})\sim 1/(6 \tilde{m}^2)$, 
where the prefactor arises for
$m_{\tilde{g}}=m_{\tilde{q}_1}=m_{\tilde{q}_2}= \tilde{m}$.
The heaviest particle in the gluino loop contribution sets the scale
for the EDM operator.  Explicitly, for an up-type quark with radiative
mass and
$m_{\tilde{B}} \ll m_{\tilde{q}} \lsim m_{\tilde{g}}$, 
the up-quark EDM is
$d_u^e \sim e (2 / 27) (\alpha^{\prime} / \alpha_s)
(m_u |m_{\tilde{B}}| / |m_{\tilde{g}}|^3 )
\sin   \left( {\rm Arg}(m_{\tilde{B}} m_{\tilde{g}}^*) \right)$,
where the specific prefactor arises for
$m_{\tilde{q}}= m_{\tilde{g}} \gg m_{\tilde{B}} $.

If the first-generation-quark masses arise radiatively, the up- and
down-quark EDMs~(\ref{quarkedm}) may be used to estimate the neutron
EDM.  In this case, the EDMs are related by 
$d_d^e = {1 \over 4} (m_d/ m_u) d_u^e$.  The valence approximation for
the neutron EDM in the
$SU(2)$ limit is $d_n^e = {4 \over 3} d_d^e - {1 \over 3}d_u^e$.
%
%
\begin{figure}[ht]
\begin{center}
\epsfxsize= 12.5cm
\leavevmode
\epsfbox{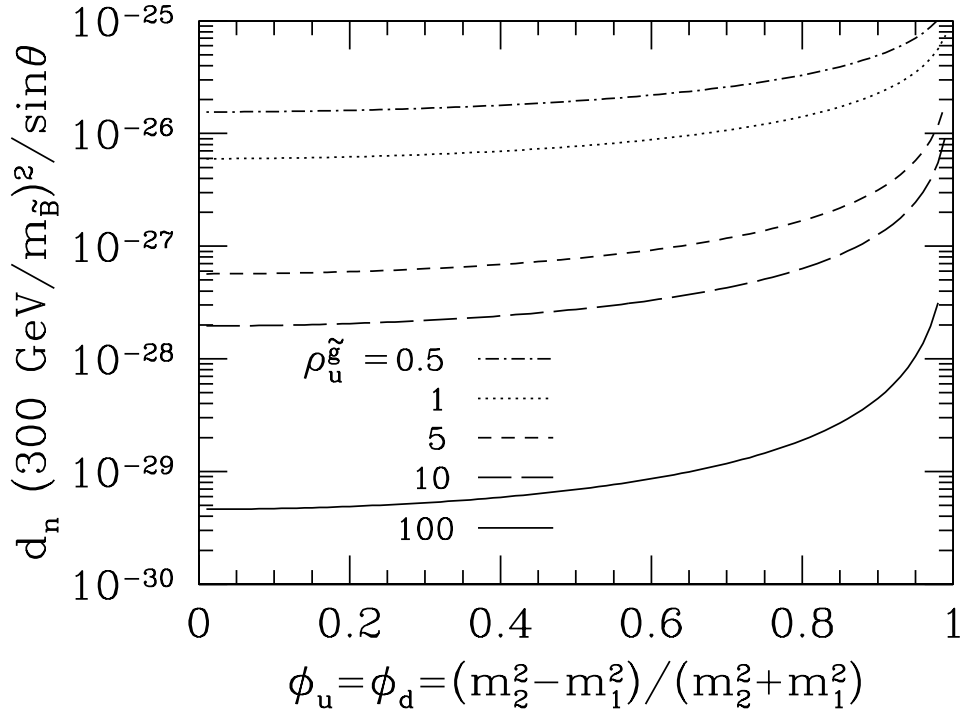}
\end{center}
\caption[f1]{Leading supersymmetric contribution to the neutron 
electric dipole moment in units of $e \cdot {\rm cm}$ with radiative
masses for the first-generation quarks as a function of the fractional
squark mass splitting, $\phi_u = \phi_d$.  The $SU(2)$ valence quark
approximation and equal up- and down-squark mass eigenvalues are
assumed.  The modulus of the Bino and gluino masses are assumed to be
related by gaugino mass unification, and the electroweak scale masses
for the valence quarks are taken to be $m_u=1$ MeV and $m_d=5$ MeV.
The intrinsic CP-violating phase is the relative phase between the
Bino and gluino masses,
$\sin \theta \equiv \sin ({\rm Arg}(m_{\tilde{B}} m_{\tilde{g}}^*))$.
}
\label{edmneut}
\end{figure}
The resulting neutron EDM from the first-generation quark EDMs is
plotted in fig.~\ref{edmneut} as a function of the fractional mass
splitting between squarks.  Gaugino mass unification is assumed for
the ratio of the Bino and the gluino mass, 
$|m_{\tilde{B}}/ m_{\tilde{g}}| =
  {5 \over 3} \alpha^{\prime} / \alpha_s$, and the electroweak scale
masses for the quarks are taken to be $m_u=1\,$MeV and $m_d=5\,$MeV.
Renormalization group running to the QCD scale would increase $d_n^e$
slightly.

If the first-generation quark masses do in fact arise radiatively, the
leading contribution to the neutron EDM in the valence approximation
given in fig. \ref{edmneut} is easily compatible with the current
experimental bound of
$ |d_n^e| < 1.0 \times 10^{-25}$ $ e \cdot {\rm cm}$, at 
the 90\% CL~\cite{nedmbound}.  Light-quark chromo-electric dipole
moments (CDMs) can also contribute to the neutron EDM~\cite{chromo}.
However, the phases of the chromo-dipole coefficients are also
naturally nearly aligned with that of the mass, just as for the EDMs,
resulting in a suppression very similar to that of the light-quark
EDMs given in~(\ref{quarkedm}) and discussed above.  With this,
estimates of the light-quark CDM contribution to the neutron
EDM~\cite{chromo} are very similar to that of the light-quark EDMs.
The strange sea content of the neutron may also allow comparable
contributions to the neutron EDM from the strange quark EDM and
CDM~\cite{strange}.  If the strange quark mass arises radiatively,
however, these moments are also significantly suppressed.  So if the
three lightest quark masses arise radiatively from soft chiral flavor
breaking, the present SUSY CP problem for the neutron EDM is
essentially eliminated by the natural phase alignment.  But, depending
on the superpartner mass spectrum and relative phase of the gluino and
Bino masses, ${\rm Arg}(m_{\tilde{B}} m_{\tilde{g}}^*)$, the neutron
EDM could be not too far below the current experimental bound.

Of course, if strict gaugino mass unification holds, in which the
phases of the gaugino masses are correlated so that 
${\rm Arg}(m_{\tilde{B}} m_{\tilde{g}}^*) =0$ at lowest order, the
leading light-quark contributions to the neutron EDM given above
vanish, just as for the Higgs--quark CP-violating couplings.  In this
case the neutron EDM receives contributions from light-quarks with
radiative masses mainly from neutralino mixing effects, which can be
much smaller than the leading contribution discussed above.  There are
also potentially small contributions from pure glue operators arising
from integrating out heavy quarks~\cite{GGG,GGGRG}.

For radiative lepton masses, only the neutralinos contribute at one
loop to both the dipole moment and mass.  Since both the electric and
magnetic dipole moments are related to the coefficient of the
electromagnetic dipole operator by eqs.~(\ref{dfsin})
and~(\ref{amusin}), it is convenient in this case to relate the EDM to
the supersymmetric contribution to the anomalous magnetic moment and
intrinsic CP-violating phases.  For the lepton EDM, the full one-loop
EDM with a radiative mass, using the general expression for a
radiative fermion mass~(\ref{mass}) and electromagnetic dipole
moment~(\ref{rawgmtwo}), may be written
\begin{equation}
d_l^e = {e ~ a_l^{\rm SUSY} \over 2 m_l} ~
  \tan \left[ {\rm Arg} \left(
  \sum_j K_l^j m_{\tilde{\chi}_j^0}
  I_{g-2}(m_{\tilde{l}_1},m_{\tilde{l}_2},m_{\tilde{\chi}_j^0})
    \over
        \sum_j K_l^j m_{\tilde{\chi}_j^0}
  I(m_{\tilde{l}_1},m_{\tilde{l}_2},m_{\tilde{\chi}_j^0})  \right)
    \right]
\label{leptonedm}
\end{equation}
where $a_l^{\rm SUSY}$ is the supersymmetric contribution to the
lepton anomalous magnetic moment, given in eq.~(\ref{mammform}).  In
the pure gaugino limit, only the Bino contributes to both the mass and
dipole moment coefficient.  The phases appearing in~(\ref{leptonedm})
are then aligned, and the one-loop lepton EDMs vanish.  In general,
however, gaugino--Higgsino mixing through electroweak symmetry
breaking introduces non-vanishing phases.

In a region of neutralino parameter space in which there is strong
mixing between gaugino and Higgsino states, the phases that appear
in~(\ref{leptonedm}) are generally not suppressed, potentially leading
to a sizeable lepton EDM. For example, if both the electron and muon
masses arise radiatively, the anomalous magnetic moments are related
by
$a_e^{\rm SUSY} = (m_e^2 / m_{\mu}^2) a_{\mu}^{\rm SUSY}$, assuming
$m_{\tilde{e}_L} = m_{\tilde{\mu}_L}$ and
$m_{\tilde{e}_R} = m_{\tilde{\mu}_R}$.
In this case the current experimental upper limit on the muon
anomalous magnetic moment of
$a_{\mu}^{\rm new} < 220 \times 10^{-10}$~\cite{MUONBOUND,MUONCALC} 
gives an upper limit on the prefactor in~(\ref{leptonedm}) for the
electron EDM of $a_e^{\rm SUSY} e/(2m_e) < 10^{-23}~e \cdot {\rm cm}$.
This is to be compared with the current experimental bound on the
electron EDM of roughly $|d_e^e| < 3 \times 10^{-27} ~e \cdot {\rm
cm}$ \cite{eedmbound}.  Therefore in the well-mixed region of
neutralino parameter space, compatibility of the electron EDM with the
current bound would require either small relative intrinsic phases or
a much smaller value of $a_e^{\rm SUSY}$ than implied by the current
bound on $a_{\mu}^{\rm new}$.  The latter would in fact occur with
very heavy selectrons.

In the mostly gaugino or Higgsino region of neutralino parameter
space, however, the physical CP-violating phase appearing in a lepton
EDM can be significantly suppressed by phase alignment.  In this case
the mostly Bino state dominates the contribution to both the mass and
dipole moment coefficient.  The general expression for the physical
phase~(\ref{dm}) is therefore applicable.  The lepton
EDM~(\ref{leptonedm}) in this limit then reduces to
$$
d_l^e \simeq {e ~a_l^{\rm SUSY} \over 2 m_l}  ~ \sum_{j=2}^4
   \left| { K_l^j m_{\tilde{\chi}_j^0}
        \over
                K_l^1 m_{\tilde{\chi}_1^0} } \right|
    \left(
 {I_{g-2}(m_{\tilde{l}_1},m_{\tilde{l}_2},m_{\tilde{\chi}_j^0}) \over
  I_{g-2}(m_{\tilde{l}_1},m_{\tilde{l}_2},m_{\tilde{\chi}_1^0}) }
     -
 {I(m_{\tilde{l}_1},m_{\tilde{l}_2},m_{\tilde{\chi}_j^0}) \over
  I(m_{\tilde{l}_1},m_{\tilde{l}_2},m_{\tilde{\chi}_1^0}) }
   \right)
$$
\begin{equation}
   ~~~~~~~~~~\times  ~
 \sin \left( {\rm Arg}(K_l^j m_{\tilde{\chi}_j^0}
   K_l^{1*} m_{\tilde{\chi}_1^0}^* )
     \right).
         \label{eedm}
\end{equation}
Just as in the case of the Higgs--lepton couplings, neutralino mixing
induced through electroweak symmetry breaking may be treated
perturbatively in this limit, with the coupling coefficients $K_l^j$
unmodified at first order.  At second order in the mixing, a
non-trivial phase dependence is introduced by the mixing between the
Bino and Higgsino states, with a small coupling
$K^j_l \sim {\cal O}(m_Z^2/\tilde{m}^2)$ for $j$ the mostly
Higgsino states, where $\tilde{m}={\rm max}(m_{\tilde{B}},\mu)$.
The leading intrinsic phase that appears on the right-hand side
of~(\ref{eedm}) is then the phase between the masses of the mostly
Bino and Higgsino states, ${\rm Arg}(m_{\tilde{B}} \mu (m_{12}^2)^*)$.
The suppression for the relative similarity of the mostly Higgsino
contributions to the mass and dipole coefficient, contained in the
difference of ratios of loop functions in the parenthesis
in~(\ref{eedm}), although not necessarily so, can be significant in
certain regions of parameter space.  For example, in the limit
$m_{\tilde{B}} \lsim m_{\tilde{l}} \ll \mu$ the difference factor 
$(\cdots )$ in~(\ref{eedm}) is dominated by the ratio of mass loop
functions since the non-renormalizable dipole operator decouples more
quickly than the renormalizable effective Yukawa coupling.  The
relative contribution of the mostly Higgsino state to the dipole
coefficient is therefore insignificant, and the phase misalignment
comes mainly from the mostly Bino and Higgsino contributions to the
radiative mass.  Parametrically, in this case, the term in parenthesis
in~(\ref{eedm}) is
$( \cdots ) \sim {\cal O} (m_{\tilde{l}}^2/ \mu^2)$.
Altogether then, in this limit the lepton EDM is
$$
d_l^e \sim { e ~a_l^{\rm SUSY} \over 2 m_l} ~
 {\cal O}( m_Z^2 m_{\tilde{l}}^2 / (\mu^3 m_{\tilde{B}}) ) ~
 \sin \left( {\rm Arg}(m_{\tilde{B}} \mu (m_{12}^2)^*) \right).
$$
As another example, in the limit $m_{\tilde{B}} \lsim \mu \ll
m_{\tilde{l}}$, the slepton masses set the scale for all the loop
functions, and the ratios of loop functions approach unity.  The
difference of loop functions in~(\ref{eedm}) may be obtained from the
results of appendices~\ref{sec:defI} and~\ref{sec:muon}.
Parametrically, in this case, $( \cdots ) \sim {\cal O} ( \mu^2 /
m_{\tilde{l}}^2)$.  The lepton EDM in this limit is then
$$
d_l^e \sim { e ~a_l^{\rm SUSY} \over 2 m_l} ~
 {\cal O}( m_Z^2  \mu / ( m_{\tilde{B}} m_{\tilde{l}}^2 )) ~
 \sin \left( {\rm Arg}(m_{\tilde{B}} \mu (m_{12}^2)^*) \right).
$$
Note in addition that $a_l^{\rm SUSY}$ is itself also suppressed in
this limit, $a_l^{\rm SUSY} \simeq m_l^2 / m_{\tilde{l}}^2$, for
$m_{\tilde{l}_L} = m_{\tilde{l}_L} =m_{\tilde{l}}
  \gg m_{\tilde{B}}$.
Obtaining a radiative mass for the electron is in fact possible with
very heavy selectrons, since the effective renormalizable Yukawa
coupling need not vanish in the superpartner decoupling limit.  As
discussed in section~\ref{sec:possible}, very heavy first-generation
scalar partners could partially or fully account for the smallness of
the first-generation fermion masses.  Depending on the specific
superpartner masses and mixings, it is possible for the physical phase
appearing in the electron EDM to be suppressed sufficiently to be
compatible with the current experimental bound in the radiative
electron mass scenario.

In summary, if the light-quark masses arise radiatively from soft
chiral flavor breaking, the supersymmetric contribution to the neutron
EDM is significantly suppressed because of the natural near-alignment
of the phases of the dipole operators and fermion masses.  This
suppression is numerically significant even though the EDM is not
suppressed by a relative loop factor.  The leading contribution is
proportional to the intrinsic CP-violating phase 
$\sin ({\rm Arg}(m_{\tilde{B}} m_{\tilde{g}}^*) $, and is easily
compatible with the current experimental bound.  Depending on the
superpartner mass spectrum and
${\rm Arg}(m_{\tilde{B}} m_{\tilde{g}}^*)$, the neutron EDM
could be not too far below the current experimental bound.  However,
with strict gaugino mass unification, 
${\rm Arg}(m_{\tilde{B}} m_{\tilde{g}}^*)= 0$, and the neutron EDM
is likely to be well beyond the reach of currently anticipated
experiments.  The electron EDM is also suppressed in the mostly
gaugino or Higgsino region of neutralino parameter space, where the
leading contribution is proportional to the intrinsic phase
$\sin ({\rm Arg}(m_{\tilde{B}} \mu (m_{12}^2)^*))$. Unlike the 
neutron EDM, the magnitude of the electron EDM in terms of this
intrinsic phase is a very model-dependent function of the neutralino
mixings and superpartner masses.  In plausible regions of neutralino
and scalar partner parameter space it can be consistent with current
experimental limits.  CP-violating Higgs--fermion couplings are
negligible with radiative masses.

%
%

\section{Probing the Stiffness of Yukawa Couplings}
\label{sec:probing}

Radiative fermion masses generated at the superpartner mass scale
require a relatively large breaking of chiral flavor symmetries in
scalar tri-linear terms, and imply that the effective Yukawa couplings
are soft, as discussed in section~\ref{sec:soft}.  This leads to
various distinctive effects, which can in principle be probed directly
in high energy experiments, including:
\begin{enumerate}
\item
Form factor effects for the effective Higgs Yukawa couplings,
summarized in the ratios $r_{f,H}(q^2)$ and $r_{f,A}(q^2)$.
\item
``Wrong Higgs'' couplings for radiative masses arising from
non-holomorphic $A^{\prime}$ operators.
\item
Large left--right mixing for second-generation scalar partners.
\item
Apparent, hard supersymmetry-breaking effects in
Higgsino--matter couplings.
\end{enumerate}
In this section the possibilities for detecting and measuring these
effects are outlined.  As discussed below, one of the most useful tools
would be the proposed $\mu^+ \mu^-$ collider, which can produce the
neutral Higgs bosons as $s$-channel resonances and make precision
measurements of Higgs--fermion couplings~\cite{MUON}.

The most obvious arena for direct experimental probes of radiatively
generated effective Yukawa couplings are precision measurements of the
physical Higgs--fermion couplings.  These couplings are given by
$\lambda_{h^0,H^0} =
\Theta \{ \cos \beta , \sin \beta \}\bar{h}_{f,H} / \sqrt{2}$
with $H=h^0,H^0$, and
$\lambda_{A^0}=
\Theta \{ \cos \beta , \sin \beta \}\bar{h}_{f,A} / \sqrt{2}$,
where the first (second) terms in curly brackets corresponds to
$\alpha=1$ $(2)$, and $\Theta$ represents the Higgs mixing matrix
between physical and interaction eigenstates~\cite{HG,HHGUIDE}
relevant to the fermion in question.  The mixing coefficients $\Theta$
are given in table~\ref{table:t3}, where $\alpha$ is the
$h^0$--$H^0$ mixing angle, and
$\tan \beta = \langle H_2 \rangle / \langle H_1 \rangle$.
{}From~(\ref{higgsratio}) and~(\ref{pseudoscratio}),
$\bar{h}_{f,H} = r_{f,H} h_{f,m}$, and
$\bar{h}_{f,A} = r_{f,A} h_{f,m}$, where $r_{f,H}$, $r_{f,A}$ are the
magnitude of the effective Yukawa couplings relative to a tree-level
coupling.  The form factor effect is contained in the magnitude and
momentum dependence of the ratios $r_{f,H}$, $r_{f,A}$.  For
$m_{\tilde{f}}^2, m_{\lambda}^2 \sim m_H^2$ the ratio $r_f(q^2)$ has a
non-trivial momentum dependence in addition to the logarithmic
momentum dependence from renormalization group evolution.  The
logarithmic dependence may be subsumed in the definition
$m_f=m_f(q^2)$ or $h_{f,m}=h_{f,m}(q^2)$ above.  For 
$m_{\tilde{f}}^2, m_{\lambda}^2 \gsim m_H^2$, the scalar 
Higgs--fermion radiative coupling ratios are greater than unity,
$r_{f,H}(m_{h^0}^2),r_{f,H}(m_{H^0}^2) \geq 1$, from intrinsic
coupling effects in the one-loop diagram, as described in
section~\ref{sec:higgs}.
Thus, even if the momentum dependence cannot be determined
experimentally, the over-all magnitude of the effective Yukawa
coupling can differ from the minimal case with tree-level masses.  In
contrast, the pseudoscalar Higgs--fermion radiative coupling ratios
deviate from unity only from neutralino mixing in the one-loop
diagram, and through non-trivial momentum dependence.  
Comparison of scalar and pseudoscalar Higgs couplings, as discussed
below, might then in principle allow the momentum dependence of the
Yukawa couplings, or equivalently of the finite Yukawa mass radii, to be
disentangled from intrinsic coupling effects.

The Higgs mixing effects in the Higgs--fermion couplings,
contained in $\Theta$, depend on the projection of the physical Higgs
bosons onto the Higgs doublet, which gives rise to the fermion mass.
In any supersymmetric theory with tree-level fermion masses and a
single pair of Higgs doublets, $H_1$ with $U(1)_Y$ hypercharge $Y=-1$,
and $H_2$ with $Y=+1$, gauge invariance and holomorphy of the
superpotential guarantee that, at tree level, up-type quarks receive
mass from $H_2$ while down-type quarks and leptons receive mass from
$H_1$.  The coupling coefficients of the physical scalar Higgs bosons
$h^0$ and $H^0$, and physical pseudoscalar $A^0$, are then fixed to be
those given in table~\ref{table:t3}.
%
%
\begin{table}
\caption{The neutral Higgs boson mixing coefficients $\Theta$
 appearing in the effective Higgs--fermion Yukawa coupling.
 The coefficient for a particular fermion depends on whether
 the fermion mass originates from a holomorphic
 $A$-term, non-holomorphic $A^{\prime}$-term, or tree-level
 superpotential Yukawa coupling.
}
\label{table:t3}
\begin{tabular}{lll}
Higgs boson & Radiative up-type $\propto A$
            & Radiative up-type $\propto A^{\prime}$
 \\
            & Radiative down-type/lepton  $\propto A^{\prime}$
            & Radiative down-type/lepton $\propto A$
 \\
            & Tree-level up-type
            & Tree-level down-type/lepton
 \\ \hline
$h^{0}$ & $\phantom{-}\cos\alpha/\sin\beta$
        & $-\sin\alpha/\cos\beta$ \\
$H^{0}$ & $\phantom{-}\sin\alpha/\sin\beta$
        & $\phantom{-}\cos\alpha/\cos\beta$ \\
$A^{0}$ & $\phantom{-}i\cot\beta$
        & $\phantom{-} i\tan\beta$
\end{tabular}
\end{table}
For radiative masses arising from holomorphic chiral flavor breaking,
the fermions receive mass from the same type of Higgs doublets, and
the coupling coefficients are the same as with tree-level masses. 
In this case the Higgs--fermion couplings differ from those in 
the minimal case
with tree-level masses only through the magnitude of the
momentum-dependent form factors $r_{f,H}(q^2)$, $r_{f,A}(q^2)$.
However, with non-holomorphic chiral flavor breaking, radiative
up-type quark masses arise from $H_1$, while radiative down-type quark
and lepton masses arise from $H_2$.  The resulting ``wrong Higgs''
coupling coefficients to physical Higgs bosons in this case are given
in table~\ref{table:t3}.  
Depending on $\alpha$ and $\beta$, these coupling coefficients
can drastically differ from those in the minimal case with 
tree-level masses or those obtained with holomorphic chiral flavour 
breaking.
The ``wrong Higgs'' modifications of the
couplings persist in the strict superpartner decoupling limit of
$m_{\tilde{f}}^2, m_{\lambda}^2 \gg m_H^2$, even though 
$r_{f,H}, r_{f,A} \to 1$ in this limit.

Because the Higgs coupling coefficients depend on two mixing
parameters, even in the minimal model, multiple measurements are
necessary in order to discern any radiative contribution to a fermion
mass.  Fortunately, it is very unlikely that the top quark or
$\tau$-lepton masses arise radiatively, as discussed in
section~\ref{sec:possible}.  The couplings of these fermions to
various Higgs bosons can be used as ``standard candles'' by which to
compare other fermion couplings.  For the lightest Higgs boson, $h^0$,
the theoretical upper bound in supersymmetric theories on the Higgs
mass and the measured top-quark mass imply $m_{h^0} < m_t$.  The
channel $h^0 \to tt$ is therefore closed, so $h^0 \to \tau \tau$ must
be used as a ``standard candle''.  In a minimal theory with tree-level
masses, all the down-type quarks and leptons gain mass from a single
Higgs doublet, $H_1$.  For $h^0$ couplings, the most useful quantity
to consider is therefore the ratio of branching ratios for a down-type
quark or lepton to that of the $\tau$-lepton
\begin{equation}
{ m_{\tau}^2 \over N_c m_f^2}
{ {\rm Br}(h^0 \to ff) \over {\rm Br}(h^0 \to \tau \tau) }\ = \
  r^2_{f,H}(m_{h^0}^2)  ~
 \left\{ 1 ~,~ \cot^2 \alpha \cot^2 \beta \right\} \,,
\label{hratio}
\end{equation}
where $N_c$ is a color factor, $N_{c}=3$ for quarks and $N_{c}=1$ for
leptons.  
Throughout this section the first term in curly brackets refers 
to tree-level masses or radiative masses from holomorphic A-terms; 
the second term to ``wrong Higgs'' couyplings for radiative masses 
from non-holomorphic A'-terms.
Finite fermion mass effects in the final-state phase space are ignored
throughout, except for the top quark.  Analogous relations hold for
the heavy Higgs scalar $H^0$:
\begin{equation}
{ m_{\tau}^2 \over N_c m_f^2}
{ {\rm Br}(H^0 \to ff) \over {\rm Br}(H^0 \to \tau \tau) }\ = \
  r^2_{f,H}(m_{H^0}^2)  ~
 \left\{ 1 ~,~ \tan^2 \alpha \cot^2 \beta \right\} \,,
\label{hhratio}
\end{equation}
and for the pseudoscalar $A^0$:
\begin{equation}
{ m_{\tau}^2 \over N_c m_f^2}
{ {\rm Br}(A^0 \to ff) \over {\rm Br}(A^0 \to \tau \tau) }\ = \
  r^2_{f,A}(m_{A^0}^2)  ~
 \left\{ 1 ~,~ \cot^4 \beta \right\} \,,
\label{Aratio}
\end{equation}
where $f$ is a down-type quark or lepton.  In the Higgs decoupling
limit of $m_{A^0}^2 \gg m_Z^2$, the light-Higgs projection onto the
Higgs doublets aligns with the expectation values and the $h^0$
couplings become standard-model-like, giving $\cot^2 \alpha \to \tan^2
\beta$.  This occurs even for radiative Yukawa couplings, and implies
that all $h^0$ branching ratios approach standard-model values in the
Higgs decoupling limit, up to possible form-factor effects.  Thus in
this limit, independently of whether the fermion masses are tree-level
or radiative from either holomorphic or non-holomorphic chiral flavor
breaking, the ratio~(\ref{hratio}) for $h^0$ just depends on the form
factor.  However, in the case of non-holomorphic breaking, the
ratios~(\ref{hhratio}) and~(\ref{Aratio}) for $H^0$ and $A^0$ can
differ drastically from unity because of the ``wrong Higgs'' coupling.
In particular, in the ``wrong Higgs'' case, it holds:
$${\rm Br}(A^0 \to \tau \tau) > {\rm Br}(A^0 \to bb) 
 {\rm for }  
 \tan \beta > \sqrt{ \sqrt{3} m_b / m_{\tau}}  \simeq 1.7.$$
The momentum dependence of the effective Higgs Yukawa coupling can in
principle be probed by comparing $r_{f,H}(m_{h^0}^2)$ with
$r_{f,H}(m_{H^0}^2)$ or $ r_{f,A}(m_{A^0}^2)$ for a given final state.

It is important to emphasize that any supersymmetric theory with
tree-level Yukawa couplings and only a pair of Higgs doublets is
guaranteed by holomorphy of the superpotential to give a ratio of
unity for~(\ref{hratio})--(\ref{Aratio}) up to small and calculable
quantum corrections, as described above. Deviations of the
lowest-order ratios would therefore be an indication of either
additional Higgs doublets or, with only a single pair of doublets,
radiative contributions to the masses -- through form-factor effects
and/or non-holomorphic ``wrong Higgs'' couplings.

For the heavy-Higgs pseudoscalar, if kinematically open, $A^0 \to tt$
can compete with $A^0 \to bb$ depending on the value of $\tan \beta$.
For reference, with a tree-level or radiative mass for the $b$-quark
with holomorphic chiral flavor breaking, the $A^0tt$ coupling is
larger than the $A^0bb$ coupling for $\tan \beta < \sqrt{m_t/m_b}
\simeq 7.5$.  In this case the (tree-level) top quark coupling can be
used as a ``standard candle'' with which to compare the charm quark
coupling through the ratio
\begin{equation}
{ m_{t}^2 \over  m_c^2}
{ {\rm Br}(A^0 \to cc) \sqrt{1 - 4 m_t^2/m_{A^0}^2} \over
  {\rm Br}(A^0 \to tt) } \ = \
  r^2_{c,A}(m_{A^0}^2)  ~
 \left\{ 1 ~,~ \tan^4 \beta  \right\}  \,,
\label{Act}
\end{equation}
where $\sqrt{1-4 m_t^2/m_{A^0}^2}$ is a kinematic correction for the
$S$-wave decay of a pseudoscalar. In a theory with tree-level Yukawa
couplings, or radiative masses with holomorphic flavor breaking, if
$A^0 \to tt$ is open, ${\rm Br}(A^0 \to cc)$ is unobservably small.
However, from~(\ref{Act}), if the charm quark gains a mass from the
``wrong Higgs'' through non-holomorphic flavor breaking, 
${\rm Br}(A^0\to cc)$ can be non-negligible, and even dominate ${\rm
Br}(A^0 \to tt)$ if $\tan \beta \gsim \sqrt{m_t / m_c} \simeq 16$.  In
this case ${\rm Br}(A^0 \to cc)$ can be small or large, depending on
the origin of the $b$-quark mass.  If the $b$-quark mass is tree-level
or radiative through a holomorphic $A$-term, 
${\rm Br}(A^0 \to cc) \leq m_c^2 / (m_b^2+m_c^2) \simeq 0.05$
(neglecting the form factors).  However, if the $b$-quark mass arises
radiatively from a non-holomorphic $A'$-term, 
${\rm Br}(A^0 \to cc) \simeq 1$ in the large $\tan \beta$ limit, even
if $A^0 \to tt$ is open.  An analogous relation to~(\ref{Act}) holds
for the heavy Higgs scalar $H^0$ with $P$-wave kinematic correction
and with $\tan^4 \beta$ replaced by $\cot^2 \alpha \tan^2 \beta$.
With $\tan \beta \gg 1$, ${\rm Br}(H^0,A^0 \to cc)$ are therefore very
sensitive to possible ``wrong Higgs'' couplings arising from a
radiative charm quark mass.  This is in contrast to 
${\rm Br}(h^0 \to cc)$, which, in the Higgs decoupling limit,
approaches the standard model value up to form factor effects, as
discussed above.

Measurements of the above ratios of branching ratios requires
flavor identification of the final states.  With adequate $\tau$- and
$b$-tagging at future lepton and possibly hadron colliders, the
ratios~(\ref{hratio})--(\ref{Aratio}) can be used to probe a radiative
$b$-quark mass.  The ratio~(\ref{Act}) also requires $c$-tagging and
top identification to probe a radiative charm-quark mass.  Extension
to other fermions is problematic since in most scenarios the branching
ratios will be small.  However, the proposed $\mu^+ \mu^-$ collider,
which can produce the Higgs bosons as $s$-channel resonances, provides
the possibility of measuring the muon--Higgs couplings~\cite{MUON}.
For $h^0$, with width less than the beam width,
$\Gamma_{h^0}^{\rm tot} \lsim \Gamma_{\rm beam}$, the total cross
section
$\sigma(\mu^+ \mu^- \to h^0 \to X)$ gives a measure of
$\Gamma(h^0 \to \mu \mu) {\rm Br}(h^0 \to X)$. Independent 
measurements of 
${\rm Br}(h^0 \to X)$ can then give $\Gamma(h^0 \to \mu \mu)$ to a few
percent precision~\cite{GUNION}.  A measurement of 
$\Gamma_{h^0}^{\rm tot}$ from scanning around the $h^0$ resonance,
with a beam energy resolution better than the beam width
$ \delta \sqrt{s} \ll \Gamma_{\rm beam} $, then yields 
${\rm Br}(h^0 \to \mu \mu) =
\Gamma(h^0 \to \mu \mu) / \Gamma_{h^0}^{\rm tot}$~\cite{MUON}.
For $H^0$ and $A^0$ with 
$\Gamma_{H^0,A^0}^{\rm tot} \gsim \Gamma_{\rm beam}$, the peak 
cross sections $\sigma(\mu^+ \mu^- \to H^0,A^0 \to X)$ at the 
center of the resonances give a direct measure of 
${\rm Br}(H^0,A^0 \to \mu \mu) {\rm Br}(H^0,A^0 \to X)$.  
Thus, a measurement of these ratios can also test whether 
the muon mass is generated radiatively. Note,
however, that
$r_{\mu,H}(0)$ is already bounded by muon magnetic moment
measurements, and will be very well bounded or determined by such
measurements by the time the $\mu^+ \mu^-$ collider is in operation.
In this case, the ratios~(\ref{hratio})--(\ref{Aratio}) for the muon
will provide a test for possible ``wrong Higgs'' couplings.

In addition to the branching ratios, a $\mu^+ \mu^-$ collider also
allows the possibility of determining absolute widths, as described
above.  The absolute magnitude of the Higgs--muon coupling obtained
from $\Gamma(h^0 \to \mu \mu)$ cannot in general be interpreted
directly because of Higgs mixing effects and the $\tan \beta$ dependence
of the $h^0 \mu \mu$ coupling.  In the Higgs decoupling limit of
$m_{A^0}^2 \gg m_Z^2$, however, the light Higgs couplings for either
tree-level or radiative fermion masses become standard-model-like, up
to the form-factor ratios, as discussed above.  So in the Higgs
decoupling limit
\begin{equation}
\left.
{ \Gamma(h^0 \to ff) \over  \Gamma(\phi^0 \to ff) }
\right|_{m_{A^0}^2 \to \infty} =\  r^2_{f,H}(m_{h^0}^2) \,,
\label{hdecoup}
\end{equation}
where $\Gamma(\phi^0 \to ff)$ is the width for the decay of the
standard model Higgs boson, $\phi^0$, which is calculable in terms of
the fermion mass.  However, away from the decoupling limit, deviations
of the $h^0$--fermion couplings only vanish like 
${\cal O}(m_{Z}^2 / m_{A^0}^2)$, and can be significant for finite
$m_{A^0}$.  Therefore, 
the ratio~(\ref{hdecoup}) of absolute to standard-model
widths is not as useful as ratios of physical Higgs boson branching
ratios for disentangling the effects of the form factor and possible
``wrong Higgs'' coupling from Higgs mixing effects.

The absolute widths for $\Gamma(H^0,A^0 \to ff)$ can also be
determined at a $\mu^+ \mu^-$ collider.  Given measurements of
${\rm Br}(H^0,A^0 \to ff)$, either independently or as described above,
measurements of $\Gamma^{\rm tot}_{H^0,A^0}$ by precision scans of the
line shapes yield the widths
$\Gamma(H^0,A^0 \to ff) = 
{\rm Br}(H^0,A^0 \to ff) \Gamma^{\rm tot}_{H^0,A^0}$.
Again, the absolute magnitude of the $H^0$--fermion couplings suffer
from mixing effects.  The pseudoscalar couplings are independent of
$h^0$--$H^0$ mixing, but do depend on $\tan \beta$ as shown in
table~\ref{table:t3}.  The normalized $A^0$ partial widths for
down-type quarks or leptons are
\begin{equation}
{ \Gamma(A^0 \to ff) \over  \Gamma(\phi^0 \to ff) } \ = \
 r^2_{f,A}(m_{A^0}^2)
\left\{ \tan^2 \beta ~,~ \cot^2 \beta \right\} \,.
\label{Awidth}
\end{equation}
Making use of these ratios requires an independent measurement of
$\tan \beta$.  Alternatively, if perturbativity of the top-quark Yukawa
coupling up to a large scale is imposed, implying
$\tan \beta \gsim 1.8$, a value of the normalized widths $\leq 3.2$
could be taken as evidence of form-factor effects or ``wrong Higgs''
couplings.

At a $\mu^+ \mu^-$ collider, with the possibility of measuring
absolute decay widths, also tests of supersymmetric theories with
tree-level Yukawa couplings and a single pair of Higgs doublets can be
made.  Another way to eliminate Higgs mixing effects in the minimal
case is to sum the normalized decay widths of both $h^0$ and $H^0$ to
a given fermion final state.  For down-type quarks or leptons with
tree-level or radiative masses from holomorphic flavor breaking, or
for up-type quarks with radiative masses from non-holomorphic flavor
breaking, this sum gives:
\begin{equation}
{\Gamma(\phi^0 \to ff) \over \Gamma(h^0 \to ff) } +
{\Gamma(\phi^0 \to ff) \over \Gamma(H^0 \to ff) } \ = \
{ \cos^2 \beta \over
 r^2_{f,H}(m_{h^0}^2)\sin^2 \alpha +
 r^2_{f,H}(m_{H^0}^2) \cos^2 \alpha }\,.
 \label{gamhsum}
\end{equation}
For up-type quarks with tree-level or radiative masses from
holomorphic $A$-terms, or for down-type quarks or leptons with
radiative masses from non-holomorphic $A^{\prime}$-terms,
(\ref{gamhsum})~holds with the substitutions 
$\cos^2 \beta \to \sin^2 \beta$ and 
$\cos^2 \alpha \leftrightarrow \sin^2 \alpha$.  With tree-level masses
the denominator on the right-hand side of~(\ref{gamhsum}) sums to
unity, up to quantum corrections.  Thus, with a radiative 
fermion mass, ~(\ref{gamhsum}) can be sensitive to the momentum 
dependence of the form factor, if an independent 
measurement of $\tan \beta$ is available.  Alternatively, with
tree-level masses the sum of~(\ref{gamhsum}) for a down-type quark or
lepton and the analogous relation for an up-type quark is unity.  This
quantity for $bb$ and $cc$ widths or $\mu \mu$ and $cc$ widths would
provide an interesting test for form-factor effects for these
fermions.  Making use of the latter sums with holomorphic flavor
breaking in general requires that $H^0 \to tt$ be closed so that 
${\rm Br}(H^0 \to cc)$ is non-negligible.

Another quantity that eliminates all Higgs mixing effects, including
$\tan \beta$ dependence, is the product of normalized partial widths
for $A^0$ decay to up-type quarks and down-type quarks or leptons.  If
$A^0 \to tt$ is open,
\begin{equation}
{ \Gamma(A^0 \to tt) \Gamma(A^0 \to ff) (1 - 4m_t^2/m_{A^0}^2) \over
   \Gamma(\phi^0 \to tt) \Gamma(\phi^0 \to ff) } \ = \
 r^2_{f,A}(m_{A^0}^2) ~
  \left\{ 1 ~,~ \cot^4 \beta \right\} \,,
\label{Aprod}
\end{equation}
where $f=b,\tau,\mu$, and $(1 - 4m_t^2/m_{A^0}^2)$ is a kinematic
correction factor for the $S$-wave decay of a pseudoscalar, compared
with the $P$-wave decay of a scalar.  If open, ${\rm Br}(A^0 \to tt)$
depends on $\tan \beta$, as discussed above.  If ${\rm Br}(A^0 \to tt)$
is non-negligible, the product~(\ref{Aprod}) with $f = \mu$ from
measurements at a $\mu^+ \mu^-$ collider provides an interesting test
of form factors or ``wrong Higgs'' couplings to the muon.  If 
${\rm Br}(A^0 \to tt)$ and ${\rm Br}(A^0 \to bb)$ are comparable, the
$b$-quark coupling can be probed.  Note that with $f = \tau$ the
product~(\ref{Aprod}) is unity for a single pair of Higgs doublets
with tree-level top quark and $\tau$-lepton Yukawa couplings.  This
may therefore be used as a good test for multiple pairs of Higgs
doublets.  If $A^0 \to tt$ is closed, analogous relations may be
applied with $\Gamma(A^0 \to cc)$.

A final interesting quantity for testing $A^0$ couplings 
is the sum of normalized
decay widths to an up-type quark and a down-type quark or lepton.  If
$A^0 \to tt$ is open, 
\begin{eqnarray}
\lefteqn{
{\Gamma( A^0 \to ff) \over \Gamma(\phi^0 \to ff)} +
{\Gamma( A^0 \to tt) (1 - 4 m_t^2/m_{A^0}^2)
       \over \Gamma(\phi^0 \to tt)} \ = \
} \nonumber \\
 & &
\hspace*{3truecm}
\left\{
   r^2_{f,A}(m_{A^0}^2) \tan^2 \beta + \cot^2 \beta ~,~
 ( r^2_{f,A}(m_{A^0}^2) +1) \cot^2 \beta \right\} \,,
\label{Asum}
\end{eqnarray}
with $f=b,\tau,\mu$.  In the minimal case with tree-level masses the
sum~(\ref{Asum}) is strictly~$\geq 2$, up to calculable quantum
corrections.  With a radiative mass for $f$ from holomorphic chiral
flavor breaking this bound can be modified by the magnitude of the
form factor ratio $r^2_{f,A}(m_{A^0}^2)$.  The effect is most dramatic
when $f$ obtains a mass radiatively from non-holomorphic breaking, in
which case the sum~(\ref{Asum}) is in principle arbitrary.  If
perturbativity of the top Yukawa up to a large scale is imposed,
implying $\tan \beta \gsim 1.8$, the sum~(\ref{Asum}) can be
significantly lower than the bound implied by tree-level masses.
Again a violation of the lower bound for~(\ref{Asum}) would be a clear
signal in a supersymmetric theory for either more than a single pair
of Higgs doublets or, with a single pair, of radiative masses.

The requirement of large tri-linear terms for radiative
second-generation masses leads to large left--right mixing for
second-generation scalars, as mentioned at the beginning of this
section.  This may be probed in a number of ways in high-energy
experiments.  The large tri-linear terms coupling a Higgs doublet to
left- and right-handed scalars can lead directly to enhanced decays
involving Higgs bosons and second-generation mass eigenstate scalars,
$\tilde{f}_1$ and $\tilde{f}_2$.  For example, if open, 
${\rm Br}(\tilde{f}_{2} \rightarrow \tilde{f}_{1}h^{0})$ or
${\rm Br}(H^0,A^0 \to \tilde{f}_{1} \tilde{f}_{2})$ can be
non-negligible.  In contrast, with tree-level masses, these are
expected to be insignificant for second-generation scalars.  In
addition, left--right mixing can also directly affect production cross
sections.  For example, the polarized cross sections
$\sigma(e_{L,R}^+e_{L,R}^- \to \tilde{f}_i \tilde{f}_i)$ etc.,
depend sensitively on the gauge couplings of $\tilde{f}_i$.  In the
minimal case, second generation scalar mass eigenstates are expected
to be nearly pure gauge eigenstates.  But with radiative masses, since
the tri-linear terms are so large, the mass eigenstates are very likely
to be well-mixed combinations of left- and right-handed scalars.

Finally, with radiative masses from scalar chiral flavor breaking, the
radiatively generated Higgsino coupling can differ drastically from
the radiative Higgs coupling, with even different parametric
dependence on the underlying couplings, as described in
section~\ref{sec:higgsino}.  This is a clear prediction of the present
scenario for radiative fermion masses.  To an electroweak scale
observer who assumes tree-level masses, this appears as a hard
violation of the supersymmetric relation between the Higgs and
Higgsino couplings.  Unfortunately, because of complicated
gaugino--Higgsino mixing effects, disentangling the Higgsino--matter
couplings from couplings of physical neutralinos and charginos in, for
example,
$\chi_i^{0} \to \tilde{f} f$ and
$\chi_i^{\pm} \to \tilde{f} f^\prime$, or
$\tilde{f} \to \chi_i^{0} f$ and
$\tilde{f} \to \chi_i^{\pm} f^\prime $,
is much more difficult than disentangling Higgs--matter couplings.
Unless some of the neutralino and chargino states are very nearly pure
Higgsino-like, this would require precision branching-ratio
measurements for many final states.

%
%

\section{Softly broken lepton number}
\label{leptonsec}

In addition to chiral flavor symmetries, it is also possible that
other global chiral symmetries such as lepton or baryon number or
matter parity~\cite{FF} are broken in the low-energy theory
predominantly by auxiliary rather than scalar expectation values.
Consider the case, for example, of lepton-number violation.  Soft
lepton-number violation involving the bilinear terms 
${\cal L} \supset m^2 L H_2$ was considered in ref.~\cite{LEE}.  Here
we discuss the consequence of lepton-number violating tri-linear
scalar terms analogous to~(\ref{ophol}) and~(\ref{opnonhol}) in the
holomorphic operators
\begin{equation}
{\cal L} \supset \widehat{A} LQD + \widehat{A}LLE
\label{lophol}
\end{equation}
and the non-holomorphic operator
\begin{equation}
{\cal L}  \supset \widehat{A}^{\prime} L^* QU
\label{lopnonhol}
\end{equation}
where the flavor structure is suppressed.  Note that the
non-holomorphic operator involves right-handed up-squarks.  This
differs from the standard tree-level superpotential lepton-number
violating Yukawa couplings, which are restricted to be holomorphic.
The lepton-number violating operators~(\ref{lophol}) give rise to a
radiatively generated neutrino mass.  At one loop a soft lepton-number
violating sneutrino--antisneutrino mixing term,
${\cal L} \supset \delta m^2_{\tilde{\nu}} \tilde{\nu}
 \tilde{\nu} + {\rm h.c.}$, is generated with
\begin{equation}
\delta m_{\tilde{\nu}}^{2} \sim
{ \langle H_{\alpha} \rangle^{2}
A^{2}\widehat{A}^{2} \over 16 \pi^2 m_{\tilde{f}}^{4} }
\label{msnu}
\end{equation}
and likewise for the non-holomorphic operator~(\ref{lopnonhol}).
Gauge invariance implies that this mixing is proportional to two
powers of both lepton-number violating and lepton-number conserving
tri-linear scalar terms.  The induced sneutrino--antisneutrino mixing
in turn gives rise to a neutrino mass at two loops
\begin{equation}
 m_{\nu} \sim
{\alpha_{2} m_{\widetilde{W}}
\delta m_{\tilde{\nu}}^{2} \over 4 \pi m_{\tilde{\nu}}^{2} } \,,
\label{mnu}
\end{equation}
\begin{figure}[ht]
\begin{center}
\epsfxsize= 7.5 cm
\leavevmode
\epsfbox[180 565 440 680]{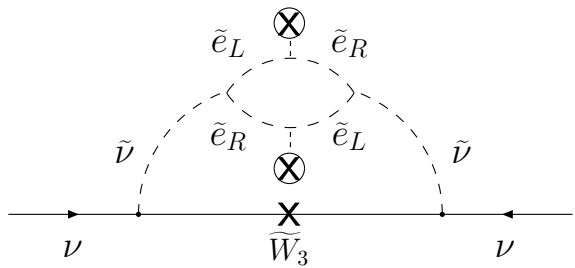}
\end{center}
\caption[f1]{Two-loop contribution to the neutrino mass from soft 
lepton-number violation.}
\label{neudiag}
\end{figure}
through diagrams such as in fig.~\ref{neudiag}.  Since the neutrinos
are left-handed, only the $\widetilde{W}_3$ chirality-violating mass
contributes to the radiatively induced neutrino mass in the pure
gaugino limit.  The magnitude of the tri-linear scalar terms for a
given neutrino mass is numerically, very roughly,
$A \widehat{A} / \widetilde{m}^{\,2} \sim (m_{\nu}/3~{\mbox{MeV}})^{1/2}
(\widetilde{m}/100~{\mbox{GeV}})^{1/2}$,
assuming 
$\widetilde{m} \simeq m_{\widetilde{W}} \simeq m_{\tilde{\nu}}$.  The
specific pattern of neutrino masses and mixings of course depends on
the flavor structure of both the lepton-number violating and
lepton-number conserving scalar tri-linear terms.  Notice that, unlike
quark and lepton masses, radiatively induced neutrino masses arise
from non-renormalizable operators in the low-energy theory, and
therefore vanish in the superpartner decoupling limit in which all
soft supersymmetry-breaking parameters are taken simultaneously large.

One generic consequence of this mechanism for soft radiative neutrino
masses is the relatively large ratio
$\delta m_{\tilde{\nu}}^{2}/m_{\tilde{\nu}}^{2} \sim (4\pi/\alpha_{2})
(m_{\nu}/m_{\widetilde{W}})$.
This is to be compared with a standard scenario in which neutrino
masses arise directly from tree-level superpotential couplings.  In
this case, a  sneutrino--antisneutrino mixing arises radiatively from the
neutrino mass (rather than the other way around) yielding
$\delta m_{\tilde{\nu}}^{2}/m_{\tilde{\nu}}^{2}\sim (\alpha_2/4 \pi) 
(m_{\nu} m_{\widetilde{W}} / m_{\tilde{\nu}}^2)$.
For a massive $\tau$-neutrino, the relatively large $\delta
m_{\tilde{\nu}}^2 / m_{\tilde{\nu}}^2$ associated with a soft neutrino
mass might be probed in
$\tilde{\nu}_{\tau}$--$\tilde{\bar{\nu}}_{\tau}$
oscillation experiments~\cite{OSC}.  Lastly, it is interesting to note
that the simple assumption of degeneracy among respective tri-linear
couplings corresponds to the case of large neutrino mixing angles.

%
%

\section{Summary and discussion}
\label{sec:summary}

Supersymmetric theories have the property that the chiral flavor
symmetries required for quark and lepton masses may be broken either
in hard renormalizable terms, or in soft supersymmetry-breaking terms.
This interesting feature arises because the squark and slepton
superpartners necessarily carry the same flavor symmetries as the
quarks and leptons.  In this paper we have investigated the
possibility that some of the fermion masses arise predominantly
radiatively from chiral flavor violation in soft
supersymmetry-breaking terms.  The breaking of such symmetries
exclusively in soft and not in hard terms is technically natural and
may be enforced by horizontal $R$-symmetries.  In this scenario some
of the flavor symmetries are broken either explicitly in the messenger
sector or spontaneously in the supersymmetry-breaking sector.  In the
latter case this amounts to an auxiliary field version of the
Froggatt--Nielsen mechanism.

Soft chiral flavor violation can in principle be the source of
radiative masses for all the first- and second-generation quarks and
leptons and the $b$-quark.  Since radiative masses are intrinsically
suppressed by a loop factor, the smallness of the first generation
masses can be due in part to this suppression.  The remaining
suppression in this radiative scenario can be obtained in part from a
hierarchy between the gaugino and scalar superpartner masses, or from
a hierarchy between the supersymmetry-breaking and flavor-breaking
scales.  The loop factor is sufficient to account for essentially all
of the suppression of second-generation masses with respect to the
electroweak scale.

Radiative fermion masses, especially for the second generation or
$b$-quark, require significant left--right scalar superpartner mixing
from scalar tri-linear terms.  Such large mixings introduce
potentially dangerous directions in field space, along which charge
and/or color are broken.  However, metastability of the charge- and
color-preserving vacuum on cosmological time scales is possible in
many models.  In addition, some classes of models with non-holomorphic
tri-linear terms or mixing with mirror matter at the
supersymmetry-breaking scale contain stabilizing terms in the
potential, which eliminate these dangerous directions and render the
charge- and color-preserving vacuum absolutely stable.

Since the dominant source of flavor violation for fermions with soft
radiative masses resides in supersymmetry-breaking terms, interesting
levels of low-energy flavor violation may arise.  These are functions
of the specific textures for the scalar masses and tri-linear terms
and therefore very model-dependent. The study of such flavor changing
associated with soft fermion masses is outside the scope of this work,
but should be considered in any future work on specific models of the
soft and hard textures.  In some classes of models certain relations
or symmetries among the soft terms can lead to partial or complete
alignment in flavor space of non-renormalizable chirality-violating
operators with the effective Yukawa couplings and reduce or
eliminate observable flavor breaking.  It is noteworthy that when all
the soft supersymmetry-violating parameters are taken simultaneously
large, radiative Yukawa couplings approach finite limits, whereas
flavor violation induced by virtual superpartners in
non-renormalizable operators is suppressed.  This is particularly
relevant in schemes with heavy first- and second-generation scalars.

Radiative fermion masses lead to a number of striking phenomenological
consequences.  Chief among these is that new contributions to
chirality-violating operators are effectively not suppressed by a loop
factor compared with the fermion mass, which itself arises at one
loop.  This is a generic feature of any theory of radiative fermion
masses.  The most important such operators are anomalous magnetic
moments for fermions with soft radiative masses.  In this scenario the
supersymmetric correction to the anomalous moment only depends on the
superpartner mass spectrum, and is necessarily positive.  The
anomalous magnetic moment of the muon is by far the most sensitive
probe of a radiative muon mass.  The current experimental bound
already constrains part of the supersymmetric parameter space in this
scenario.  Even if the muon mass is not predominantly radiative, a
contribution to the muon anomalous magnetic moment from relatively
large chiral flavor violation in the muon tri-linear term might
provide an interesting interpretation for a non-vanishing measurement
of $a_{\mu}^{\rm new}$.

CP-violating Higgs--fermion couplings
and electric dipole moments are naturally suppressed by the alignment
of the phases of radiatively generated chirally violating operators
with the phase of a radiatively generated mass.  This alignment is 
naturally
very precise in interesting regions of parameter space.  
Thus, the standard supersymmetric CP problem, is 
mitigated or eliminated in the scenario of radiative
fermion masses studied here.

Other phenomenological consequences of radiative masses are related to
the softness of the Yukawa couplings.  Most notably, the
Higgs--fermion couplings are momentum-dependent with non-trivial form
factors.  The lowest-order operator, which represents the momentum
dependence, may be characterized in terms of a finite Higgs Yukawa
radius.  This radius, like the anomalous magnetic moment, is not
suppressed by a loop factor.  Even at zero momentum the Higgs Yukawa
coupling can differ from the mass Yukawa coupling.  In addition,
``wrong Higgs'' couplings result if the radiative mass arises from
non-holomorphic soft terms.  All these Higgs--fermion coupling effects
may be tested at future colliders by looking for deviations of various
sum rules and relations among Higgs boson decay widths and branching
ratios to fermion final states.  Many of these sum rules and relations
are guaranteed by holomorphy to be satisfied at lowest order in any
supersymmetric theory with a single pair of Higgs doublets and hard
fermion masses.

A further signal of the softness of a fermion mass is the associated
radiatively generated Higgsino coupling.  Large differences between
the couplings of Higgs and Higgsinos to matter are expected in this
scenario.  To a low-energy observer who assumes hard Yukawa couplings,
these would manifest themselves as apparent hard violations of
supersymmetry in renormalizable interactions.  Unfortunately direct
measurement of such couplings at future colliders is likely to be very
difficult.  Finally, the relatively large left--right scalar
superpartner mixing associated with second-generation or $b$-quark
radiative masses can lead to significant modifications of
supersymmetric production cross sections and branching ratios from
minimal expectations.

The spontaneous breaking of chiral flavor symmetries in the
supersymmetry-breaking sector through auxiliary rather than scalar
directions can be extended to other global chiral symmetries such as
lepton number.  Soft lepton number violation gives rise, radiatively,
to lepton-violating sneutrino masses which in turn induce
 radiative neutrino masses.  Sneutrino--antisneutrino oscillations are
relatively large in this scenario and might be probed if the
${\tau}$-neutrino is not too light.

In summary, the interesting features of radiative fermion masses
arising from soft chiral flavor violation include:
\begin{itemize}
\item
Radiative masses for first- and second-generation quarks and leptons
and for the $b$-quark can be accommodated while satisfying 
(meta)stability requirements for the charge- and color-preserving 
vacuum.
\item
Higgs Yukawa couplings are momentum-dependent with non-trivial form
factors and finite Higgs Yukawa radii.
\item
``Wrong Higgs'' radiative Yukawa couplings arise from non-holomorphic
soft chiral flavor breaking.
\item
Apparent large hard violation of supersymmetry in Higgsino couplings.
\item
Supersymmetric contributions to anomalous magnetic moments are
positive and not suppressed by a loop factor.
A radiative muon mass will be very well probed by the Brookhaven
muon $g-2$ experiment.
\item  
CP-violating electric dipole moments are suppressed by natural phase
alignment.  Both the neutron and electron EDMs could be not too far
below current experimental bounds.
\item
Left--right scalar superpartner mixing is enhanced.
\item
Radiative neutrino masses arising from soft lepton number violation
imply enhanced sneutrino--antisneutrino oscillations.
\end{itemize}
Specific models for the hard and soft textures should be developed
to further explore this interesting possibility for the origin
of some of the fermion masses.

\acknowledgements
The authors thank A.~Djouadi, C.~Greub, H.~Haber, U.~Nierste, G.J.~van
Oldenborgh, and D.~Wyler for discussions, and M. Dress for comments 
on the manuscript. F.~B. and S.~T. acknowledge
the hospitality of the high-energy group at Rutgers University and
N.~P. that of the CERN theory group. F.~B., N.~P., and S.~T. are also
grateful to the Aspen Center for Physics, where parts of this work
were done. F.~B. and G.~F. respectively acknowledge the support of the
Schweizerischer Nationalfonds and Rutgers University throughout most
of this research. This work was supported by the US Department of
Energy under grant DE-FG02-96ER40559, the US National Science
Foundation under grants PHY--94--23002 and PHY--98--70115, the Alfred
P. Sloan Foundation, and Stanford University through a Fredrick
E. Terman Fellowship.


\newpage
\appendix

\section{Definitions and integrals}
\label{sec:def}

\subsection{Scalar superpartner mass and mixing}
\label{sec:sfermion}

The $2 \times 2$ squark or slepton mass squared matrix,
\begin{equation}
{\cal M}^2 =
\left( \begin{array}{cc}
   \, m^2_{LL} \,      & (m^2_{LR})    \\[1.01ex]
   (m^2_{LR})^\ast     &  m^2_{RR}
 \end{array} \right) \,,
\label{massmatr}
\end{equation}
written here in the basis $\{ \tilde{f}_L,\tilde{f}_R\}$, is Hermitian
${\cal M}^2 = ({\cal M}^2)^{\dagger}$, with eigenvalues
\begin{equation}
 m^{2}_{\tilde{f}_{1,2}} =
 \frac{1}{2} \left\{ \left(m^2_{LL} +m^2_{RR}\right) \mp
               \sqrt{\left(m^2_{LL} -m^2_{RR}\right)^2 + 4|m^2_{LR}|^2}
             \right\} \,.
\label{eigenvalues}
\end{equation}
Notice that $\tilde{f}_R$ is here the actual right component of
$\tilde{f}$ and corresponds to the conjugate of the field $\phi_R$
used in the text.

The eigenvectors $\tilde{f}_1$ and $\tilde{f}_2$ corresponding to the
eigenvalues in~(\ref{eigenvalues}) are obtained from $\tilde{f}_L$,
$\tilde{f}_R$ through a unitary transformation :
\begin{equation}
\left( \begin{array}{c} \tilde{f}_1 \\ \tilde{f}_2
       \end{array}  \right)
 =  U
\left( \begin{array}{c}  \tilde{f}_L \\ \tilde{f}_R
       \end{array}  \right)
\equiv
\left( \begin{array}{cc}
  \cos \theta_f  e^{+i\phi}  & \ \sin \theta_f \ \ \ \     \\
  \sin \theta_f \ \ \ \ \    & - \cos \theta_f e^{-i\phi}
       \end{array} \right)
\left( \begin{array}{c}  \tilde{f}_L \\ \tilde{f}_R
       \end{array}  \right) \,,
\label{urotation}
\end{equation}
where $\phi \equiv {\rm Arg}(m_{LR}^2)
              = - {\rm Arg}(A \langle H_\alpha \rangle)$
or $ \phi = - {\rm Arg}(A^\prime \langle H_\alpha^\ast \rangle )$.
The mixing angle $\theta_f $ is defined, up to a two-fold ambiguity by
the relations
\begin{equation}
 \sin 2 \theta_f =
  -\frac{ 2 |m^2_{LR}|}{m^2_{\tilde{f}_2} -m^{2}_{\tilde{f}_1}} \,;
\hspace*{1.5truecm}
 \cos 2 \theta_f =
 \pm \frac{m^2_{LL} - m^2_{RR}}{m^2_{\tilde{f}_2} -m^2_{\tilde{f}_1}} \,.
\end{equation}
Note that in the strict superpartner decoupling limit, for which
$m_{LR}^2/m_{LL,RR}^2 \to 0$, the mixing angles approach 
$\cos \theta_f \to 0 $ and $\sin \theta_f \to 1$.

The fractional splitting of the mass squared eigenvalues used in the
text is related to the mass squared matrix by
\begin{equation}
\phi_f \equiv
 \frac{m^{2}_{\tilde{f}_2} -m^{2}_{\tilde{f}_1}}
      {m^{2}_{\tilde{f}_2} +m^{2}_{\tilde{f}_1}}
 =
 {1 \over {\rm Tr} {\cal M}^2 }~
\sqrt{ ({\rm Tr}{\cal M}^2)^2 - 4 ~{\rm Det}{\cal M}^2 }\,.
\end{equation}
Note that $\phi_f \to 1^-$ for ${\rm Det}{\cal M}^2 \to 0^+$,
corresponding to a vanishing eigenvalue in the scalar mass matrix.


\subsection{Fermion mass integral}
\label{sec:defI}

The two-point function $I(m_1^{2},m_2^{2},m_{\lambda}^2)$, introduced
in section~\ref{sec:mass}, is a completely symmetric function in its
three variables
\begin{equation}
 I(m_1^2,m_2^2,m_\lambda^2)  =   - \frac{
 m_1^2 \,m_2^2 {\rm ln} ( m_1^2/ m_2^2 ) +
 m_2^2 \,m_\lambda^2 {\rm ln} ( m_2^2/ m_\lambda^2 )  +
 m_\lambda^2 \,m_1^2 {\rm ln} ( m_\lambda^2/ m_1^2 )
                                         }
{(m_1^2 -m_2^2)(m_2^2 -m_\lambda^2)(m_\lambda^2 -m_1^2)
          }\,.
\label{copwfunc}
\end{equation}
It may be cast in the simpler form:
\begin{equation}
 I(m_1^2,m_2^2,m_\lambda^2) \ = \ \frac{1}{m_1^2-m_2^2} \left(
\frac{\ln \beta_1}{\beta_1-1} - \frac{\ln \beta_2}{\beta_2-1}
                                                       \right)\,,
\label{ibeta}
\end{equation}
where $\beta_i = m_\lambda^2 / m_i^2$, $i=1,2$, in which, however,
only the symmetry between two of the three variables is manifest. This
form is more useful to obtain some interesting limiting cases.  In the
limit of degenerate scalar-partner masses the loop
function~(\ref{ibeta}) becomes
\begin{equation}
 I(m^2,m^2,m_\lambda^2) \ = \  \frac{1}{m^2} \,
 \frac{1}{(\beta -1)^2} \left(1- \beta + \beta \ln \beta \right) \,,
\label{ione}
\end{equation}
whereas in the limit of highly non-degenerate eigenvalues, 
$m = m_2 \gg m_1 $,
\begin{equation}
 I(0,m^2,m_\lambda^2) \ = \  \frac{1}{m^2} \,
  \frac{ \ln \beta}{\beta - 1} \,.
\label{izero}
\end{equation}
The limiting behavior of these functions for either fixed gaugino mass
or fixed scalar mass is shown in table~\ref{ilim}. The loop function
is typically bounded by
$I(m_1^2, m_2^2,m^2_{\lambda}) \times \max(m_1^2,m_2^2,m_{\lambda}^2)
\lesssim {\cal O}(1)$,
but this can be enhanced in certain limits by a logarithm.
Large scalar splittings also slightly enhance the loop integral.
\begin{table}
\caption{Limiting behavior of
 $ I(m^2,m^2,m_\lambda^2)$ and  $ I(0,  m^2,m_\lambda^2)$  for
 either
 fixed gaugino mass $m_\lambda = {\overline{m_\lambda}}$ or
 fixed scalar mass $m = {\overline{m}}$.
 The limit in the first column reflects the infrared behavior of
 the loop integral for $m$ or $m_{\lambda} \to 0$.}
\label{ilim}
\begin{tabular}{llll}
   $m_{\lambda} = \overline{m_{\lambda}}$ fixed         &
 $ m \to  0 $                             &
 $ m \to {\overline{m_\lambda}}$          &
 $ m \to \infty $                   \\
\hline
 $ \quad I(m^2,m^2,{\overline{m_\lambda}}^2)$         &
 $   \frac{1}{{\overline{m_\lambda}}^2}
 \left( -\ln (\frac{m^2}{{\overline{m_\lambda}}^2}) \right)
      \to  +\infty $                                  &
 $  \frac{1}{{\overline{m_\lambda}}^2} \cdot
   \frac{1}{2} $                                      &
 $  \frac{1}{m^2}  \to 0 \quad $                   \\[1.2ex]
 $ \quad I(0,  m^2,{\overline{m_\lambda}}^2)$         &
 $  \frac{1}{{\overline{m_\lambda}}^2}
   \left( -\ln (\frac{m^2}{{\overline{m_\lambda}}^2}) \right)
     \to  +\infty   $                                 &
 $ \frac{1}{{\overline{m_\lambda}}^2}$                &
 $ \frac{1}{m^2}
    \left( \ln (\frac{m^2}{{\overline{m_\lambda}}^2}) \right)
    \to  0 \quad $                                  \\[1.3ex]
\hline
 & & & \\
 \hline
  $m = \overline{m}$ fixed                      &
 $ m_\lambda \to 0 $                             &
 $ m_\lambda \to {\overline{m}}$                 &
 $ m_\lambda \to\infty $                         \\
\hline
 $ \quad I(\overline{m}^2,\overline{m}^2,m_\lambda^2)$ &
 $ \frac{1}{\overline{m}^2} $                          &
                                                       &
 $ \frac{1}{{m_\lambda}^2}
   \left( \ln (\frac{m_\lambda^2}{{\overline{m}}^2})
   \right) \to 0 $                                    \\[1.2ex]
 $ \quad I(0,\overline{m}^2,m_\lambda^2) $             &
 $ \frac{1}{\overline{m}^2}
   \left(- \ln (\frac{m_\lambda^2}{{\overline{m}}^2})
   \right) \to +\infty $                               &
                                                       &
 $ \frac{1}{{m_\lambda}^2}
   \left( \ln (\frac{m_\lambda^2}{{\overline{m}}^2}) \right)
   \to 0 $
\end{tabular}
\end{table}
%


\subsection{Higgs vertex function}
\label{defC}

Except for a sign difference, the convention of the first reference
in~\cite{CFUNC} is followed in the definition of the three-point
function employed here:
$C_0(p_1^2,p_2^2,2p_1\! \cdot\! p_2; m_1^2,m_\lambda^2,m_2^2)$.
Explicitly,
\begin{eqnarray}
 \lefteqn{
 C_0(p_1^2,p_2^2, 2p_1\! \cdot\! p_2; m_1^2, m_\lambda^2, m_2^2)
         }           \nonumber \\[1.2ex]
  &  &
\quad \quad \quad
\ \equiv \
 \frac{i}{\pi^2}  \int d^4 k
 \frac{1}{ \bigl[       k^2 - m_1^2       + i\epsilon \bigr]
           \bigl[ (k+p_1)^2 - m_\lambda^2 + i\epsilon \bigr]
           \bigl[ (k+p_1+p_2)^2 - m_2^2   + i\epsilon \bigr]
         }           \nonumber \\[1.2ex]
  &  &
\quad \quad \quad
  \ = \ \quad
  \int_0^1 d x \int_0^x  d y
 \frac{1}{\left[ ax^2 + b y^2 + c x y + d x + e y + f \right]} \,,
\label{c0general}
\end{eqnarray}
where the $a$, $b$, $c$, $d$, $e$, and $f$ are
\begin{equation}
 a \!=\! p_2^2 \,;    \
 b \!=\! p_1^2 \,;    \
 c \!=\! 2 p_1 \!\cdot\! p_2 \,;  \
 d \!=\!-p_2^2 \!+\!m_\lambda^2 \!-\!m_2^2 \,;  \
 e \!=\!-p_1^2 \!-\!2p_1\!\cdot\! p_2 \!+\!m_1^2 \!-\!m_\lambda^2 \,; \
 f \!=\! m_2^2 -i\epsilon \,.
\label{abcsymbols}
\end{equation}
The general solution is:
\begin{eqnarray}
 \lefteqn{
 C_0(p_1^2,p_2^2, 2p_1\! \cdot\! p_2; m_1^2, m_\lambda^2, m_2^2)
         }            \nonumber \\[1.2ex]
  &  &
\quad \quad \quad
\ = \
 \frac{1}{( c + 2 \alpha b)}
 \sum_{j=1}^3 \, (-1)^{j+1} \sum_{k=1,2}
  \left[ Li_2\left(\frac{y_j}{y_j-y_{jk}}\right) -
         Li_2\left(\frac{y_j-1}{y_j-y_{jk}}\right)
  \right] \,,
\label{generalsol}
\end{eqnarray}
where the $y_j$ are
\begin{eqnarray}
 y_1 = - \frac{(d +e \alpha +2 a +c \alpha)}{c +2 \alpha b}\,,
 \quad
 y_2 = - \frac{(d +e \alpha)}{c +2 \alpha b} \
     \frac{1}{1\! -\!\alpha}\,,
 \quad
 y_3 =   \frac{(d +e \alpha)}{c +2 \alpha b} \
     \frac{1}{\alpha} \,,
\label{yjsym}
\end{eqnarray}
and $\alpha$ is one of the two solutions of the equation 
$\alpha^2 b+\alpha c +a = 0 $.  Finally, for each $j=1,2,3$, the
quantities $y_{j1}$ and $y_{j2}$ are the roots of the quadratic
equations
\begin{eqnarray}
 Q_1(y) & = &
 p_1^2 y^2      +(m_1^2 -m_\lambda^2 -p_1^2)y + m_\lambda^2 -i\epsilon
      \nonumber \\
 Q_2(y) & = &
(p_1+p_2)^2 y^2 +(m_1^2 -m_2^2 -(p_1+p_2)^2)y + m_2^2       -i\epsilon
      \nonumber \\
 Q_3(y) & = &
 p_2^2 y^2      +(m_\lambda^2 -m_2^2 -p_2^2) y + m_2^2      -i\epsilon \,.
\label{qjsym}
\end{eqnarray}

In the case of the decay $H \to f {\bar f}$, or of the 
resonant production $f {\bar f} \to H$, the momenta are such that 
$p=q$, $p_1= q_1$, $q_2 =-p_2$. All external fields are on-shell, and
the approximation $p_1^2= p_2^2 = 0 $ and 
$2 p_1\!\cdot \! p_2 =m_H^2$ is appropriate for $m_f^2 \ll m_H^2$.
Both $\alpha$'s also vanish in this limit ($a=b=0$).  An independent
integration then yields
\begin{eqnarray}
 \lefteqn{
 C_0(0,0,m_H^2;m_1^2, m_{\lambda}^2, m_2^2)
         }                  \nonumber \\[1.2ex]
  &  &
\quad
\ = \
 \frac{1}{m_H^2}          \left\{
  \sum_{i=1,2}  \left[
    Li_2 \left(\frac{x_0}{x_0\!-\!x_i}\right)
 \!-\!  Li_2 \left(\frac{x_0\!-\!1}{x_0\!-\!x_i}\right) \right]
   - \left[
    Li_2 \left(\frac{x_0}{x_0\!+\!f/d}\right)
 \!-\!  Li_2 \left(\frac{x_0\!-\!1}{x_0\!+\!f/d}\right)
   \right]
                         \right\} \,,             \nonumber \\[1.2ex]
  &  &
\label{c0mh}
\end{eqnarray}
where $x_0 $ is the root of the linear equation,
$ cx+e=0$ and $x_1, x_2$ the roots of the quadratic equation
$ c x^2 + (d+e) x + f = 0$,
\begin{equation}
x_{1,2}\ = \ \frac{m_2^2 -m_1^2 +m_H^2}{2 m_H^2}
\mp  \left[
 \left(\frac{m_2^2 - m_1^2 + m_H^2 }{2 m_H^2} \right)^2  \!\! -
 \left(\frac{m_2^2}{ m_H^2}\right)^2 \!\! + i \epsilon
    \right]^{1/2}
  \label{rooteq}
\end{equation}
and $d$ and $f$ follow from~(\ref{abcsymbols}) in this limit.

Notice that the arguments of logarithms and dilogarithms
in~(\ref{generalsol}) and~(\ref{c0mh}) are in general complex.
Moreover, depending on the relative size of the masses involved, some
arguments of $Li_2(x) $ in~(\ref{generalsol}) and~(\ref{c0mh}) may be
such that ${\rm Re}(x)>1$. Since $Li_2(x) $ has a cut on the real
axis, starting from 1, the relations
\begin{eqnarray}
 Li_2(x) & = & - Li_2(1-x) + \frac{1}{6} \pi^2 - \ln(x)\ln(1-x)
  \nonumber  \\
 Li_2(x) & = & - Li_2\left(\frac{1}{x} \right)
         - \frac{1}{6} \pi^2 - \frac{1}{2} \ln^2 (-x)
  \label{Lieq}
\end{eqnarray}
may have to be used.

With general superpartner masses, the limit of vanishing Higgs mass,
$m_H \to 0 $, reduces~(\ref{c0mh}) to the mass loop function
\begin{equation}
 C_0 (0,0,0;m_1^2,m_{\lambda}^2,m_2^2) =
I(m_1^2,m_2^2,m_{\lambda}^2) \,.
\end{equation}

If $m_H \ne 0$, the two limits of degenerate scalar masses, 
$m_1^2 =m_2^2 = m^2$, and of maximally split ones, 
$m_1^2 = 0, m_2^2 = m^2$,
explicitly analysed for the mass loop function, do not yield
considerable simplifications for
$C_0 (0,0,m_H^2 ; m^2,m_{\lambda}^2,m^2)$ and
$C_0 (0,0,m_H^2 ;   0,m_{\lambda}^2,m^2)$ with respect
to~(\ref{c0mh}). For fixed $m$ scalar mass $m= \overline{m}$,
$C_0 (0,0,m_H^2 ; \overline{m}^2,{m_\lambda}^2,\overline{m}^2)$
and
$C_0 (0,0,m_H^2 ; 0,{m_\lambda}^2,{\overline m}^2)$
acquire simple analytic expressions in
the limit $m_\lambda \to \overline{m}$.
In both cases, $a=b=d=0$. In addition, in the first case,
it is $e=-c$
and the three-point function is then
\begin{equation}
 C_0(0,0,m_H^2; \overline{m}^2, \overline{m}^2, \overline{m}^2) =
   \frac{1}{m_H^2}
 \left\{ Li_2\left(\frac{1}{x_2}\right) +
         Li_2\left(\frac{1}{x_1}\right)  \right\}\,.
\label{C01}
\end{equation}
The roots $x_{1,2}$ are in this limit
\begin{equation}
 x_{1,2}\ = \
 \frac{1}{2} \left[ 1 \mp  \sqrt{1-4 \frac{\overline{m}^2}{m_H^2}
       \! +\! i \epsilon \,}\ \, \right] \,.
  \label{rooteq1}
\end{equation}
Using~(\ref{Lieq}), $x_1+x_2=1$, and analytic continuation
\begin{equation}
 C_0(0,0,m_H^2; \overline{m}^2,\overline{m}^2,\overline{m}^2) =
  \frac{1}{2 m_H^2}
\left\{ \begin{array}{ll}
\displaystyle
  -  \left[{\rm ln}
 \left(\frac{1+\sqrt{1\!-\!4 \overline{m}^2/ m_H^2}}
            {1-\sqrt{1\!-\!4 \overline{m}^2/ m_H^2}}\right)  - i\pi
   \right]^2      &
\displaystyle  \quad  2 \frac{\overline{m}}{m_H}  < 1\,,  \\[1.9ex]
\displaystyle
+\ 4 \,\arcsin^2\left(\frac{m_H}{2 \overline{m}}\right)
                  &
\displaystyle  \quad  2 \frac{\overline{m}}{m_H} \ge 1\,.
\end{array} \right. 
\label{I1}
\end{equation}
The imaginary piece for $2 m/ m_H < 1$
corresponds to the cut for physical intermediate states.

In the case of maximal splitting, it is
\begin{eqnarray}
 \lefteqn{
  C_0 ( 0,0,m_H^2 ; 0,{\overline m}^2,{\overline m}^2)
         }                  \nonumber \\[1.2ex]
  &  &
\quad
\ = \
  \frac{1}{m_H^2}
 \left\{  Li_2\left(\frac{1+\overline{m}^2/ m_H^2}{x_2}\right) +
       \! Li_2\left(\frac{1+\overline{m}^2/ m_H^2}{x_1}\right) -
       \! Li_2\left(\frac{  \overline{m}^2/ m_H^2}{x_2}\right) -
       \! Li_2\left(\frac{  \overline{m}^2/ m_H^2}{x_1}\right)
 \right\}\,,                   \nonumber \\[1.01ex]
  &  &
\label{C02}
\end{eqnarray}
with $x_{1,2}$ given by:
\begin{equation}
 x_{1,2} \ = \ \frac{1}{2}
 \left[      \left(1+\frac{\overline{m}^2}{m_H^2}\right)
 \mp  \sqrt{ \left(1-\frac{\overline{m}^2}{m_H^2}\right)^2
 \!+\! i \epsilon\, } \ \,
 \right]\,.
  \label{rooteq2}
\end{equation}
Using again~(\ref{Lieq}) and the fact that
$x_1 +x_2 =1 +{\overline{m}^2}/m_H^2$, it can be shown
that
\begin{equation}
  C_0 ( 0,0,m_H^2 ; 0,{\overline m}^2,{\overline m}^2) =
  \frac{1}{m_H^2}
\left\{ \begin{array}{ll}
\displaystyle
 - \frac{1}{2}
  \left( {\rm ln} \frac{\overline{m}^2}{m_H^2} +i \pi \right)^2
 -\frac{\pi^2}{6}
 - Li_2 \left(\frac{\overline{m}^2}{m_H^2}\right)
         &
\displaystyle \quad \frac{\overline{m}}{m_H}  < 1\,,  \\[1.9ex]
\displaystyle
 \ + \  Li_2 \left(\frac {m_H^2}{\overline{m}^2} \right)
         &
\displaystyle \quad \frac{\overline{m}}{m_H} \ge 1 \,,
\end{array} \right.
\label{I2}
\end{equation}
where $Li_2 (x)$ is a real number for $x <1$.



\subsection{Higgsino vertex function}
\label{defCino}

The vertex function $V_{ijlkh}$ appearing in the Higgsino
coupling relevant to the decays
$ \widetilde{f_h}(q_2) \to {\widetilde \chi^0}_i(q) \, f_{L,R}(q_1)$
is given by
\begin{eqnarray}
V_{ijlkh} & = &
\left[
  m_{{\widetilde \chi}_j^0} - m_{{\widetilde \chi}_i^0}
 \left(
  \frac{ m_{{\widetilde f}_k}^2 \!- \!m_{{\widetilde \chi}_j^0}^2 }
       { m_{{\widetilde f}_h}^2 \!- \!m_{{\widetilde \chi}_i^0}^2 }
 \right) \right]
 C_0(q^2,0,q_2^2; m_{H_l^0}^2, m_{{\widetilde \chi}_j^0}^2,
                               m_{{\widetilde f}_k}^2)    \nonumber \\
     &  & \hspace*{1.8truecm} + \
\left(
 \frac{ m_{{\widetilde \chi}_i^0}}
       { m_{{\widetilde f}_h}^2 \!- \!m_{{\widetilde \chi}_i^0}^2 }
\right)
 \left[ B_0(q^2;   m_{{\widetilde \chi}_j^0}^2, m_{H_l}^2)
     -  B_0(q_2^2; m_{H_l}^2, m_{{\widetilde f}_k}^2)
 \right] \,.
\label{higgsinovert}
\end{eqnarray}
The definition of the two-point function $B_0(q^2;m_1^2,m_2^2)$
follows standard
conventions, as for example in~\cite{CFUNC}, except for a minus
sign
\begin{equation}
 B_0(q^2;m_1^2,m_2^2) =
 \left(\mu^2 \pi e^{\gamma}\right)^{(4-n)/2}
 \frac{i}{\pi^2} \int d^n k
 \frac{1}{\bigl[k^2-m_1^2+i\epsilon\bigr]
          \bigl[(k+q)^2-m_2^2+i\epsilon\bigr]}
\end{equation}
where $\gamma $ is Euler's constant.

\newpage


\section{Anomalous magnetic and electric dipole moments}
\label{sec:muon}

The electromagnetic dipole operator for a Dirac fermion is
given by the Lagrangian operator
\begin{equation}
{\cal L} \supset
-{1 \over 2} d_f ~ \overline{f}_L \sigma^{\mu \nu} f_R F_{\mu \nu}
 ~+~{\rm h.c.}
\end{equation}
where $d_f$ is the dipole moment coefficient, which is, in general
complex and $f_{L,R} = P_{L,R} f$ are the left- and right-handed
chiral components of the Dirac fermion.  In a general basis with
complex fermion mass $m_f$, the electric dipole moment is
\begin{equation}
d_f^e = |d_f| \sin  \left( {\rm Arg}( d_f m_f^*) \right)
\label{dfsin}
\end{equation}
and the anomalous magnetic moment is
\begin{equation}
a_f =   {2 m_f \over e Q_f} |d_f |
   \cos  \left( {\rm Arg}( d_f m_f^*) \right)\,,
\label{amusin}
\end{equation}
where $Q_f$ is the fermion electric charge.

In the CP-conserving case, ${\rm Arg}(d_f m_f^*)=0$, the one-loop
chirality-violating contribution of fig.~\ref{gmtwoadiag} to the muon
anomalous magnetic moment is given by
\begin{equation}
{\cal L} \supset
  + \frac{\alpha^\prime}{4\pi} \,(e Q_{\mu}) \, \
   {m^2_{LR}}
\sum_{j}K_f^{j} \, m_{\tilde{\chi}_{j}^{0}} \,
  I_{g-2} (m^{2}_{\tilde{\mu}_{1}}, m^{2}_{\tilde{\mu}_{2}},
           m^2_{\tilde{\chi}_{j}^{0}} )
  \, {\cal O}_{g-2} \,,
\label{rawgmtwo}
\end{equation}
where $Q_\mu = -1$ is the muon electric charge,
the operator
$  {\cal {O}}_{g-2} \equiv
  {\bar \mu}(p^\prime) \,  \sigma^{\mu \nu} F_{\mu \nu}
        \mu(p) $,
and the loop function is
\begin{equation}
 I_{g-2} ( m^2_1, m^2_2 m^2_\lambda )   =
  \frac{1}{m^2_\lambda} \, \frac{1}{m^2_2 \!-\! m^2_1}
 \left\{
\frac{\beta_1\left(\beta_1^2 -1 -2 \beta_1 \ln \beta_1\right) }
     { 2 \left( \beta_1 -1 \right) ^3}
  - \left(1 \to 2\right)
 \right\}\,,
\label{igmtwofun}
\end{equation}
with $\beta_i = m_\lambda^2 / m_i^2$, $i=1,2$.
\begin{table}
\caption{
Limiting behavior of $ I_{g-2}(m^2,m^2,{\overline{m_\lambda}}^2)$  and
 $ I_{g-2}(0,  m^2,{\overline{m_\lambda}}^2)$
 for either
 fixed gaugino mass $m_\lambda = {\overline{m_\lambda}}$, or
 fixed scalar mass $m = {\overline{m}}$.
 The limit in the first column reflects the infrared behavior of
 the loop integral for $m$ or $m_\lambda \to 0$.}
\label{igmtwolim}
\begin{tabular}{llll}
   $m_{\lambda} = \overline{m_\lambda}$ fixed                         &
 $ m \to  0 $                                                   &
 $ m \to {\overline{m_\lambda}}$                                      &
 $ m \to \infty $                                              \\
\hline
 $\quad I_{g-2\,}(m^2,m^2,{\overline{m_\lambda}}^2 )$                 &
 $ \frac{1}{({\overline{ m_\lambda}}^2 )^2 }
   \left( -\ln (\frac{m^2}{ { \overline {m_\lambda}}^2}   )
   \right)  \to  + \infty  $                                    &
 $ \frac{1}{({\overline{m_\lambda}}^2)^2}\cdot \frac{1}{12} $         &
 $ \frac{1}{(m^2)^2}\cdot \frac{1}{2} \to  0 \quad $           \\[1.2ex]
 $ \quad I_{g-2\,}(0,m^2,{\overline { m_\lambda } }^2 )$              &
 $ \frac{1}{({\overline{m_\lambda}}^2)^2 }
    \left( - \ln (\frac{m^2} {\overline{m_\lambda }^2})
    \right) \to  + \infty  $                                    &
 $\frac{1}{({\overline{m_\lambda}}^2)^2} \cdot \frac{1}{3} $          &
 $\frac{1}{m^2} \frac{1}{{\overline{m_\lambda}}^2}
  \cdot \frac{1}{2} \to  0  \quad  $                           \\[1.3ex]
\hline
 & & & \\
\hline
   $m = \overline{m}$ fixed                    &
 $ m_\lambda \to 0 $                                                  &
 $ m_\lambda \to {\overline{m}}$                                      &
 $ m_\lambda \to\infty $                                             \\
\hline
 $ \quad I_{g-2\,}(\overline{m}^2,\overline{m}^2,m_\lambda^2) $       &
  $ \frac{1}{(\overline{m}^2)^2} \cdot \frac{1}{2}$             &
                                                                &
 $ \frac{1}{({m_\lambda}^2)^2}
   \left( \ln (\frac{m_\lambda^2}{{\overline{m}}^2}) \right)
   \to  0  \quad $                                             \\[1.2ex]
 $ \quad I_{g-2\,}(0,\overline{m}^2,m_\lambda^2) $                    &
 $ \frac{1}{\overline{m}^2} \frac{1}{m_\lambda^2}
   \cdot \frac{1}{2}  \to + \infty $                            &
                                                                &
 $ \frac{1}{({m_\lambda}^2)^2}
   \left( \ln (\frac{m_\lambda^2}{{\overline{m}}^2}) \right)
   \to  0 \quad $
\end{tabular}
\end{table}
For comparison with the mass loop function,
the limiting behavior of~(\ref{igmtwofun}) for
degenerate scalar partners masses is
\begin{equation}
 I_{g-2\,}(m^2,m^2,m^2_\lambda) \ = \
  \frac{1}{(m^2)^2} \,
  \frac{1}{2 (\beta - 1)^4}
  \left(2 \beta (\beta + 2) \ln \beta
        -5 \beta^2 + 4 \beta + 1 \right) \,,
\label{igm2one}
\end{equation}
while for large scalar splitting
\begin{equation}
 I_{g-2\,}(0,m^2,m^2_\lambda ) \ = \
  \frac{1}{m^2} \frac{1}{m_\lambda^2} \,
  \frac{1}{2 (\beta - 1)^3}
  \left(2 \beta^2 \ln \beta
        -3 \beta^2 + 4 \beta - 1 \right) \,.
\label{igm2zero}
\end{equation}
The limiting behavior of these functions for either fixed gaugino mass
or fixed scalar mass is shown in table~(\ref{igmtwolim}).  Using the
expression for the radiatively generated mass~(\ref{mass}), the
coupling~(\ref{rawgmtwo}) may be written
\begin{equation}
{\cal L} \supset
  - \frac{e\,Q_\mu}{2}  \,
  m_\mu \,
 \frac{\quad \sum_{j}\, K_f^{j} \, m_{\tilde{\chi}_{j}^{0}} \,
  I_{g-2} ( m^{2}_{\tilde{\mu}_{1}}, m^{2}_{\tilde{\mu}_{2}},
            m^2_{\tilde{\chi}_{j}^{0}} )}
      {\sum_{j}\, K_f^{j} \, m_{\tilde{\chi}_{j}^{0}} \,
  I ( m^{2}_{\tilde{\mu}_{1}}, m^{2}_{\tilde{\mu}_{2}},
      m^2_{\tilde{\chi}_{j}^{0}} ) }
  \, {\cal O}_{g-2}
 \equiv - a_{\mu} ~ {e Q_{\mu} \over 4 m_{\mu} }  {\cal O}_{g-2}\,.
\end{equation}
The supersymmetric contribution to the muon anomalous magnetic moment,
$ a_{\mu}^{\rm SUSY}$, is then explicitly given in eq.~(\ref{mammform}).
For a quark, the dominant gluino contribution to the dipole moment
is given by the Lagrangian operator
\begin{equation}
{\cal L} \supset
   \frac{2 \alpha_s}{3\pi} \,(e Q_{q}) \, \
   {m^2_{LR}} ~
 m_{\tilde{g}} \,
  I_{g-2} (m^{2}_{\tilde{q}_{1}}, m^{2}_{\tilde{q}_{2}},
           m^2_{\tilde{g}} )
  \, {\cal O}_{g-2}
\label{gluedip}
\end{equation}

%
%
\begin{figure}[ht]
\begin{center}
\epsfxsize=9 cm
\leavevmode
\epsfbox[155 565 485 680]{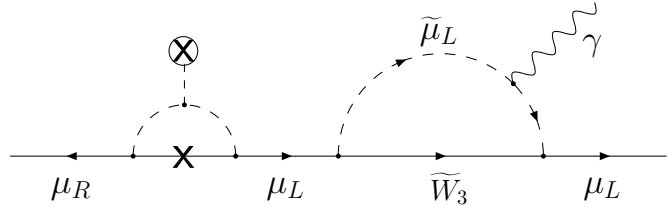}
\end{center}
\caption[f1]{Two-loop one-particle-reducible contribution to the
 muon anomalous magnetic moment.  The chirality-violating loop
 corresponds to an external mass insertion through the on-shell
 equation of motion.}
\label{gmtwocdiag}
\end{figure}
%

%
%
\begin{figure}[h]
\begin{center}
\epsfxsize=6.5 cm
\leavevmode
\epsfbox[200 562 425 680]{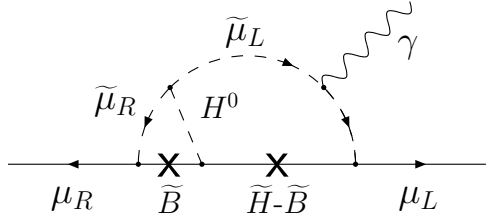}
\end{center}
\caption[f1]{Two-loop one-particle-irreducible contribution to the
 muon anomalous magnetic moment. In the heavy Higgs limit
 the one-loop sub-diagram on the left corresponds to the radiative
 Higgsino--scalar--fermion coupling of fig.~\ref{higgsinodiag}.
 There are additional contributions from chirality conserving
 gaugino propagators.}
\label{gmtwobdiag}
\end{figure}
Before concluding, it is worth noting that the result~(\ref{rawgmtwo})
obtained from the chirality-violating gaugino propagator in
fig.~\ref{gmtwoadiag} is the leading contribution to the anomalous
magnetic moment.  In models with hard Yukawa couplings, supersymmetric
one-loop contributions in general arise also from the
chirality-conserving gaugino propagator.  The fermion on-shell
equation of motion produces an external mass insertion, thereby giving
a contribution to the chirality-violating anomalous magnetic moment.
With a radiatively generated fermion mass, such contributions are
formally two-loop one-particle-reducible since the external mass
insertion is one-loop, as shown in fig.~\ref{gmtwocdiag}.  These are
of the same order as chirality-violating two-loop
one-particle-irreducible contributions, such as those arising from the
effective one-loop Yukawa coupling, as shown in fig.~\ref{gmtwobdiag}.
The result~(\ref{rawgmtwo}) therefore represents the full one-loop
supersymmetric contribution to the anomalous magnetic moment of a
fermion with a soft radiative mass.

\newpage


\newpage

\begin{thebibliography}{100}

\bibitem{FN}
 C. Froggatt and H. B. Nielsen, Nucl. Phys. {\bf B147} (1979) 277.


\bibitem{HORIZONTAL}
 M. Dine, R. Leigh, and A. Kagan, Phys. Rev. {\bf D48} (1990) 4269;   \\
 Y. Nir and N. Seiberg, Phys. Lett. {\bf B309} (1993) 337;            \\
 P. Pouliot and N. Seiberg, Phys. Lett. {\bf B318} (1993) 169;        \\
 M. Leurer, Y. Nir, and N. Seiberg, Nucl. Phys. {\bf B420} (1994) 468;\\
 D. Kaplan and M. Schmaltz, Phys. Rev. {\bf D49} (1994) 3741;         \\
 L. J. Hall and H. Murayama, Phys. Rev. Lett. {\bf 75} (1995) 3985.


\bibitem{ANOMALOUS}
 For a review of anomalous Abelian horizontal symmetries, see
  P. Ramond, hep-ph/9604251 and references therein.


\bibitem{MSSMCAL}
 L.J. Hall, R. Rattazzi, and U. Sarid, Phys. Rev. {\bf D50} (1994) 7048;\\
 R. Hempfling, Phys. Rev. {\bf D49} (1994) 6168; \\
 M. Carena, M. Olechowski, S. Pokorski, and C.E.M. Wagner,
       Nucl. Phys. {\bf B426} (1994) 269; \\
 B. Wright, hep-ph/9404217; \\
 A. Donini, Nucl. Phys. {\bf B467} (1996) 3; \\
 D.M. Pierce, J.A. Bagger, K. Matchev, R.-J. Zhang,
             Nucl. Phys. {\bf B491} (1997) 3.


\bibitem{MSSMUSE}
 F.M. Borzumati, M. Olechowski, and S. Pokorski,
             Phys. Lett. {\bf B349} (1995) 311; \\
 T. Blazek, S. Raby, and S. Pokorski,
             Phys. Rev. {\bf D52} (1995) 4151; \\
 N.  Polonsky, Phys. Rev. {\bf D54} (1996) 4537; \\
 R. Rattazzi and U. Sarid, hep-ph/9612464;       \\
 F.M. Borzumati, hep-ph/9702307.


\bibitem{NTWO}
 F. del Aguila, M. Dugan, B. Grinstein, L. Hall, G.G. Ross, and P. West,
   Nucl. Phys. {\bf B250} (1985) 225.


\bibitem{BANKS}
 T. Banks, Nucl. Phys. {\bf B303} (1988) 172.


\bibitem{MA}
 E. Ma, Phys. Rev. {\bf D39} (1989) 1922.


\bibitem{KRASNIKOV}
 N.V. Krasnikov, Phys. Lett. {\bf B302} (1993) 59.


\bibitem{AHCH}
 N. Arkani--Hamed, C.--H. Cheng, and L.J. Hall, Phys. Rev.
  {\bf D54} (1996) 2242 and Nucl. Phys. {\bf B472} (1996) 95.


\bibitem{TALKS}
 F.~Borzumati, G.~R.~Farrar, N.~Polonsky and S.~Thomas,
  hep-ph/9712428; hep-ph/9805314.


\bibitem{HR}
 L.J. Hall and L. Randall, Phys. Rev. Lett. {\bf 65} (1990) 2939.


\bibitem{CARENA}
 M. Carena {\it et al.} in~\cite{MSSMCAL}.


\bibitem{DUTCH}
 G.J.~van Oldenborgh and J.A.M.~Vermaseren,  Z. Phys. {\bf C46} (1990) 425;\\
 G.J. van Oldenborgh,  Comput. Phys. Commun. {\bf 66} (1991) 1.


\bibitem{FK}
 H. Fusaoka and Y. Koide, Phys. Rev. {\bf D57} (1998) 3986.


\bibitem{ABEL}
 S.A. Abel and J.M. Fr\`ere, Phys. Rev. {\bf D55} (1997) 1623.


\bibitem{GWCONDITION}
 S. Glashow and S. Weinberg, Phys. Rev. {\bf D15} (1977) 1958.


\bibitem{MaNgWong}
 E. Ma, D. Ng, and G.-G.Wong, Z. Phys. {\bf C47} (1990) 431.


\bibitem{CKN}
 S. Dimopoulos and  G.F.~Giudice, Phys. Lett. {\bf B357} (1995) 573; \\
 A. Pomarol and D. Tommasini, Nucl. Phys. {\bf B466} (1996) 3;\\
 A.G. Cohen, D.B. Kaplan, and A.E. Nelson, Phys. Lett.
      {\bf B388} (1996) 588.


\bibitem{GHS}
 J.F. Gunion, H.E. Haber, and M. Sher, Nucl. Phys {\bf B306} (1988) 1.


\bibitem{LP}
 See appendix~B in P. Langacker and N. Polonsky, Phys. Rev.
                                     {\bf D50} (1994) 2199.


\bibitem{STABILITY}
 M. Claudson, L. J. Hall, and I. Hinchliffe, Nucl. Phys.
                             {\bf B228} (1983) 501;  \\
 T. C. Shen, Phys. Rev. {\bf D37} (1988) 3537;       \\
 A. Kusenko, P. Langacker, and G. Segre, Phys. Rev.
                             {\bf D54} (1996) 5824.



\bibitem{SARID}
 U. Sarid, Phys. Rev. {\bf D58} (1998) 085017.



\bibitem{THINWALL}
 S. Coleman, Phys. Rev. {\bf D15} (1977) 2929;          \\
 C. Callan and S. Coleman, Phys. Rev. {\bf D16} (1977) 1762.


\bibitem{THICKWALL}
 A. D. Linde, Nucl. Phys. {\bf B216} (1983) 421.



\bibitem{MESSENGER}
 M. Dine, Y. Nir, and Y. Shirman, Phys. Rev. {\bf D55} (1997) 1501; \\
 T. Han and R.J. Zhang, Phys. Lett. {\bf B428} (1998) 120.


\bibitem{PQ}
See, for example, 
A. Pomarol and M. Quiros, Phys. Lett. B {\bf 438} (1998) 255.


\bibitem{FF}
 G. Farrar and P. Fayet, Phys. Lett. {\bf B76} (1978) 575.


\bibitem{NOLOOP}
 J.L. Hewett, T. Takeuchi, and S. Thomas, in
  {\it Electroweak Symmetry Breaking and New Physics at the TeV Scale},
  ed. by T. Barklow {\it et al.} (World Scientific, Singapore, 1996),
  p. 548, hep-ph/9603391.


\bibitem{AMU}
 P. Fayet, in {\it Unification of the Fundamental Particle
  Interactions}, eds. S. Ferrara, J. Ellis and P. van Nieuwenhuizen
  (Plenum Press, New York, 1980) p. 587;                        \\
 J.A. Grifols and A. Mendez, Phys. Rev. {\bf D26} (1982) 1809; \\
 J. Ellis, J.S. Hagelin, and D.V. Nanopoulos,  Phys. Lett.
       {\bf B116} (1982) 283;                                    \\
 R. Barbieri and L. Maiani, Phys. Lett. {\bf B117} (1982) 203; \\
 D.A. Kosower, L.M. Krauss, and N. Sakai,  Phys. Lett.
       {\bf B133} (1983) 305;                                    \\
 T.C. Yuan, R. Arnowitt, A.H. Chamseddine, and P. Nath,
   Z. Phys. {\bf C26} (1984) 407;                               \\
 J. Lopez, D.V. Nanopoulos, and X. Wang, Phys. Rev. {\bf D49}
      (1991) 366;                                               \\
  For recent calculations and estimates of the anomalous
  muon magnetic moment in typical supersymmetric models see:\\
 U. Chattopadhyay and P. Nath, Phys. Rev. {\bf D53} (1996) 1648; \\
 T. Moroi, Phys. Rev. {\bf D53} (1996) 6565, E.: {\it ibid.} 
                    {\bf D56} (1997) 4424;      \\
 M. Carena, G.F. Giudice, and C. E. M. Wagner, Phys. Lett.
       {\bf B390 } (1997) 234.


\bibitem{MUONBOUND}
 J. Bailey {\it et al.}, Nucl. Phys. {\bf B150} (1979) 1.


\bibitem{MUONCALC}
 T. Kinoshita and W. Marciano, in {\it Quantum Electrodynamics},
  eds. T. Hasegawa {\it et al.} (Universal Academy Press, Tokyo, 1992),
  p. 419; \\
 A. Czarnecki and W. Marciano, hep-ph/9810512; \\
 G.~Degrassi and G.F.~Giudice, Phys. Rev. {\bf D58} (1998) 053007.


\bibitem{BROOKHAVEN}
 V. Hughes, in {\it Frontiers of High Energy Physics}, eds. 
  T. Hasagawa {\it et al.} (Universal Academy Press, Tokyo, 1992), p. 717.


\bibitem{oldneutron}
 J. Ellis, S. Ferrara, and D. Nanopoulos,
   Phys. Lett. B {\bf 114} (1982) 231;  \\
 W. Buchm\"uller and D. Wyler, Phys. Lett. B {\bf 121} (1983) 321;  \\
 J. Polchinski and M. Wise, Phys. Lett. B {\bf 125} (1983) 393.


\bibitem{phasebounds}
 For a discussion of bounds on supersymmetric phases implied
  by EDM measurements, see
 W. Fischler, S. Paban, and S. Thomas, Phys. Lett. B {\bf 289} (1992) 373.


\bibitem{MaNg}
 E. Ma and D. Ng, Phys. Rev. Lett. {\bf 65} (1990) 2499.


\bibitem{nedmbound}
 I .S. Altarev {\it et al.}, Yad. Fiz. {\bf 59} (1996) 1204,
  Phys. Atom. Nucl. {\bf 59} (1996) 1152.


\bibitem{chromo}
 R. Arnowitt, J. Lopez, and D. Nanopoulos,
  Phys. Rev. D {\bf 42} (1990) 2423;   \\
 R. Arnowitt, M. Duff, and K. Stelle,
  Phys. Rev. D {\bf 43} (1991) 3085;   \\
 Y. Kizikuri and N. Oshimo, Phys. Rev. D {\bf 45} (1992) 1806.


\bibitem{strange} 
 A. Zhitnitsky and I. Khriplovich,
  Yad. Fiz. {\bf 34} (1981) 167
  [Sov. J. Nucl. Phys. {\bf 34} (1981) 95];   \\
 V. M. Khatsymovsky, I. Khriplovich, and A. Zhitnitsky,
   Z. Phys. C {\bf 36} (1987) 455;     \\
 X.-G. He, B. Mckellar, and S. Pakvasa,  
   Phys. Lett. B {\bf 254} (1991) 231; \\
 J. Ellis and R. Flores, Phys. Lett. B {\bf 377} (1996) 83.


\bibitem{GGG} 
 S. Weinberg, Phys. Rev. Lett. {\bf 63} (1989) 2333;  \\
 D. Chang, J. Kephart, W.-Y. Keung, and T. C. Yuan, Phys. Lett. B
   {\bf 68} (1992) 439.


\bibitem{GGGRG}
 A. Morozov, Yad. Fiz. {\bf 40} (1984) 788
  [Sov. J. Nucl. Phys. {\bf 40} (1984) 505];   \\
 E. Braaten, C. S. Li, and T. C. Yuan,
  Phys. Rev. Lett. {\bf 64} (1990) 1709,   
  Phys. Rev. D {\bf 42} (1990) 276;            \\ 
 D. Chang, W.-Y. Keung, C. S. Li, and T. C. Yuan,
  Phys. Lett. B {\bf 241} (1990) 589.


\bibitem{eedmbound}
 E. Commins, S. Ross, D. DeMille, and B. Regan,
  Phys. Rev. A {\bf 50} (1994) 2960.


\bibitem{MUON}
 V. Barger, M.S. Berger, J.F. Gunion, and T. Han, Phys. Rev. Lett.
  {\bf 75} (1995) 1462; Phys. Rep. {\bf 286} (1997) 1; \\
 V. Barger, talk presented at the {\it 4th International
  Conference on the Physics Potential and Development
  of $\mu^+ \mu^-$ Colliders}, Dec. 1997, hep-ph/9803480.


\bibitem{HG}
 J.F. Gunion and H.E. Haber, Nucl. Phys. {\bf B272} (1986) 1;
   E: {\it ibid.} {\bf 402} (1993) 567 and University of California,
   Davis preprint, UCD-92-31, 1992, hep--ph/9301205


\bibitem{HHGUIDE}
 J. Gunion, H. Haber, G. Kane, and S. Dawson,
  {\it The Higgs Hunter's Guide} (Addison--Wesley, New York, 1990).


\bibitem{GUNION}
 J.F. Gunion, talk presented at the {\it 4th International Conference
  on the Physics Potential and Development of $\mu^+ \mu^-$ Colliders},
  Dec. 1997, hep-ph/9804358.


\bibitem{LEE}
 I. Lee, Nucl. Phys. {\bf B246} (1984) 120.


\bibitem{OSC}
 M. Hirsch, H.V. Klapdor--Kleingrothaus, and S.G. Kovalenko,
             Phys. Lett. {\bf B398} (1997) 311;  \\
 Y. Grossman and H.E. Haber, Phys. Rev. Lett. 78 (1997) 3438.


\bibitem{CFUNC}
 U. Nierste, D. M\"uller, and M. B\"ohm, Z. Phys. {\bf C57} (1993) 605;\\
 G.~'t~Hooft and M. Veltman, Nucl. Phys. {\bf B153} (1979) 365.


\end{thebibliography}
\end{document}